\documentclass[preprint,aps,showpacs,preprintnumbers,amsmath,amssymb,superscriptaddress]{revtex4}

\usepackage{graphicx}% Include figure files
\usepackage{color}% color
\usepackage{dcolumn}% Align table columns on decimal point
\usepackage{bm}% bold math

\let\rr\relax  %% should actually remove \rr commands from files

\newcommand\as{\alpha_S}

\def\bit{\begin{itemize}}
\def\eit{\end{itemize}}

\def\what{\widehat}
\def\fbi{~{\rm fb}^{-1}}
\def\call{{\cal L}}
\def\lyear{L_{\rm year}}
\def\sig{\sigma}
\def\gamhsmtot{\Gamma_{\hsm}^{\rm tot}}
\def\br{{\rm BF}}
\def\gamhtot{\Gamma_{\h}^{\rm tot}}

\def\hl{h^0}
\def\ha{A^0}
\def\hh{H^0}

\def\mha{m_{\ha}}
\def\mhh{m_{\hh}}

\def\h{h}
\def\mh{m_{\h}}
\def\epem{e^+e^-}
\def\mupmum{\mu^+\mu^-}
\def\anti{\overline}
\def\tanb{\tan\beta}
\def\what{\widehat}
\def\rts{\sqrt s}
\def\hsm{h_{SM}}
\def\mhsm{m_{\hsm}}

\def\gev{~{\rm GeV}}

\def\srts{\sigma_{\!\!\!\sqrt s}^{\vphantom y}}
\def\lsim{\alt}
\def\gsim{\agt}
\def\beq{\begin{equation}}
\def\eeq{\end{equation}}
\def\bea{\begin{eqnarray}}
\def\eea{\end{eqnarray}}

\begin{document}

\title{
\textbf{Recent Progress in Neutrino Factory and Muon Collider 
Research within the Muon Collaboration}
}
\author{Mohammad~M.~Alsharo'a}\affiliation{Illinois Institute of Technology, Physics Div., Chicago, IL 60616}
\author{Charles~M.~Ankenbrandt}\affiliation{Fermi National Accelerator Laboratory, P. O. Box 500, Batavia, IL 60510}
\author{Muzaffer~Atac}\affiliation{Fermi National Accelerator Laboratory, P. O. Box 500, Batavia, IL 60510}
\author{Bruno~R.~Autin}\affiliation{CERN, 1211 Geneva 23, Switzerland}
\author{Valeri~I.~Balbekov}\affiliation{Fermi National Accelerator Laboratory, P. O. Box 500, Batavia, IL 60510}
\author{Vernon~D.~Barger}\affiliation{Department of Physics, University of Wisconsin, Madison, WI 53706}
\author{Odette~Benary}\affiliation{Tel~Aviv~University, Tel Aviv 69978, Israel}
\author{J.~Roger~J.~Bennett}\affiliation{Rutherford~Appleton~Laboratory, Chilton, Didcot, United Kingdom}
\author{Michael~S.~Berger}\affiliation{Indiana University, Physics Department, Bloomington, IN 47405}
\author{J.~Scott~Berg}\affiliation{Brookhaven National Laboratory, Upton, NY 11973}
% ...
%\author{Max~Zolotorev}\affiliation{Lawrence Berkeley National Laboratory, 1 Cyclotron Rd., Berkeley, CA 94720}

\date{\today}

\begin{abstract}
We describe  the status of our effort to realize a first neutrino
factory and the progress made in understanding the problems associated
with the collection and cooling of muons towards that end.  We
summarize  the physics that can be done with neutrino factories as well
as  with intense cold beams of muons.  The
physics potential of muon colliders is reviewed, both as Higgs
Factories and compact high energy lepton colliders.  The status and timescale 
of our research and development effort is reviewed as well as  the
latest designs in cooling channels including the promise of ring coolers in
achieving longitudinal and transverse cooling simultaneously. We detail the
efforts being made to mount an international cooling experiment to
demonstrate the ionization cooling of muons.
\end{abstract}

\thispagestyle{empty}
\maketitle
%\tableofcontents
%\newpage  % TURN THIS ON

%\listoffigures    % TURN THIS ON
%\newpage % TURN THIS ON

%\listoftables   %TURN THIS ON
%\newpage% TURN THIS ON

%\input introduction 
 %% introduction.tex, Mike's new version, received Jan.9

\section{Introduction\label{intro}}

Recent results from the SNO collaboration~\cite{snolatest} coupled with data
from the SuperK collaboration~\cite{superk} have provided convincing
evidence that neutrinos oscillate and that they very likely do so among the
three known neutrino species. Experiments currently under way or planned in
the near future will shed further light on the nature of these mixings among
neutrino species and the magnitudes of the mass differences between them.
Neutrino oscillations and the implied non-zero masses and mixings represent
the first experimental evidence of effects beyond the Standard Model, and as
such are worthy of vigorous scientific study.

This document indicates our progress along a path toward establishing an
ongoing program of research in accelerator and experimental physics based on
muon beams, and neutrino beams derived therefrom, that can proceed in an
incremental fashion. At each step, new physics vistas open, leading
eventually to a Neutrino Factory and possibly a Muon Collider. This concept
has aroused significant interest throughout the world scientific community.
In the U.S., a formal collaboration of some 110 scientists, the Neutrino
Factory and Muon Collider Collaboration, 
also known as the Muon Collaboration (MC)~\cite{EPP:collaboration}, has
undertaken the study of designing a Neutrino Factory, along with R\&D
activities in support of a Muon Collider design. The MC comprises three
sponsoring national laboratories (BNL, FNAL, LBNL) along with groups from
other U.S. national laboratories and universities and individual members
from non-U.S. institutions.

One of the first steps toward a Neutrino Factory is a proton driver that can
be used to provide intense beams of conventional neutrinos in addition to
providing the intense source of low energy muons (from pion decay) that must
first be ``cooled'' before being accelerated and stored. 
Our vision is that
while a proton driver is being constructed, R\&D on collecting and cooling
muons would continue. A source of intense cold muons could be immediately
used for physics measurements, such as determining the electric and magnetic
dipole moments of the muon to higher precision, muonium-antimuonium
oscillations, muon spin rotation experiments and rare muon decays. 
Once the capability of cooling and
accelerating muons is fully developed, a storage ring for such muons would
serve as the first Neutrino Factory. Its specific beam energy and its
distance from the long-baseline experiment will be chosen using the
knowledge of neutrino oscillation parameters gleaned from the present
generation of solar and accelerator experiments (Homestake, Kamiokande,
SuperKamiokande, SAGE, GALLEX, K2K, SNO), the next generation experiments
(MiniBooNE, MINOS, CNGS, KamLAND, Borexino), and the high-intensity
conventional beam experiments that would already have taken place.

A Neutrino Factory provides both $\nu _{\mu }$ and $\anti\nu _{e}$ beams of
equal intensity from a stored $\mu ^{-}$ beam, and their charge-conjugate
beams for a stored $\mu ^{+}$ beam. Beams from a Neutrino Factory are
intense compared with today's neutrino sources. In addition, they have
smaller divergence than conventional neutrino beams of comparable energy.
These properties permit the study of non-oscillation physics at near
detectors, and the measurement of structure functions and associated
parameters in non-oscillation physics, to unprecedented accuracy. Likewise,
they permit long-baseline experiments that can determine oscillation
parameters to unprecedented accuracy.

Depending on the value of the parameter $\sin ^{2}2\theta _{13}$ in the
three-neutrino oscillation formalism, the oscillation $\nu _{e}\rightarrow
\nu _{\mu }$ is expected to be measurable. By comparing the rates for this
channel with its charge-conjugate channel $\anti\nu _{e}\rightarrow \anti\nu
_{\mu }$, the sign of the leading mass difference in neutrinos, $\delta
m_{32}^{2}$, can be determined by observing the passage through matter of
the neutrinos in a long-baseline experiment. Such experiments can also shed
light on the CP-violating phase, $\delta $, in the lepton mixing matrix and
enable the study of CP violation in the lepton sector. (It is known that CP
violation in the quark sector is insufficient to explain the baryon
asymmetry of the Universe; lepton sector CP violation possibly played a
crucial role in creating this asymmetry during the initial phases of the Big
Bang.)

While the Neutrino Factory is being constructed, R\&D aimed at making the
Muon Collider a reality would be performed.
The Muon Collider will require muon beams that are more intensely cooled 
and have generally more challenging properties than those for a 
Neutrino Factory, so the latter forms a practical goal en route to the former.
A Muon Collider, if realized,
provides a tool to explore Higgs-like objects by direct $s$-channel fusion,
much as LEP explored the $Z$. It also provides a potential means to reach
higher energies (3--4~TeV in the center of mass) using relatively compact
collider rings.

\subsection{History}

The concept of a Muon Collider was first proposed by Budker~\cite
{PREFACE:budker} and by Skrinsky~\cite{PREFACE:skrinsky} in the 60s and
early 70s. However, additional  substance to the concept had to wait 
until the idea
of ionization cooling was developed by Skrinsky and Parkhomchuk~\cite
{INTRO:ref3}. The ionization cooling approach was expanded by Neuffer~\cite
{INTRO:ref4} and then by Palmer~\cite{PREFACE:palmer}, whose work led to the
formation of the Neutrino Factory and Muon Collider Collaboration (MC)~\cite
{EPP:collaboration} in 1995\footnote{
A good summary of the Muon Collider concept can be found in the Status
Report of 1999~\cite{INTRO:ref5}; an earlier document~\cite{INTRO:ref6},
prepared for Snowmass-1996, is also useful reading. MC Notes prepared by the
Collaboration are available on the web~\cite{INTRO:ref11}}.

The concept of a neutrino source based on a pion storage ring was originally
considered by Koshkarev~\cite{INTRO:ref7}. However, the intensity of the
muons created within the ring from pion decay was too low to provide a
useful neutrino source. The Muon Collider concept provided a way to produce
a very intense muon source. The physics potential of neutrino beams produced
by high-intensity  muon storage rings was briefly investigated in 1994 by King~\cite{king94}and in more detail by Geer in 1997 at a Fermilab
workshop~\cite{rajageer,geer} where it became evident that the neutrino
beams produced by muon storage rings needed for the Muon Collider were
exciting in their own right. As a result, the MC realized that a Neutrino
Factory could be an important first step toward a Muon Collider. With this
in mind, the MC has shifted its primary emphasis toward the issues relevant
to a Neutrino Factory. The Neutrino Factory concept quickly captured the
imagination of the particle physics community, driven in large part by the
exciting atmospheric neutrino deficit results from the SuperKamiokande
experiment. The utility of non-oscillation neutrino physics from neutrinos 
produced by muon storage rings has been studied in detail from 1997 
onwards~\cite{non-osc}.

There is also considerable international activity on Neutrino
Factories, with international conferences held at Lyon in
1999~\cite{INTRO:ref13}, Monterey in 2000~\cite{INTRO:ref14}, Tsukuba
in 2001~\cite{nufact01}, London in 2002~\cite{nufact02} and another
planned in New York in 2003~\cite{nufact03}. There are also efforts in Europe~\cite{europenf} and Japan~\cite{japannf} to study different approaches to realizing the neutrino factory.
Recently a proposal has
been submitted to perform an International Muon Ionization Cooling
Experiment (MICE) to the Rutherford Appleton
Laboratory~\cite{mice-prop}.

\subsection{Feasibility Studies}

Complementing the MC experimental and theoretical R\&D program, which
includes work on targetry, cooling, rf hardware (both normal conducting and
superconducting), high-field solenoids, liquid hydrogen absorber design, 
muon scattering experiments, theory,
simulations, parameter studies, and emittance exchange~\cite{INTRO:ref12},
the Collaboration has participated in several paper studies of a complete
Neutrino Factory design.

In the fall of 1999, Fermilab, with help from the MC, undertook a
Feasibility Study (``Study-I'') of an entry-level Neutrino Factory~\cite
{INTRO:ref1}. Study-I showed that the evolution of the Fermilab accelerator
complex into a Neutrino Factory was clearly possible. The performance
reached in Study-I, characterized in terms of the number of 50-GeV muon
decays aimed at a detector located 3000 km away from the muon storage ring,
was $N$ = 2 $\times $ 10$^{19}$ decays per ``Snowmass year'' (10$^{7}$ s)
per MW of protons on target.

Simultaneously, Fermilab launched a study of the physics that might be
addressed by such a facility~\cite{INTRO:ref9} and, more recently, initiated
a study to compare the physics reach of a Neutrino Factory with that of
conventional neutrino beams~\cite{superbeams} powered by a high-intensity
proton driver (referred to as ``superbeams''). As will be described later in
this paper, a steady and diverse physics program will result from following
the evolutionary path from a superbeam to a full-fledged Neutrino Factory.

Subsequently, BNL organized a follow-on study 
(``Study-II'')~\cite{EPP:studyii} on a
high-performance Neutrino Factory, again in collaboration with the MC.
Study-II demonstrated that BNL was likewise a suitable site for a Neutrino
Factory. Based on the improvements in Study-II, the number of 20-GeV muon
decays aimed at a detector located 3000 km away from the muon storage ring,
was $N$ = 1.2 $\times $ 10$^{20}$ decays per Snowmass year per MW of protons
on target. Thus, with an upgraded 4 MW proton driver, the muon decay
intensity would increase to 4.8 $\times $ $10^{20}$ decays per Snowmass
year. (R\&D to develop a target capable of handling this beam power would be
needed.) Though these numbers of neutrinos are potentially available for
experiments, in the current storage-ring design the angular divergence at
both ends of the production straight section is higher than desirable for
the physics program. In any case, we anticipate that storage-ring designs
are feasible that would allow 30--40\% of the muon decays to provide useful
neutrinos.

\begin{figure}[tbh]
\centerline{\includegraphics[width=4.0in]{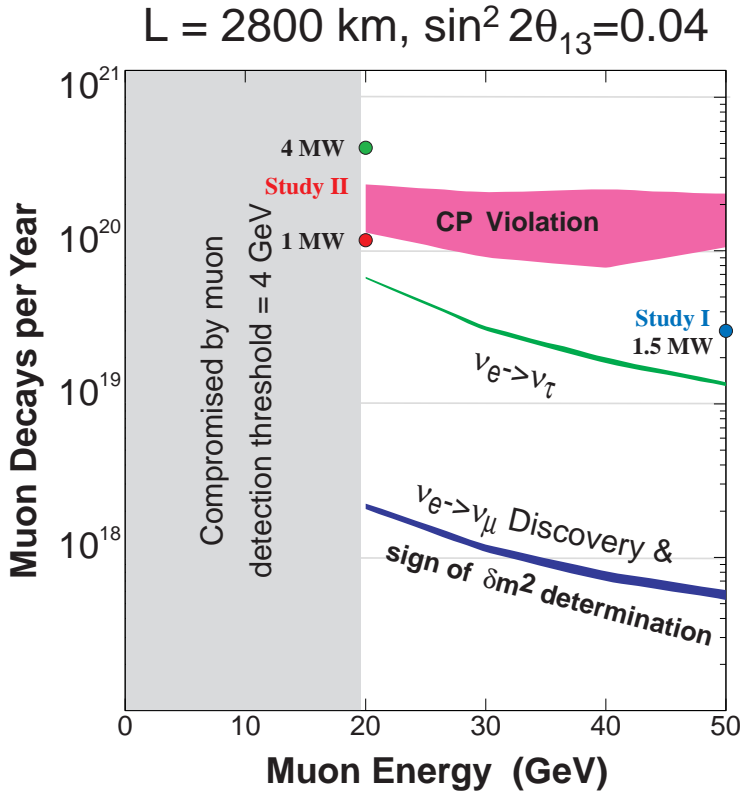}}
\caption[Muon decays in a straight section \textit{vs.} muon energy]
{(Color) Muon decays in a straight section per $10^{7}\,$s \textit{vs.} muon energy, with
fluxes required for different physics searches assuming a 50~kT detector.
Simulated performance of the two studies is indicated.}
\label{studycomp}
\end{figure}

Both Study-I and -II are site specific in that each has a few site-dependent
aspects; otherwise, they are generic. In particular, Study-I assumed a new
Fermilab booster to achieve its beam intensities and an underground storage
ring. Study-II assumed BNL site-specific proton driver specifications
corresponding to an upgrade of the 24-GeV AGS complex and a BNL-specific
layout of the storage ring, which is housed in an above-ground berm to avoid
penetrating the local water table. The primary substantive difference
between the two studies is that Study-II aimed at a lower muon energy (20
GeV), but higher intensity (for physics reach) than Study-I. Taking the two
Feasibility Studies together, we conclude that a high-performance Neutrino
Factory could easily be sited at either BNL or Fermilab. Figure \ref
{studycomp} shows a comparison of the performance of the Neutrino Factory
designs in Study-I and Study-II~\cite{INTRO:ref9} with the physics
requirements.

To put the above performance figures in context, it is important to note
that a $\mu ^{+}$ storage ring with an average neutrino energy of 15~GeV and 
$2\times 10^{20}$ useful muon decays would yield (in the absence of
oscillations) $\approx $30,000 charged-current events in the $\nu _{e}$
channel per kiloton-year in a detector located 732~km away. In comparison, a
1.6~MW superbeam~\cite{superbeams} from the Fermilab Main Injector with an
average neutrino energy of 15~GeV would yield only $\approx $13,000 $\nu_\mu$ charged-current events per kiloton-year. In addition to having
lower intensity than a Neutrino Factory beam, a superbeam would have
significant $\nu_e$ contamination, which will be the major background in 
$\nu_\mu\rightarrow \nu_e$ appearance searches. That is, it will be
much easier to detect the oscillation $\nu_e\rightarrow \nu_\mu$ from
a muon storage ring neutrino beam than to detect the oscillation $\nu_\mu
\rightarrow \nu_e$ from a conventional neutrino beam, because the
electron final state from the conventional beam has significant background
contribution from $\pi^{0}$'s produced in the events.

\subsection{Neutrino Factory Description\label{NFsection}}

\begin{figure}[tbh]
\centerline{\includegraphics[width=4.0in,angle=-90]{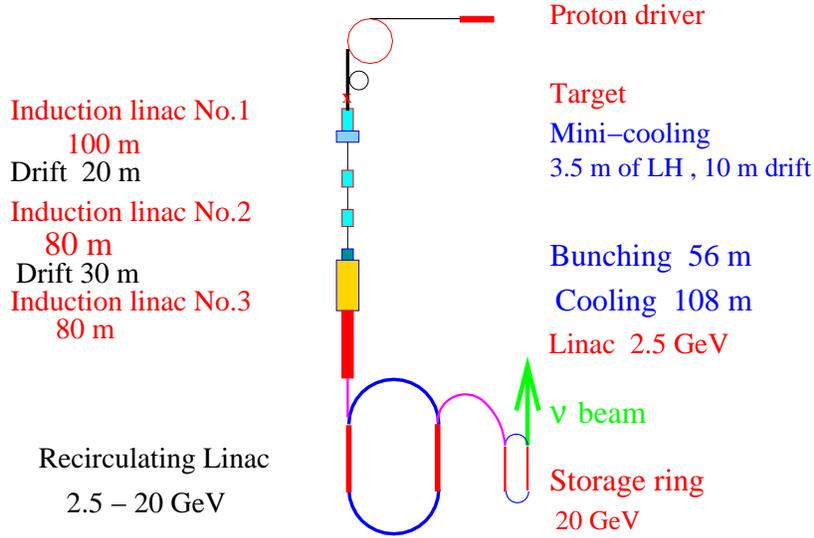}}
\caption[Schematic of the Neutrino Factory Study-II version]{(Color)Schematic of
the Neutrino Factory Study-II version.}
\label{nufact-scheme-bnl}
\end{figure}

The muons we use result from decays of pions produced when an intense proton
beam bombards a high-power production target. The target and downstream
transport channel are surrounded by superconducting solenoids to contain the
pions and muons, which are produced with a larger spread of transverse and
longitudinal momenta than can be conveniently transported through an
acceleration system. To prepare a beam suitable for subsequent acceleration,
we first perform a ``phase rotation,'' during which the initial large energy
spread and small time spread are interchanged using induction linacs. Next,
to reduce the transverse momentum spread, the resulting long bunch, with an
average momentum of about 250 MeV/c, is bunched into a 201.25-MHz bunch
train and sent through an ionization cooling channel consisting of LH$_{2}$
energy absorbers interspersed with rf cavities to replenish the energy lost
in the absorbers. The resulting beam is then accelerated to its final energy
using a superconducting linac to make the beam relativistic, followed by one
or more recirculating linear accelerators (RLAs). Finally, the muons are
stored in a racetrack-shaped ring with one long straight section aimed at a
detector located at a distance of roughly 3000 km. A schematic layout is
shown in Fig.~\ref{nufact-scheme-bnl}.

\subsection{Detector}

Specifications for the long-baseline Neutrino Factory detector are rather
typical for an accelerator-based neutrino experiment. However, because of
the need to maintain a high neutrino rate at these long distances ($\approx $%
3000 km), the detectors considered here are 3--10 times more massive than
those in current neutrino experiments.

Several detector options could be considered for the far detector:

\begin{itemize}
\item  A 50 kton steel--scintillator--proportional-drift-tube (PDT) detector

\item  A large water-Cherenkov detector, similar to SuperKamiokande but with
either a magnetized water volume or toroids separating smaller water tanks~\cite{DET:uno}.

\item  A massive liquid-argon magnetized detector~\cite{landd}.
\end{itemize}

For the near detector, a compact liquid-argon TPC (similar to the ICARUS
detector~\cite{ICARUS}) could be used. An experiment with a relatively thin
Pb target (1~$L_{rad}$), followed by a standard fixed-target spectrometer
could also be considered.

\subsection{Staging Scenario}

If desired by the particle physics community, a fast-track plan leading
directly to a Neutrino Factory could be executed. On the other hand, the
Neutrino Factory offers the distinct advantage that it can be built in
stages. This could satisfy both programmatic and cost constraints by
allowing an ongoing physics program while reducing the annual construction
funding needs. Depending on the results of our technical studies and the
results of ongoing searches for the Higgs boson, it is hoped that the
Neutrino Factory is really the penultimate stage, to be followed later by a
Muon Collider (e.g., a Higgs Factory). Such a collider offers the potential
of bringing the energy frontier in particle physics within reach of a
moderate-sized machine. Possible stages for the evolution of a muon beam
facility are described in Section~\ref{StagingOps}.

\subsection{R\&D Program\label{RDprog}}

Successful construction of a muon storage ring to provide a copious source
of neutrinos requires development of many novel approaches; construction of
a high-luminosity Muon Collider requires even more. It was clear from the
outset that the breadth of R\&D issues to be dealt with would be beyond the
resources available at any single national laboratory or university. For
this reason, in 1995, interested members of the high-energy physics and
accelerator physics communities formed the MC to coordinate the required
R\&D efforts nationally. The task of the MC is to define and carry out R\&D
needed to assess the technical feasibility of constructing initially a muon
storage ring that will provide intense neutrino beams aimed at detectors
located many thousands of kilometers from the accelerator site, and
ultimately a $\mu ^{+}\mu ^{-}$ collider that will carry out fundamental
experiments at the energy frontier in high-energy physics.

The MC also serves to coordinate muon-related R\&D activities of the
NSF-sponsored University Consortium (UC) and the state-sponsored Illinois
Consortium for Accelerator Research (ICAR), and is the focal point for
defining the needs of muon-related R\&D to the managements of the sponsoring
national laboratories and to the funding agencies (both DOE and NSF). As
already noted, though the MC was formed initially to carry out R\&D that
might lead eventually to the construction of a Muon Collider, more recently
its focus has shifted mainly, but not exclusively, to a Neutrino Factory.

The MC maintains close contact with parallel R\&D efforts under way in
Europe (centered at CERN) and in Japan (centered at KEK). Through its
international members, the MC also fosters coordination of the international
muon-beam R\&D effort. Two major initiatives, a Targetry Experiment (E951)
in operation at BNL and a Muon Cooling R\&D program (MUCOOL), have been
launched by the MC. In addition, the Collaboration, working in conjunction
with the UC and ICAR in some areas, coordinates substantial efforts in
accelerator physics and component R\&D to define and assess parameters for
feasible designs of muon-beam facilities.

\subsection{Outline of Report}

In what follows, we give the motivation and a scenario for a staged approach
to constructing a Neutrino Factory and eventually a Muon Collider. 
Section~\ref{physics} discusses the physics opportunities, starting from
conventional ``superbeams'' and going to cold muon beams, then a Neutrino
Factory with its near and far detectors, and finally a Muon Collider. In
Section~\ref{neufact}, we describe the components of a Neutrino Factory,
based on the Study-II design, and indicate a scientifically productive
staged path for reaching it. Section~\ref{higgsfact} covers our present
concept of an entry-level Higgs Factory Muon Collider. In support of the
construction of a Neutrino Factory, an R\&D program is already under way to
address various technical issues. A description of the status and plans for
this program is presented in Section~\ref{r_and_d}. Section~\ref{mice}
describes current thinking about a cooling demonstration experiment that
would be carried out as an international effort. Finally, in Section~\ref{Summary} we provide a brief summary of our work.

\section{Physics Motivation}
\label{physics}
In this Section we cover the physics potential of the Neutrino Factory
accelerator complex, which includes superbeams of conventional
neutrinos that are possible using  the proton driver needed for the
factory,  and intense beams of cold
muons that become available once the muon cooling and collection
systems for the factory are in place. Once the cold muons are
accelerated and stored in the muon storage ring, we realize the full
potential of the factory in both neutrino oscillation and
non-oscillation physics.
Cooling muons will be a learning experience. We hope that the
knowledge gained in constructing a Neutrino Factory can be used to
cool muons sufficiently to produce the first muon collider operating
as a Higgs factory. We examine the physics capabilities of  such a
collider, which if realized, will invariably lead to higher energy
muon colliders with exciting physics opportunities.
\subsection{Neutrino Oscillation Physics}
Here we discuss~\cite{study2} the current evidence for neutrino 
oscillations, and hence
neutrino masses and lepton mixing, from solar and atmospheric data.  A review
is given of some theoretical background including models for neutrino masses
and relevant formulas for neutrino oscillation transitions.  We next mention
the near-term and mid-term experiments in this area and comment on what they
hope to measure.  We then discuss the physics potential of a muon storage ring
as a Neutrino Factory in the long term. 
\subsubsection{Evidence for Neutrino Oscillations} 
In a modern theoretical context, one generally expects nonzero neutrino masses
and associated lepton mixing.  Experimentally, there has been accumulating
evidence for such masses and mixing.  All solar neutrino experiments
(Homestake, Kamiokande, SuperKamiokande, SAGE, GALLEX and SNO) 
show a significant
deficit in the neutrino fluxes coming from the Sun~\cite{sol}. This deficit
can be explained by oscillations of the $\nu_e$'s into other weak
eigenstate(s), with $\Delta m^2_{\rm sol}$ of the order $10^{-5}$ eV$^2$ for
solutions involving the Mikheyev-Smirnov-Wolfenstein (MSW) resonant matter
oscillations~\cite{wolf}--\cite{ms} 
or of the order of $10^{-10}\,$eV$^2$ for vacuum
oscillations~\cite{just-so}.  Accounting for the data with vacuum oscillations (VO) requires
almost maximal mixing.  The MSW solutions include one for small mixing angle
(SMA) and one for large mixing angle (LMA).
Another piece of evidence for neutrino oscillations is the atmospheric neutrino
anomaly, observed by Kamiokande~\cite{kam}, IMB~\cite{imb}, SuperKamiokande~\cite{sk} with the highest statistics, and by Soudan~\cite{soudan2} and MACRO~\cite{macro}.  These data can be fit by the inference of $\nu_{\mu} \rightarrow
\nu_x$ oscillations with $\Delta m^2_{\rm atm}\sim 3 \times 10^{-3}\rm\,eV^2$~\cite{sk} and maximal mixing $\sin^2 2 \theta_{\rm atm} = 1$.  The identification
$\nu_x = \nu_\tau$ is preferred over $\nu_x=\nu_{sterile}$, and the
identification $\nu_x=\nu_e$ is excluded by both the Superkamiokande data and
the Chooz experiment~\cite{chooz}.
In addition, the LSND experiment~\cite{lsnd} has reported 
$\bar\nu_\mu \to \bar \nu_e$ and $\nu_{\mu} \to \nu_e$ oscillations with
$\Delta m^2_{\rm LSND} \sim 0.1\mbox{--}1\rm~eV^2$ and a range of possible mixing angles.
This result is not confirmed, but also not completely ruled out, by a similar
experiment, KARMEN~\cite{karmen}.  The miniBOONE experiment at Fermilab is
designed to resolve this issue, as discussed below.
If one were to try to fit all of these experiments, then, since they involve
three quite different values of $\Delta m^2_{ij}=m(\nu_i)^2-m(\nu_j)^2$, which
could not satisfy the identity for three neutrino species, 
\begin{equation}
\Delta m^2_{32} + \Delta m^2_{21} + \Delta m^2_{13}=0 \,,
\label{mident}
\end{equation}
it would follow that one would have to introduce at least one further
 neutrino.
Since it is known from the measurement of the $Z$ width that there are
only three leptonic weak doublets with associated light neutrinos, it
follows that such further neutrino weak eigenstate(s) would have to be
electroweak singlet(s) (``sterile'' neutrinos).  Because the LSND
experiment has not been confirmed by the KARMEN experiment, we choose
here to use only the (confirmed) solar and atmospheric neutrino data
in our analysis, and hence to work in the context of three active
neutrino weak eigenstates.

\subsubsection{Neutrino Oscillation Formalism} 
     In this theoretical context, consistent with solar and atmospheric data,
there are three electroweak-doublet neutrinos and the neutrino mixing 
matrix is described by
\begin{equation}
U=\left(
\begin{array}{ccc}
c_{12} c_{13}&c_{13} s_{12}&s_{13} e^{-i\delta}\\
-c_{23}s_{12}-s_{13}s_{23}c_{12}e^{i\delta}
&c_{12}c_{23}-s_{12}s_{13}s_{23}e^{i\delta}&c_{13}s_{23}\\
s_{12}s_{23}-s_{13}c_{12}c_{23}e^{i\delta}
&-s_{23}c_{12}-s_{12}c_{23}s_{13}e^{i\delta}&c_{13}c_{23}
\end{array}
\right)K^\prime \,,
\end{equation}
where $c_{ij}=\cos\theta_{ij}$, $s_{ij}=\sin\theta_{ij}$, and $K^\prime =
{\rm diag}(1,e^{i\phi_1},e^{i\phi_2})$.  The phases $\phi_1$ and $\phi_2$ do not affect neutrino oscillations.  Thus, in this framework, the neutrino mixing
relevant for neutrino oscillations depends on the four angles $\theta_{12}$,
$\theta_{13}$, $\theta_{23}$, and $\delta$, and on two independent differences
of squared masses, $\Delta m^2_{\rm atm}$, which is $\Delta m^2_{32} =
m(\nu_3)^2-m(\nu_2)^2$ in the favored fit, and $\Delta m^2_{\rm sol}$, which may be taken to be $\Delta m^2_{21}=m(\nu_2)^2- m(\nu_1)^2$.  Note that these
$\Delta m^2$ quantities involve both magnitude and sign; although in a two-species neutrino
oscillation in vacuum the sign does not enter, in the 
three-species-oscillation, which includes  both matter effects and $CP$ violation,
the signs of the $\Delta m^2$ quantities enter and can, in principle, be
measured.
For our later discussion it will be useful to record the formulas for the
various neutrino-oscillation transitions.  In the absence of any matter effect, the probability that a (relativistic) weak neutrino eigenstate
$\nu_a$ becomes $\nu_b$ after propagating a distance $L$ is
\begin{eqnarray}
P(\nu_a \to \nu_b) &=& \delta_{ab} - 4 \sum_{i>j=1}^3
Re(K_{ab,ij}) \sin^2  \Bigl ( \frac{\Delta m_{ij}^2 L}{4E} \Bigr ) +
\nonumber\\
&& {}+ 4 \sum_{i>j=1}^3 Im(K_{ab,ij})
 \sin \Bigl ( \frac{\Delta m_{ij}^2 L}{4E} \Bigr )
\cos \Bigl ( \frac{\Delta m_{ij}^2 L}{4E} \Bigr )
\label{pab}
\end{eqnarray}
where
\begin{equation}
K_{ab,ij} = U_{ai}U^*_{bi}U^*_{aj} U_{bj}
\label{k}
\end{equation}
and
\begin{equation}
\Delta m_{ij}^2 = m(\nu_i)^2-m(\nu_j)^2 \,.
\label{delta}
\end{equation}
Recall that in vacuum, $CPT$ invariance implies
$P(\bar\nu_b \to \bar\nu_a)=P(\nu_a \to \nu_b)$ and hence, for $b=a$,
$P(\bar\nu_a \to \bar\nu_a) = P(\nu_a \to \nu_a)$.  For the
CP-transformed reaction $\bar\nu_a \to \bar\nu_b$ and the T-reversed
reaction $\nu_b \to \nu_a$, the transition probabilities are given by the
right-hand side of (\ref{pab}) with the sign of the imaginary term reversed.
(Below we shall assume $CPT$ invariance, so that $CP$ violation is equivalent to $T$ violation.) 
In most cases there is only one mass scale
relevant for long-baseline neutrino oscillations, $\Delta m^2_{\rm atm} \sim {\rm few} \times 10^{-3}\rm\,eV^2$, and one possible neutrino mass spectrum is the hierarchical one 
\begin{equation} 
\Delta m^2_{21}
= \Delta m^2_{\rm sol} \ll \Delta m^2_{31} \approx \Delta m^2_{32}=\Delta m^2_{\rm atm} \,.
\label{hierarchy}
\end{equation}
In this case, $CP$ $(T)$ violation effects may be negligibly small, so that in
vacuum
\begin{equation}
P(\bar\nu_a \to \bar\nu_b) = P(\nu_a \to \nu_b)
\label{pcp}
\end{equation}
and
\begin{equation}
P(\nu_b \to \nu_a) = P(\nu_a \to \nu_b) \,.
\label{pt}
\end{equation}
In the absence of $T$ violation, the second equality (\ref{pt}) would still hold in uniform matter, but even in the absence of $CP$ violation, the first equality
(\ref{pcp}) would not hold.  With the hierarchy (\ref{hierarchy}), the
expressions for the specific oscillation transitions are
\begin{eqnarray}
P(\nu_\mu \to \nu_\tau) & = & 4|U_{33}|^2|U_{23}|^2
\sin^2 \Bigl ( \frac{\Delta m^2_{\rm atm}L}{4E} \Bigr ) \nonumber\\
& = & \sin^2(2\theta_{23})\cos^4(\theta_{13})
\sin^2 \Bigl (\frac{\Delta m^2_{\rm atm}L}{4E} \Bigr ) \,,
\label{pnumunutau}
\end{eqnarray}
\begin{eqnarray}
P(\nu_e \to \nu_\mu) & = & 4|U_{13}|^2 |U_{23}|^2
\sin^2 \Bigl ( \frac{\Delta m^2_{\rm atm}L}{4E} \Bigr ) \nonumber\\
& = & \sin^2(2\theta_{13})\sin^2(\theta_{23})
\sin^2 \Bigl (\frac{\Delta m^2_{\rm atm}L}{4E} \Bigr ) \,,
\label{pnuenumu}
\end{eqnarray}
\begin{eqnarray}
P(\nu_e \to \nu_\tau) & = & 4|U_{33}|^2 |U_{13}|^2
\sin^2 \Bigl ( \frac{\Delta m^2_{\rm atm}L}{4E} \Bigr ) \nonumber\\
& = & \sin^2(2\theta_{13})\cos^2(\theta_{23})
\sin^2 \Bigl (\frac{\Delta m^2_{\rm atm}L}{4E} \Bigr ) \,.
\label{pnuenutau}
\end{eqnarray}
In neutrino oscillation searches using reactor antineutrinos,
i.e,\ tests of $\bar\nu_e \to \bar\nu_e$, the two-species mixing hypothesis used to fit the data is
\begin{eqnarray}
P(\nu_e \to \nu_e) & = & 1 - \sum_x P(\nu_e \to \nu_x) \nonumber\\
                   & = & 1 - \sin^2(2\theta_{\rm reactor})
\sin^2 \Bigl (\frac{\Delta m^2_{\rm reactor}L}{4E} \Bigr ) \,,
\label{preactor}
\end{eqnarray}
where $\Delta m^2_{\rm reactor}$ is the squared mass difference relevant for
$\bar\nu_e \to \bar\nu_x$.  In particular, in the upper range of values of
$\Delta m^2_{\rm atm}$, since the transitions $\bar\nu_e \to \bar\nu_\mu$ and
$\bar\nu_e \to \bar\nu_\tau$ contribute to $\bar\nu_e$ disappearance, one has
\begin{equation}
P(\nu_e \to \nu_e) = 1 - \sin^2(2\theta_{13})\sin^2 \Bigl
(\frac{\Delta m^2_{\rm atm}L}{4E} \Bigr ) \,,
\label{preactoratm}
\end{equation}
i.e., $\theta_{\rm reactor}=\theta_{13}$, and, for the value $|\Delta m^2_{32}| = 3 \times 10^{-3}\rm\,eV^2$ from SuperK, the CHOOZ experiment on $\bar\nu_e$ disappearance
yields the upper limit~\cite{chooz}
\begin{equation}
\sin^2(2\theta_{13}) < 0.1 \,,
\label{chooz}
\end{equation}
which is also consistent with conclusions from the SuperK data analysis~\cite{sk}.
Further, the quantity ``$\sin^2(2\theta_{\rm atm})$'' often used to fit
the data on atmospheric neutrinos with a simplified two-species mixing
hypothesis, is, in the three-generation case,
\begin{equation}
\sin^2(2\theta_{\rm atm}) \equiv \sin^2(2\theta_{23})\cos^4(\theta_{13}) \,.
\label{thetaatm}
\end{equation}
The SuperK experiment finds that the best fit to their data is 
$\nu_\mu \to \nu_\tau$ oscillations with maximal mixing, and hence
$\sin^2(2\theta_{23})=1$ and $|\theta_{13}| \ll 1$.  The various solutions of
the solar neutrino problem involve quite different values of $\Delta m^2_{21}$
and $\sin^2(2\theta_{12})$: (i)~large mixing angle solution, LMA: $\Delta
m^2_{21} \simeq {\rm few} \times 10^{-5}\rm\,eV^2$ and $\sin^2(2\theta_{12})
\simeq 0.8$; (ii)~small mixing angle solution, SMA: $\Delta m^2_{21} \sim
10^{-5}\rm\,eV^2$ and $\sin^2(2\theta_{12}) \sim 10^{-2}$, (iii)~LOW: $\Delta m^2_{21}
\sim 10^{-7}\rm\,eV^2$, $\sin^2(2\theta_{12}) \sim 1$, and (iv)~``just-so'': $\Delta
m^2_{21} \sim 10^{-10}\rm\,eV^2$, $\sin^2(2\theta_{12}) \sim 1$.  The SuperK experiment
favors the LMA solutions~\cite{sol}; for other global fits, see, e.g.,
Ref.~\cite{sol}. 
We have reviewed the three neutrino oscillation phenomenology that is
consistent with solar and atmospheric neutrino oscillations. In what
follows, we will examine the neutrino experiments planned for the
immediate future that will address some of the relevant physics. We
will then review the physics potential of the Neutrino Factory.
\subsubsection{Relevant Near- and Mid-Term Experiments} 
There are currently intense efforts to confirm and extend the evidence for
neutrino oscillations in all of the various sectors --- solar, atmospheric, and
accelerator.  Some of these experiments are running; in addition to
SuperKamiokande and Soudan-2, these include the Sudbury Neutrino Observatory,
SNO, and the K2K long baseline experiment between KEK and Kamioka.  Others are
in development and testing phases, such as 
miniBOONE, MINOS, the CERN--Gran Sasso
program, KamLAND, Borexino, and MONOLITH~\cite{anl}.  
Among the long baseline neutrino
oscillation experiments, the approximate distances are $L \simeq 250$~km for
K2K, 730~km for both MINOS (from Fermilab to Soudan) and the proposed CERN--Gran Sasso experiments.  
K2K is a $\nu_\mu$ disappearence experiment with a
conventional neutrino beam having a mean energy of about 1.4~GeV, going from
KEK 250~km to the SuperK detector.  It has a near detector for beam
calibration.  It has obtained results consistent with the SuperK experiment,
and has reported that its data disagree by $2\sigma$ with the no-oscillation
hypothesis~\cite{k2k}.  
MINOS is another conventional neutrino beam experiment
that takes a beam from Fermilab 730~km to a detector in the Soudan mine in
Minnesota.  It again uses a near detector for beam flux measurements and has
opted for a low-energy configuration, with the flux peaking at about 3~GeV.
This experiment is scheduled to start taking data in 2005 and, after some
years of running, to obtain higher statistics than the K2K experiment and to
achieve a sensitivity down to the level $|\Delta m^2_{32}| \sim
10^{-3}\rm\,eV^2$.  
The CERN--Gran Sasso program will also come on in
2005.  It will use  a higher-energy neutrino beam, $E_\nu\sim17$~GeV,
 from CERN to the
Gran Sasso deep underground laboratory in Italy.  This program will emphasize
detection of the $\tau$'s produced by the $\nu_\tau$'s that result from the
inferred neutrino oscillation transition $\nu_\mu \to \nu_\tau$.  The OPERA
experiment will do this using emulsions~\cite{opera}, while the ICARUS proposal
uses a liquid argon chamber~\cite{icanoe}.  For the joint capabilities of
MINOS, ICARUS and OPERA experiments see Ref.~\cite{minicop}.
Plans for the Japan Hadron Facility (JHF),
also called the High Intensity Proton Accelerator (HIPA), include the use of 
a 0.77~MW
proton driver to produce a high-intensity conventional neutrino beam with a
path length of 300~km to the SuperK detector~\cite{jhf}.  Moreover, at Fermilab,
the miniBOONE experiment is scheduled 
to start data taking in the near future and to confirm or
refute the LSND claim after a few years of running.
There are several neutrino experiments relevant to the solar neutrino anomaly.  The SNO experiment is currently running and has recently reported their first
results that confirm solar neutrino oscillations~\cite{snolatest}.
These involve measurement of the solar neutrino flux and energy
distribution using the charged current reaction on heavy water, $\nu_e
+ d \to e + p + p$.  They are expected to report on the neutral
current reaction $\nu_e + d \to \nu_e + n + p$ shortly. The neutral
current rate is unchanged in the presence of oscillations that involve
standard model neutrinos, since the neutral current channel is equally
sensitive to all the three neutrino species.  If however, sterile
neutrinos are involved, one expects to see a depletion in the neutral
current channel also. However, the uncertain normalization of the $^8$B flux makes it difficult to constrain a possible sterile neutrino component in the oscillations~\cite{unknowns}.
The KamLAND experiment~\cite{kamland} in Japan started taking data
in January 2002.  This is a reactor antineutrino experiment using
baselines of  100--250 km. It will search for $\bar\nu_e$ disappearance
and is sensitive to the solar neutrino oscillation scale.
KamLAND can provide precise measurements of the LMA solar parameters~\cite{bmw-kamland}
and recently the first results from
KamLAND have confirmed the LMA solution~\cite{kamland1}.  A global
analysis of the KamLAND and solar neutrino data has further restricted
the solar $\delta m^2$ range and the best fit value currently is
$7\times10^{-5}$~eV$^2$~\cite{bargermarf,fogli}.
On a similar time scale, the Borexino experiment in Gran Sasso is scheduled 
to turn
on and  measure the $^7$Be neutrinos from the sun.  These experiments
should help us determine which of the various solutions to the solar neutrino
problem is preferred, and hence the corresponding values of $\Delta m^2_{21}$
and $\sin^2(2\theta_{12})$.
This, then, is the program of relevant experiments during the period
2000--2010.  By the end of this period, we may expect that much will be learned
about neutrino masses and mixing.  However, there will remain several
quantities that will not be well measured and which can be measured by a
Neutrino Factory. 
\subsubsection{Oscillation Experiments at a Neutrino Factory } 
\label{neuf}
Although a Neutrino Factory based on a muon storage ring will turn on
several years after this near-term period in which K2K, MINOS, and the
CERN-Gran Sasso experiments will run, it has a
valuable role to play, given the very high-intensity neutrino beams of
fixed flavor-pure content, including, uniquely, $\nu_e$ and
$\bar\nu_e$ beams in addition to $\nu_\mu$ and $\bar\nu_\mu$ beams. A
conventional positive charge selected neutrino beam  is primarily
$\nu_\mu$ with some admixture of $\nu_e$'s and other flavors from $K$
decays ({\cal O}(1\%) of the total charged current rate) and  the fluxes of these neutrinos can only be fully understood after measuring the charged
particle spectra from the target with high accuracy.  In contrast,
the potential of
the neutrino beams from a muon storage ring is that the neutrino beams
would be of extremely high purity: $\mu^-$ beams would yield 50\%
$\nu_\mu$ and 50\% $\bar\nu_e$, and $\mu^+$ beams, the charge
conjugate neutrino beams.  Furthermore, these could be produced with
high intensities and low divergence that make it possible to go 
to longer baselines.

\begin{figure}[tbh!]
\centerline{\includegraphics[width=4.0in]{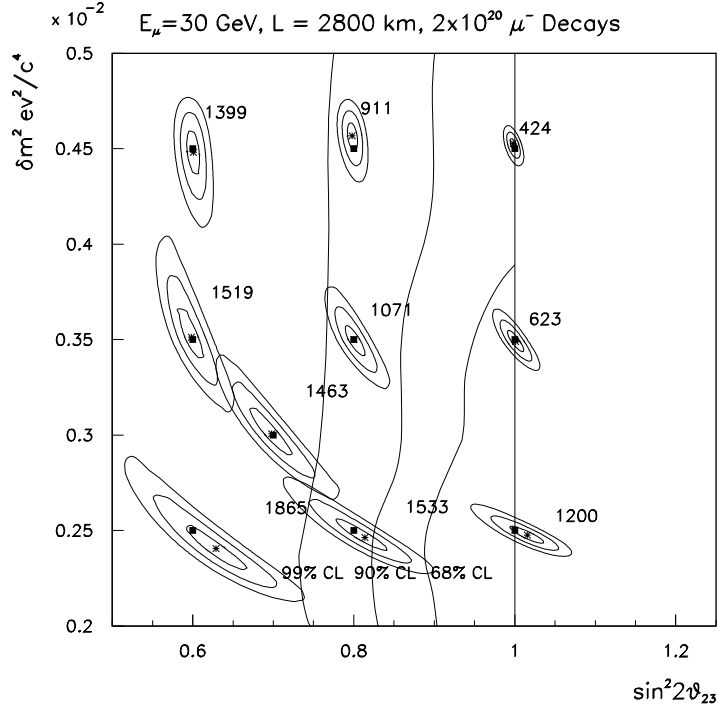}}
\bigskip
\caption[Error ellipses in  $\delta m^2$ sin$^2 2\theta$ space for a Neutrino Factory]
{ \label{fig:30gev_disap_fit}
Fit to muon neutrino survival distribution for $E_\mu=30$ GeV and $L=2800$~km for 10
pairs of sin$^2 2\theta$, $\delta m^2$ values. For each fit, the
1$\sigma$,\ 2$\sigma$
and 3$\sigma$ contours are shown. The generated points are indicated by the
dark
rectangles and the fitted values by stars. The SuperK 68\%, 90\%, and 99\% 
confidence
levels are superimposed. Each point is labelled by the predicted number of 
signal events for that point.}
\end{figure}

In what follows, we shall take the design values from Study-II
of $10^{20}$ $\mu$ decays per ``Snowmass year'' ($10^7$ sec) as being typical.
The types of neutrino oscillations that can be searched for with the Neutrino
Factory based on the muon storage ring are listed in 
Table~\ref{tab:nu-osc-ratings} for the 
case of $\mu^-$ which decays to $ \nu_\mu e^- \bar\nu_e$: 
\begin{table}
\caption[Neutrino Oscillation Modes]{Neutrino-oscillation modes that can be studied with conventional
neutrino beams or with beams from a Neutrino Factory, with ratings as to
degree of difficulty in each case; * = well or easily measured, $\surd$ =
measured poorly or with difficulty, --- = not
measured.\label{tab:nu-osc-ratings}}
\begin{center}
\begin{tabular}{|llcc|}
\hline
& & Conventional & Neutrino \\[-2ex]
\raisebox{1ex}[0pt]{Measurement } & \raisebox{1ex}[0pt]{Type} & beam &
Factory \\
\hline
$\nu_\mu\to\nu_\mu,\,\nu_\mu\to\mu^-$ & survival & $\surd$ & *\\
$\nu_\mu\to\nu_e,\,\nu_e\to e^-$ & appearance & $\surd$ & $\surd$\\
$\nu_\mu\to\nu_\tau,\,\nu_\tau\to\tau^-,\,\tau^-\to(e^-,\mu^-)...$  &
appearance & $\surd$ & $\surd$ \\ \hline
$\bar \nu_e\to\bar \nu_e,\,\bar\nu_e\to e^+$ & survival  & --- & $*$\\
$\bar\nu_e\to\bar\nu_\mu,\,\bar\nu_\mu\to\mu^+$  & appearance & --- &  $*$
\\
$\bar\nu_e\to\bar\nu_\tau,\,\bar\nu_\tau\to\tau^+,\,\tau^+\to(e^+,\mu^+)...$
& appearance & ---& $\surd$ \\
\hline
\end{tabular}
\end{center}
\end{table}
It is clear from the 
processes listed that since the beam contains both neutrinos and
antineutrinos, the only way to determine  the flavor of the 
parent neutrino  is to
determine the identity of the final state charged lepton and measure its
charge.  
A capability unique to the Neutrino Factory will be the 
measurement of the oscillation $\bar\nu_e \to \bar\nu_\mu$,
giving a wrong-sign $\mu^+$.  Of greater difficulty would be the measurement of
the transition $\bar\nu_e \to \bar\nu_\tau$, giving a $\tau^+$ which will decay
part of the time to $\mu^+$.  These physics goals mean that a detector must
have excellent capability to identify muons and measure their charges.
Especially in a steel-scintillator detector, the oscillation $\nu_\mu \to
\nu_e$ would be difficult to observe, since it would be 
difficult to distinguish
an electron shower from a hadron shower.  From the above formulas for
oscillations, one can see that, given the knowledge of $|\Delta m^2_{32}|$ and
$\sin^2(2\theta_{23})$ that will be available by the time a Neutrino Factory is
built, the measurement of the $\bar\nu_e \to \bar\nu_\mu$ transition yields the
value of $\theta_{13}$.
To get a rough idea of how the sensitivity of 
an oscillation experiment would scale with energy and baseline length, 
recall that the event rate in the absence of oscillations is 
simply the neutrino flux times the cross section.  
First of all, neutrino cross sections in the region above
about 10 GeV (and slightly higher for $\tau$ production) grow linearly with
the neutrino energy.  Secondly, the beam divergence is
a function of the initial muon storage ring energy; 
this divergence yields a flux, as a
function of $\theta_d$, the angle of deviation from the forward direction, that
goes like $1/\theta_d^2 \sim E^2$.  Combining this with the linear $E$
dependence of the neutrino cross section 
and the overall $1/L^2$ dependence of the flux far from the
production region, one finds that the event rate goes like 
\begin{equation} 
\frac{dN}{dt} \sim \frac{E^3}{L^2} \,.
\label{eventrate}
\end{equation}
We base our discussion on the event rates given in the 
Fermilab Neutrino Factory study~\cite{INTRO:ref9}. For
a stored muon energy of 20~GeV,  and a distance of 
$L=2900$ to the WIPP Carlsbad site in New Mexico, these event rates amount to 
several thousand events per kton of detector per year, i.e,\ they are
satisfactory for the physics program.   This is also true for the other
path lengths under consideration, namely $L=2500$~km from BNL to Homestake and
$L=1700$~km to Soudan.  A usual racetrack design would only allow a single
pathlength $L$, but a bowtie design could allow two different path lengths
(e.g.,~\cite{zp}). 
We anticipate that at a time when the Neutrino Factory turns on, $|\Delta
m^2_{32}|$ and $\sin^2(2\theta_{23})$ would be known at perhaps the 10\% level
(while recognizing that future projections such as this are obviously uncertain).
The Neutrino Factory will significantly improve precision in these parameters,
as can be seen from Fig.~\ref{fig:30gev_disap_fit} which shows the error ellipses possible for a 30~GeV muon storage ring.
In addition, the Neutrino Factory can contribute to the measurement of: (i) 
$\theta_{13}$, as discussed above; (ii) measurement of the sign of $\Delta
m^2_{32}$ using matter effects; and (iii) possibly a measurement of $CP$
violation in the leptonic sector, if $\sin^2(2\theta_{13})$,
$\sin^2(2\theta_{21})$, and $\Delta m^2_{21}$ are sufficiently large.  To
measure the sign of $\Delta m^2_{32}$, one uses the fact that matter effects
reverse sign when one switches from neutrinos to antineutrinos, and carries out
this switch in the charges of the stored $\mu^\pm$.  We elaborate on this next.

\subsubsection{Matter Effects} 
\begin{figure}[tbh!]
\centerline{\includegraphics[width=4.0in]{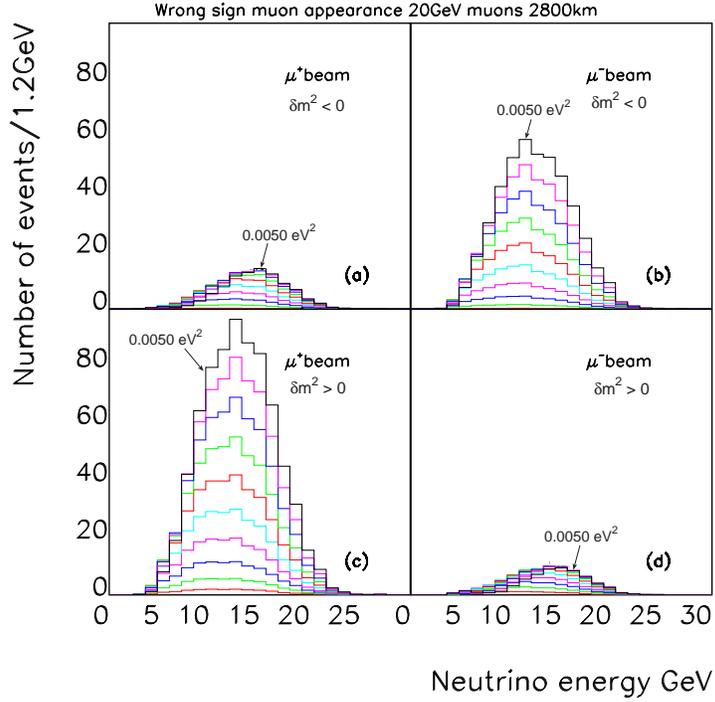}}
\caption[Wrong sign muon appearance rates and sign of  $\delta m^2_{32}$]
{(Color)The wrong sign muon appearance rates for a 20 GeV muon storage ring at
a baseline of 2800~km with 10$^{20}$ decays and a 50 kiloton detector
for (a)~$\mu^+$ stored and negative $\delta m^2_{32}$\,, (b)~$\mu^-$ stored
and negative $\delta m^2_{32}$\,, (c)~$\mu^+$ stored and positive $\delta
m^2_{32}$\,,
(d)~$\mu^-$ stored and positive $\delta m^2_{32}$. The values of $|\delta
m^2_{32}|$ range from 0.0005 to 0.0050 eV$^2$ in steps of 0.0005~eV$^2$.  
Matter enhancements are evident in (b) and (c).
\label{fig:hists}}
\end{figure}

\begin{figure}[tbh!]
\centerline{\includegraphics[width=4.0in]{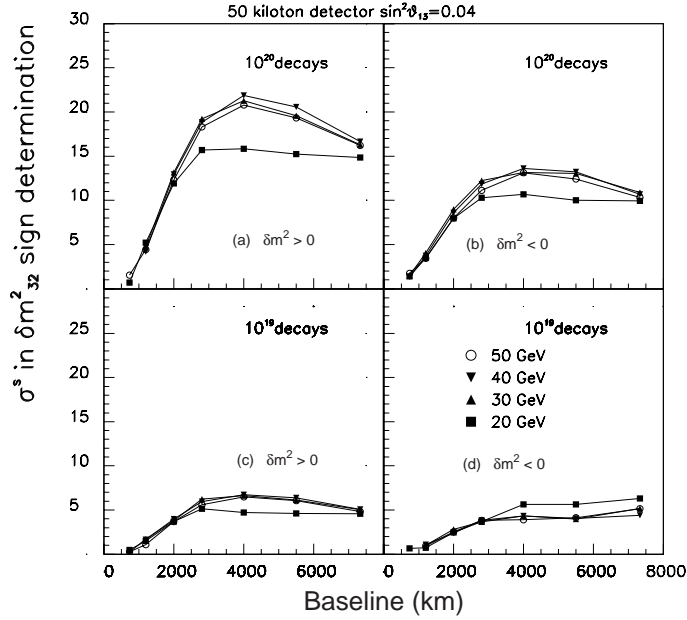}}
\caption[$\delta m_{32}^2$ sign determination at a Neutrino Factory]
{The statistical significance (number of standard deviations) 
with which the
sign of $\delta m_{32}^2$ can be determined versus baseline length for
various muon storage ring energies. The results are shown for 
a 50~kiloton detector, and (a)~10$^{20}$
$\mu^+$ and $\mu^-$ decays and positive values of $\delta m_{32}^2$;
(b)~10$^{20}$ $\mu^+$ and $\mu^-$ decays and
negative values of $\delta m_{32}^2$; (c)~10$^{19}$ $\mu^+$ and 
$\mu^-$ decays and positive values of $\delta
m_{32}^2$; (d)~10$^{19}$ $\mu^+$ and $\mu^-$
decays and negative values of $\delta m_{32}^2$.
\label{fig:sigmas}}
\end{figure}

With the advent of the muon storage ring, the distances at which one 
can place detectors
are large enough so that for the first time matter effects can be exploited in
accelerator-based oscillation experiments.  Simply put, matter effects are the
matter-induced oscillations that neutrinos undergo along their flight path
through the Earth from the source to the detector.  Given the typical density
of the earth, matter effects are important for the neutrino energy range $E
\sim {\cal O}(10)$ GeV and $\Delta m^2_{32} \sim 10^{-3}$~eV$^2$, values relevant for
the long baseline experiments. Matter effects in neutrino propagation were 
first pointed out by Wolfenstein~\cite{wolf} and Barger, Pakvasa, Phillips and Whisnant~\cite{bppw-1980}. 
(See the papers~\cite{dgh}--\cite{cpv} for details  of the matter effects
and their relevance to neutrino factories.) In brief,
assuming a normal hierarchy, the transition
probabilities for propagation through
matter of constant density are~\cite{golden,formcon}
\begin{eqnarray}
P(\nu_e \to \nu_\mu) &=&
x^2 f^2 + 2 x y f g (\cos\delta\cos\Delta + \sin\delta\sin\Delta)
+ y^2 g^2\,,\\
P(\nu_e \to \nu_\tau) &=& {\rm cot}^2 \theta_{23} x^2 f^2 - 2 x y f g
(\cos\delta\cos\Delta + \sin\delta\sin\Delta)
+ {\rm tan}^2 \theta_{23} y^2 g^2\,,\\
P(\nu_\mu \to \nu_\tau) &=& \sin^2 2\theta_{23} \sin^2\Delta
 \\ & + &\alpha
 \sin 2\theta_{23} \sin 2\Delta \bigg({\hat A \over 1-\hat A}
\sin \theta_{13} \sin 2\theta_{12} \cos 2\theta_{23} \sin\Delta-\Delta
\cos^2 \theta_{12} \sin 2\theta_{23}\bigg)\,, \nonumber
\end{eqnarray}
where
\begin{eqnarray}
\Delta &\equiv& |\delta m_{31}^2| L/4E_\nu
= 1.27 |\delta m_{31}^2/{\rm eV^2}| (L/{\rm km})/ (E_\nu/{\rm GeV}) \,,
\label{eq:D}\\
\hat A &\equiv& |A/\delta m_{31}^2| \,,
\label{eq:Ahat}\\
\alpha &\equiv& |\delta m^2_{21}/\delta m^2_{31}|  \,,\\
x &\equiv& \sin\theta_{23} \sin 2\theta_{13} \,,
\label{eq:x}\\
y &\equiv& \alpha \cos\theta_{23} \sin 2\theta_{12} \,,
\label{eq:y}\\
f &\equiv& \sin((1\mp\hat A)\Delta)/(1\mp\hat A) \,,
\label{eq:f}\\
g &\equiv& \sin(\hat A\Delta)/\hat A \,.
\label{eq:alpha}
\end{eqnarray}
The amplitude $A$ for $\nu_e e$ forward scattering in matter is given
by
\begin{equation}
A = 2\sqrt2 G_F N_e E_\nu = 1.52 \times 10^{-4}{\rm\,eV^2} Y_e
\rho({\rm\,g/cm^3}) E({\rm\,GeV}) \,.
\label{eq:A}
\end{equation}
Here $Y_e$ is the electron fraction and $\rho(x)$ is the matter
density. For neutrino trajectories that pass through the earth's
crust, the average density is typically of order 3~gm/cm$^3$ and $Y_e
\simeq 0.5$.  
For neutrinos with $\delta
m^2_{31} > 0$ or anti-neutrinos with $\delta m^2_{31} < 0$, $\hat A =
1$ corresponds to a matter resonance.
Thus, for a Neutrino Factory operating
with positive stored muons (producing a $\nu_e$ beam) one expects an
enhanced production of opposite sign ($\mu^-$) charged-current events
as a result of the oscillation $\nu_e\to \nu_\mu$ if $\delta m^2_{32}$
is positive and vice versa for stored negative
beams.
Figure~\ref{fig:hists}~\cite{barger-raja} shows the wrong-sign 
muon appearance spectra 
 as function of $\delta m^2_{32}$ for both $\mu^+$ and
$\mu^-$ beams for both signs of $\delta m^2_{32}$ at a baseline of
2800~km. The resonance enhancement in wrong sign muon production is
clearly seen in Fig.~\ref{fig:hists}(b) and (c).
By comparing these (using first a stored $\mu^+$ beam and then a stored $\mu^-$
beam) one can thus determine the sign of $\Delta m^2_{32}$ as well as the value
of $\sin^2(2\theta_{13})$.  
Figure~\ref{fig:sigmas}~\cite{barger-raja} shows the difference in negative
log-likelihood between a
correct and wrong-sign mass hypothesis expressed as a number of
equivalent Gaussian standard deviations versus baseline length for
muon storage ring energies of 20, 30, 40 and 50~GeV. The values of the
oscillation parameters are for the LMA scenario with 
$\sin^22\theta_{13}=0.04$. 
Figure~\ref{fig:sigmas}(a) is for 10$^{20}$ decays
for each sign of stored energy and a 50 kiloton detector and positive
$\delta m^2_{32}$ , (b) is for negative $\delta m^2_{32}$ for various
values of stored muon energy. Figures~\ref{fig:sigmas} (c) and (d)
show the corresponding curves for 10$^{19}$ decays and a 50 kiloton
detector. An entry-level machine would permit one to perform a
5$\sigma$ differentiation of the sign of $\delta m^2_{32}$ at a
baseline length of $\sim$2800~km.
%3
For the Study II design, in accordance with the previous
Fermilab study~\cite{INTRO:ref9}, 
one estimates  that it is possible to determine the sign of $\delta m^2_{32}$
even if 
$\sin^2(2\theta_{13})$ is as small as $\sim 10^{-3}$.
\subsubsection{CP Violation}
\begin{figure}[tbh!]
\centerline{\includegraphics[width=4.0in]{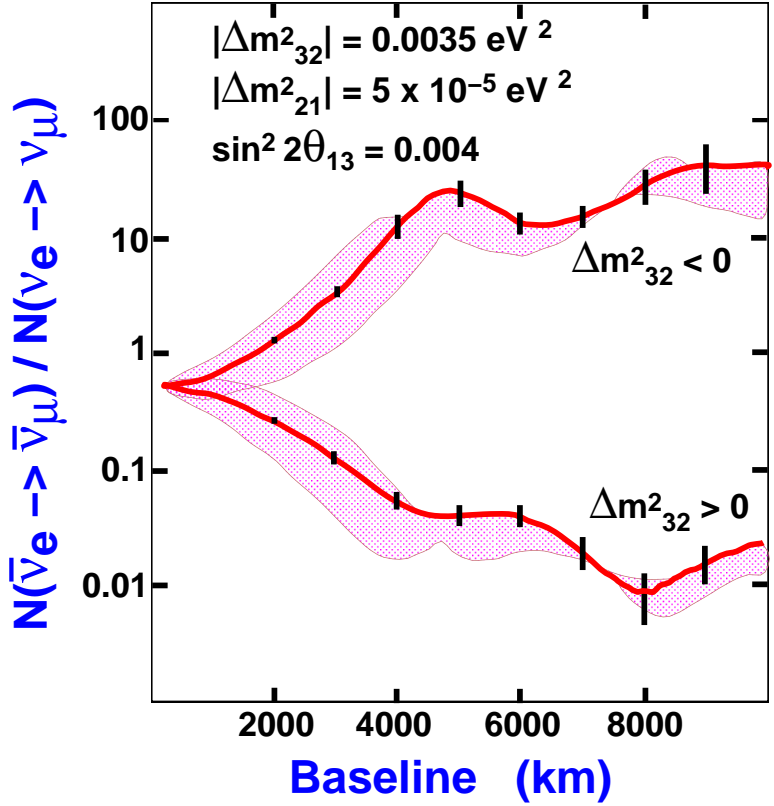}}
\bigskip
\caption[CP violation effects in a Neutrino Factory]
{ \label{cpfig}
(Color)Predicted ratios of wrong-sign muon event rates when positive and
negative muons are stored in a 20~GeV Neutrino Factory, shown as a
function of baseline.  A muon measurement threshold of 4~GeV is
assumed. The lower and upper bands correspond, respectively, to negatve
and positive $\delta m^2_{32}$. The widths of the bands show how the
predictions vary as the $CP$ violating phase $\delta$ is varied from
$-\pi$/2 to $\pi$/2, with the thick lines showing the predictions for
$\delta=0$. The statistical error bars correspond to a
high-performance Neutrino Factory yielding a data sample of 10$^{21}$
decays with a 50~kiloton detector. The curves are based on calculations
presented in~\cite{barger-entry}. }
\end{figure}

$CP$ violation is measured by the (rephasing-invariant) product
\begin{eqnarray}
J & =& Im(U_{ai}U_{bi}^* U_{aj}^* U_{bj}) \cr\cr
&  = &\frac{1}{8}
\sin(2\theta_{12})\sin(2\theta_{13})\cos(\theta_{13})\sin(2\theta_{23})\sin
\delta  \,.
\end{eqnarray}
Leptonic CP violation also requires that each of the leptons in each charge
sector be nondegenerate with any other leptons in this sector; this is,
course, true of the charged lepton sector and, for the neutrinos, this requires
$\Delta m^2_{ij} \ne 0$ for each such pair $ij$.  In the quark sector, $J$ is 
known to be small: $J_{\rm CKM} \sim {\cal O}(10^{-5})$.  
A promising asymmetry to measure is $P(\nu_e \to \nu_\mu)-P(\bar\nu_e - 
\bar\nu_\mu)$.  As an illustration, in the absence of matter effects, 
\begin{eqnarray}
P(\nu_e \to \nu_\mu) - P(\bar\nu_e \to \bar\nu_\mu) & = & 4J(\sin 2\phi_{32}+
\sin 2\phi_{21} + \sin 2\phi_{13}) \cr
& = & -16J \sin \phi_{32} \sin \phi_{13} \sin \phi_{21} \,,
\label{pnuenumudif}
\end{eqnarray}
where
\begin{equation}
\phi_{ij} = \frac{\Delta m^2_{ij}L}{4E} \,.
\label{phiijdef}
\end{equation}
In order for the $CP$ violation in Eq.~(\ref{pnuenumudif}) to be large enough to measure, it is necessary that $\theta_{12}$, $\theta_{13}$, and $\Delta
m^2_{\rm sol} = \Delta m^2_{21}$ not be too small. From atmospheric neutrino data,
we have $\theta_{23}\simeq \pi/4$ and $\theta_{13} \ll 1$.  If LMA describes
solar neutrino data, then $\sin^2(2\theta_{12}) \simeq 0.8$, so $J \simeq
0.1\sin(2\theta_{13})\sin \delta$.  For example, if
$\sin^2(2\theta_{13})=0.04$, then $J$ could be $\gg J_{CKM}$.  Furthermore, for
 parts of  the LMA phase space where  
$\Delta m^2_{\rm sol} \sim 4 \times 10^{-5}$ eV$^2$ 
the CP violating effects might be observable. In the absence of matter, one
would measure the asymmetry
\begin{equation}
\frac{P(\nu_e \to \nu_\mu) - P(\bar\nu_e \to \bar\nu_\mu)}{
P(\nu_e \to \nu_\mu) + P(\bar\nu_e \to \bar\nu_\mu)} =
\frac{\sin(2\theta_{12})\cot(\theta_{23})\sin\delta \sin \phi_{21}}{
\sin \theta_{13}}
\end{equation}
However, in order to optimize this ratio, because of the smallness of $\Delta
m^2_{21}$ even for the LMA, one must go to large pathlengths $L$, and here
matter effects are important.  These make leptonic $CP$ violation challenging to measure, because, even in the absence of any intrinsic $CP$ violation, these
matter effects render the rates for $\nu_e \to \nu_\mu$ and $\bar\nu_e \to
\bar\nu_\mu$ unequal since the matter interaction is opposite in sign for $\nu$
and $\bar\nu$.  One must therefore subtract out the matter effects in order to
try to isolate the intrinsic $CP$ violation.  Alternatively, one might think of
comparing $\nu_e \to \nu_\mu$ with the time-reversed reaction $\nu_\mu \to
\nu_e$.  Although this would be equivalent if $CPT$ is valid, as we assume, and
although uniform matter effects are the same here, the detector response is
quite different and, in particular, it is quite difficult to identify $e^\pm$.
Results from SNO and KamLAND testing the LMA~\cite{bmw-kamland} 
will help further planning.
The Neutrino Factory provides an ideal set of controls to measure $CP$
violation effects since we can fill the storage ring with either $\mu^+$
or $\mu^-$ particles and measure the ratio of the number of events
$\bar\nu_e\rightarrow \bar\nu_\mu$/$\nu_e\rightarrow\nu_\mu$.
Figure~\ref{cpfig} shows this ratio for a Neutrino Factory with
10$^{21}$ decays and a 50~kiloton detector as a function of the
baseline length. The ratio depends on the sign of $\delta
m^2_{32}$. The shaded band around either curve shows the variation of
this ratio as a function of the $CP$-violating phase $\delta$. The
number of decays needed to produce the error bars shown is directly
proportional to $\sin^2\theta_{13}$, which for the present example is
set to 0.004. Depending on the magnitude of $J$, one may be driven to
build a Neutrino Factory just to understand $CP$ violation in the lepton
sector, which could have a significant role in explaining the baryon
asymmetry of the Universe~\cite{yanag}.

\subsection{Physics Potential of Superbeams}
It is possible to extend the reach of the current conventional
neutrino experiments by enhancing the capabilities of the proton
sources that drive them. These enhanced neutrino beams have been
termed ``superbeams'' and form an intermediate step on the way to a
Neutrino Factory. Their capabilities have been explored in  recent
papers~\cite{superbeams,bargersuperbeam,superbeam-peak}. These articles consider
the capabilities of enhanced proton drivers at (i) the proposed
0.77~MW 50~GeV proton synchrotron at the Japan Hadron Facility
(JHF)~\cite{jhf}, (ii) a 4~MW upgraded version of the JHF, (iii) a
new $\sim 1$~MW 16~GeV proton driver~\cite{brighter} that would replace
the existing 8~GeV Booster at Fermilab, or (iv) a fourfold intensity
upgrade of the 120~GeV Fermilab Main Injector (MI) beam (to 1.6~MW)
that would become possible once the upgraded (16~GeV) Booster was
operational.  Note that the 4~MW 50~GeV JHF and the 16~GeV upgraded
Fermilab Booster are both suitable proton drivers for a neutrino
factory. The conclusions of both reports are that superbeams will
extend the reaches in the oscillation parameters of the current
neutrino experiments but ``the sensitivity at a Neutrino Factory to
$CP$ violation and the neutrino mass hierarchy extends to values of
the amplitude parameter $\sin^2 2\theta_{13}$ that are one to two
orders of magnitude lower than at a superbeam''~\cite{bargersuperbeam,superbeam-peak}.

\begin{figure}[tbh!]
\centerline{\includegraphics[width=4.0in]{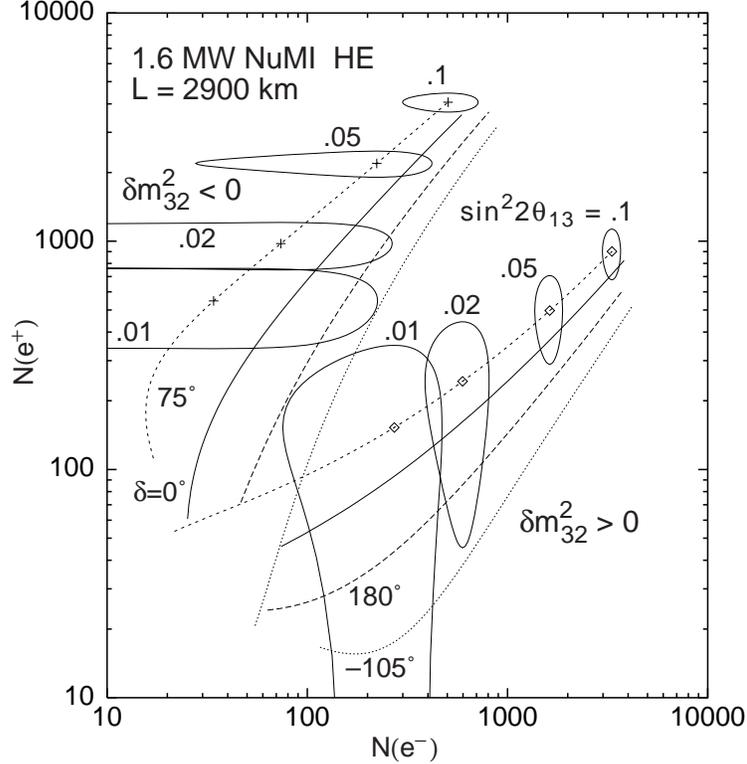}}
\caption[Error ellipses for superbeams for electron appearance]
{Three-sigma error ellipses in the 
$\left(N(e^+), N(e^-)\right)$ plane, shown for 
$\nu_\mu \to \nu_e$ and $\bar\nu_\mu \to \bar\nu_e$ oscillations 
in a NuMI-like 
high energy neutrino beam driven by a 1.6~MW proton driver. 
The calculation assumes a liquid argon detector with the parameters 
listed in \cite{superbeams}, a baseline of 2900~km, 
and 3~years of running with neutrinos, 6~years running 
with antineutrinos. 
Curves are shown for different CP phases $\delta$ (as labelled), and 
for both signs of $\delta m^2_{32}$ with 
$|\delta m^2_{32}| = 0.0035$~eV$^2$, and 
the sub-leading scale $\delta m^2_{21} = 10^{-4}$~eV$^2$. 
Note that $\sin^22\theta_{13}$ varies along the curves from
0.001 to 0.1, as indicated~\cite{bargersuperbeam}.
}
\label{fig:signdm2}
\end{figure}

\begin{figure}[tbh!]
\centerline{\includegraphics[width=4.0in]{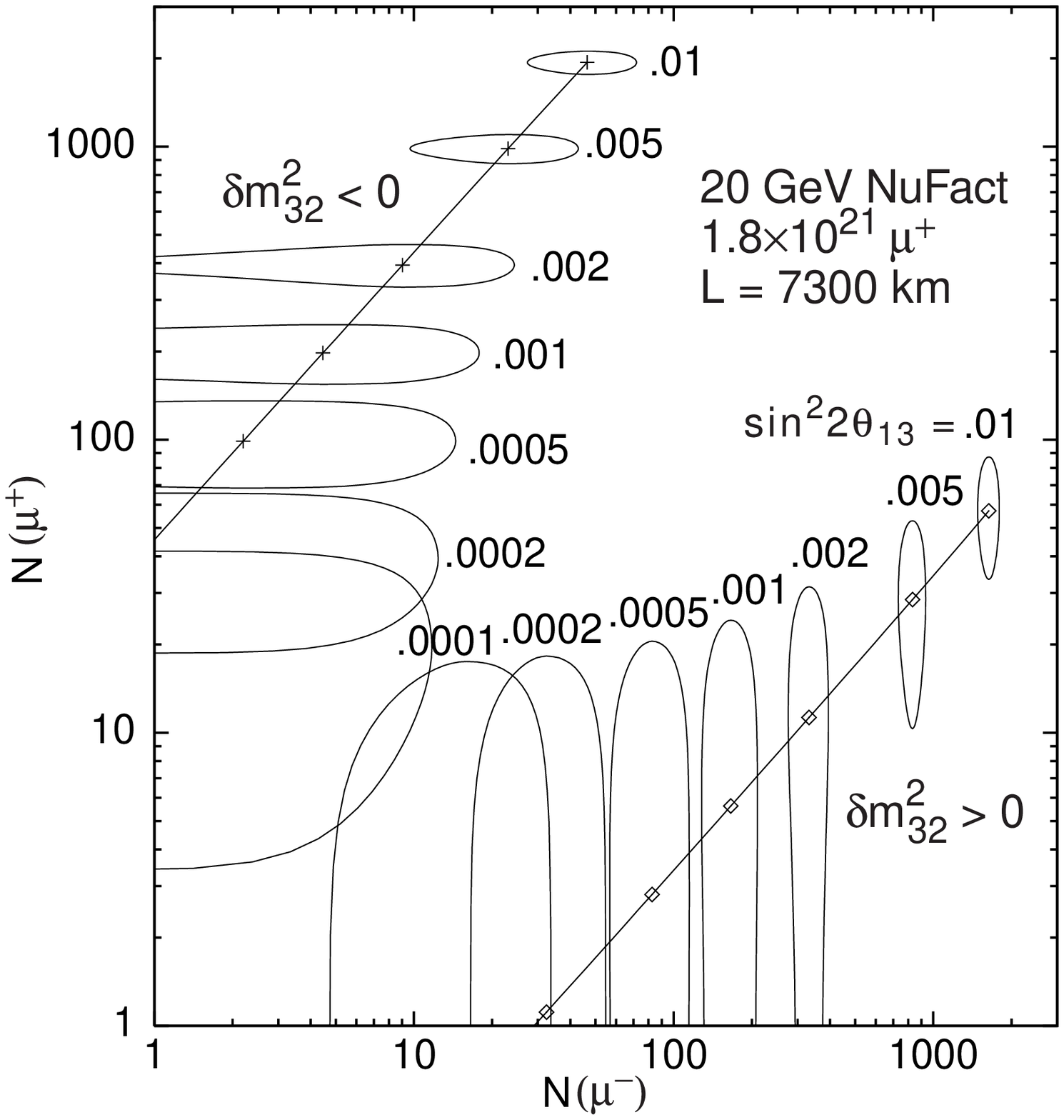}}
\caption[Error ellipses for Neutrino Factory for muon appearance]
{Three-sigma error ellipses in the 
$\left(N(\mu+), N(\mu-)\right)$ plane, shown for a 20~GeV neutrino 
factory delivering $3.6\times10^{21}$ useful muon decays and
$1.8\times10^{21}$ antimuon decays, with a 50~kt
detector at $L = 7300$~km, $\delta m^2_{21} = 10^{-4}$~eV$^2$, 
and $\delta = 0$. Curves are shown for both signs of
$\delta m^2_{32}$; $\sin^22\theta_{13}$ varies along the curves from
0.0001 to 0.01, as indicated~\cite{bargersuperbeam}.
}
\label{fig:nufact}
\end{figure}

To illustrate these points, we choose one of the most favorable superbeam 
scenarios 
studied: a 1.6~MW NuMI-like high energy beam with $L = 2900$~km, detector 
parameters corresponding to the liquid argon scenario in~\cite{bargersuperbeam,superbeam-peak}, and oscillation parameters 
$|\delta m^2_{32}| = 3.5 \times 10^{-3}$~eV$^2$ and 
$\delta m^2_{21} = 1 \times 10^{-4}$~eV$^2$. 
The calculated three-sigma error ellipses in the 
$\left(N(e^+), N(e^-)\right)$ plane are shown in Fig.~\ref{fig:signdm2}
for both signs of $\delta m^2_{32}$, with the curves corresponding to 
various $CP$ phases $\delta$ (as labeled). The magnitude of 
the $\nu_\mu \to \nu_e$ oscillation amplitude parameter 
$\sin^2 2\theta_{13}$ varies along each curve, as indicated. The 
two groups of curves, which correspond to the two signs of $\delta m^2_{32}$, 
are separated by more than $3\sigma$ provided 
$\sin^2 2\theta_{13} \gsim 0.01$. Hence the mass heirarchy can be determined 
provided the $\nu_\mu \to \nu_e$ oscillation amplitude is not more than an 
order of magnitude below the currently excluded region. Unfortunately, within 
each group of curves, the $CP$-conserving predictions are separated from the 
maximal $CP$-violating predictions by at most $3\sigma$. Hence, it will 
be difficult to conclusively establish $CP$ violation in this scenario.
Note for comparison that a very long baseline experiment at a neutrino 
factory would be able to observe $\nu_e \to \nu_\mu$ oscillations and 
determine the sign of $\delta m^2_{32}$ for values of $\sin^2 2\theta_{13}$ 
as small as ${\cal O}(0.0001)$. This is illustrated in Fig.~\ref{fig:nufact}.
A Neutrino Factory thus outperforms a conventional superbeam in its ability to
determine the sign of  $\delta m^2_{32}$.
Comparing Fig.~\ref{fig:signdm2} and Fig.~\ref{fig:nufact} one sees that 
the value of   $\sin^2 2\theta_{13}$, which has yet to be measured, 
will determine the parameters of the first Neutrino Factory.

\begin{figure}[tbh!]
\centerline{\includegraphics[width=4.0in]{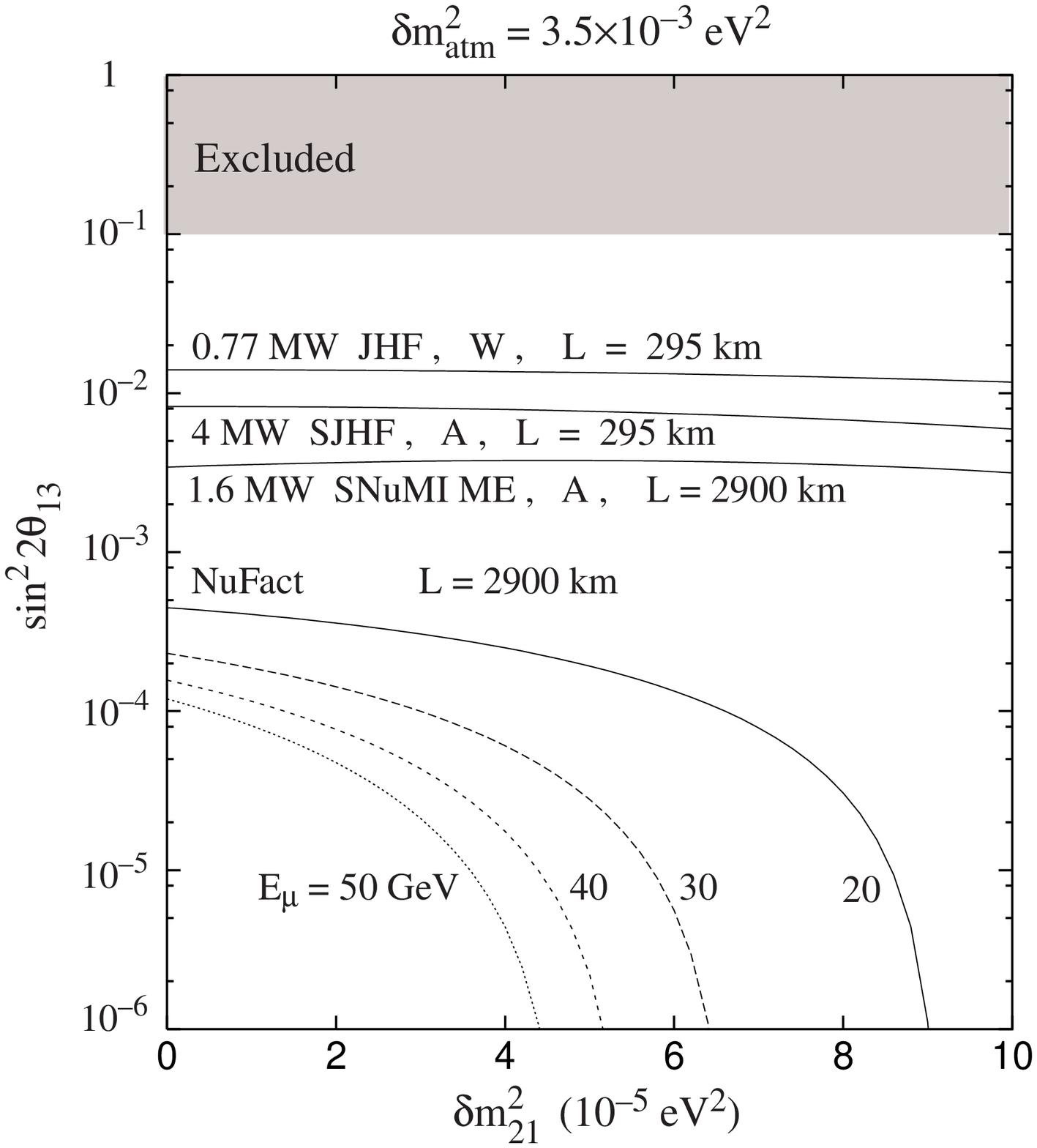}}
\caption[Comparison of superbeams and Neutrino Factories]
{Summary of the $3\sigma$ level sensitivities for the 
observation of $\nu_\mu \to \nu_e$ at various MW-scale superbeams 
(as indicated) with liquid argon ``A'' and water cerenkov ``W'' detector 
parameters, and the observation of $\nu_e \to \nu_\mu$ in a 50~kt detector 
at 20, 30, 40, and 50~GeV neutrino factories delivering $2 \times 10^{20}$ 
muon decays in 
the beam-forming straight section. The limiting $3\sigma$ contours are 
shown in the ($\delta m^2_{21}, \sin^2 2\theta_{13}$) plane. All curves 
correspond to 3~years of running. The grey shaded 
area is already excluded by current experiments.
}
\label{fig:reach} 
\end{figure}

Finally, we compare the superbeam $\nu_\mu \to \nu_e$ reach with the 
corresponding Neutrino Factory $\nu_e \to \nu_\mu$ reach in 
Fig.~\ref{fig:reach}, which shows the $3\sigma$ sensitivity contours in 
the $(\delta m^2_{21}, \sin^2 2\theta_{13})$ plane. The superbeam 
$\sin^2 2\theta_{13}$ reach of a few $\times 10^{-3}$ is almost independent 
of the sub-leading scale $\delta m^2_{21}$. However, since the neutrino 
factory probes oscillation amplitudes $O(10^{-4})$ the sub-leading effects 
cannot be ignored, and $\nu_e \to \nu_\mu$ events 
would be observed at a Neutrino Factory 
over a significant range 
of $\delta m^2_{21}$ even if $\sin^2 2\theta_{13} = 0$.
%% restart here 1.16.02, noon
\subsection{Non-oscillation physics at a Neutrino Factory} The study
of the utility of intense neutrino beams from a muon storage ring in
determining the parameters governing non-oscillation physics was begun
in 1997~\cite{rajageer}. More complete studies can be found
in~\cite{INTRO:ref9} and recently a European group has brought out an
extensive study on this topic~\cite{cern-nonosc}.  A Neutrino Factory
can measure individual parton distributions within the proton for all
light quarks and anti-quarks.  It could improve valence distributions
by an order of magnitude in the kinematical range $x\gsim 0.1$ in the
unpolarized case.  The individual components of the sea ($\bar{u}$,
$\bar{d}$, ${s}$ and $\bar{s}$), as well as the gluon, would be
measured with relative accuracies in the range of 1--10\%, for
$0.1\lsim x \lsim 0.6$. A full exploitation of the Neutrino Factory
potential for polarized measurements of the shapes of individual
partonic densities requires an {\it a priori} knowledge of the
polarized gluon density.  The forthcoming set of polarized deep
inelastic scattering experiments at CERN, DESY and RHIC may provide
this information.  The situation is also very bright for measurements
of $C$-even distributions. Here, the first moments of singlet, triplet
and octet axial charges can be measured with accuracies that are up to
one order of magnitude better than the current uncertainties. In
particular, the improvement in the determination of the singlet axial
charge would allow a definitive confirmation or refutation of the
anomaly scenario compared to the `instanton' or `skyrmion' scenarios,
at least if the theoretical uncertainty originating from the small-$x$
extrapolation can be kept under control. The measurement of the octet
axial charge with a few percent uncertainty will allow a determination
of the strange contribution to the proton spin better than 10\%, and
allow stringent tests of models of $SU(3)$ violation when compared to
the direct determination from hyperon decays.  A measurement of
$\as(M_Z)$ and $\sin^2\theta_W$ will involve different systematics
from current measurements and will therefore provide an important
consistency check of current data, although the accuracy of these
values is not expected to be improved.  The weak mixing angle can be
measured in both the hadronic and leptonic modes with a precision of
approximately $2\times 10^{-4}$, dominated by the statistics and the
luminosity measurement.  This determination would be sensitive to
different classes of new-physics contributions.  Neutrino interactions
are a very good source of clean, sign-tagged charm particles. A
Neutrino Factory can measure charm production with raw event rates up
to 100 million charm events per year with $\simeq$ 2 million
double-tagged events.  (Note that charm production becomes significant
for storage ring energies above 20~GeV).  Such large samples are
suitable for precise extractions of branching ratios and decay
constants, the study of spin-transfer phenomena, and the study of
nuclear effects in deep inelastic scattering.  The ability to run with
both hydrogen and heavier targets will provide rich data sets useful
for quantitative studies of nuclear models.  The study of $\Lambda$
polarization both in the target and in the fragmentation regions will
help clarify the intriguing problem of spin transfer.

Although the neutrino beam energies are well below any reasonable
threshold for new physics, the large statistics makes it possible to
search for physics beyond the Standard Model. The high intensity
neutrino beam allows a search for the production and decay of neutral
heavy leptons with mixing angle sensitivity two orders of magnitude
better than present limits in the 30--80 MeV range.  The exchange of
new gauge bosons decoupled from the first generation of quarks and
leptons can be seen via enhancements of the inclusive charm production
rate, with a sensitivity well beyond the present limits.  A novel
neutrino magnetic moment search technique that uses oscillating
magnetic fields at the neutrino beam source could discover large
neutrino magnetic moments predicted by some theories.  Rare
lepton-flavor-violating decays of muons in the ring could be tagged in
the deep inelastic scattering final states through the detection of
wrong-sign electrons and muons, or of prompt taus.
%
% below modified K.J. 28.jul.2002
%
\subsection{Physics that can be done with Intense Cold Muon Beams}
Experimental studies of muons at low and medium energies have had a
long and distinguished history, starting with the first search for
muon decay to electron plus gamma-ray~\cite{Hincks-Pontecorvo}, and
including along the way the 1957 discovery of the nonconservation of
parity, in which the $g$ value and magnetic moment of the muon were
first measured~\cite{Garwinetal}.  The years since then have brought
great progress: limits on the standard-model-forbidden decay $\mu\to
e\gamma$ have dropped by nine orders of magnitude, and the muon
anomalous magnetic moment $a_\mu=(g_\mu-2)/2$ has yielded one of the
more precise tests ($\approx1$ ppm) of physical theory~\cite{BNLg-2}.
The front end of a Neutrino Factory has the potential to provide
$\sim10^{21}$ muons per year, five orders of magnitude beyond the most
intense beam currently available\footnote{The $\pi$E5 beam at PSI,
Villigen, providing a maximum rate of $10^9$
muons/s~\cite{Edgecock}.}.

Such a facility could enable a rich variety of precision
measurements. In the area of low energy muon physics a majority of
experiments with a high physics potential is limited at present by
statistics.  The list of conceivable projects includes 
(see Table~\ref{tab:LEexpts}):
\begin{itemize}
\item 
precise determinations of the properties characterizing the muon,
which are the mass $m_{\mu}$, magnetic moment $\mu_{\mu}$, magnetic
anomaly $a_{\mu}$, charge $q_{\mu}$ and lifetime $\tau_{\mu}$,
\item 
measurements the muon decay parameters (Michel parameters),
\item
CPT tests from a comparison of $\mu^-$ and $\mu^+$ properties,
\item 
measurements of fundamental constants of general importance (e.g. the
electromagnetic fine structure constant $\alpha$ or the weak
interaction Fermi constant $G_F$)
\item sensitive searches for physics beyond the Standard Model either
through measuring differences of muon parameters from Standard Model
predictions or in dedicated searches for rare and forbidden processes,
such as $\mu \rightarrow e \gamma$, $\mu \rightarrow eee$, $\mu^-N
\rightarrow e^-N$ conversion and muonium-antimuonium (${\rm
M}-\overline{\rm M}$) conversion or searches for a permanent electric
dipole moment $d_{\mu}$ of the particle,
\item searches for $P$ and $T$ violation in muonic atoms,
\item precise determinations of nuclear properties in muonic
(radioactive) atoms,
\item  applications in condensed matter, thin films and at surfaces,
\item  applications in life sciences, and
\item  muon catalyzed fusion($\mu$CF).
\end{itemize}

A detailed evaluation of the possibilities has recently been made by a
CERN study group, which assumed a facility  with a 4 MW proton 
driver\cite{Aysto_01}.

In the search for ``forbidden'' decays,
Marciano~\cite{Marciano97} has suggested that muon Lepton Flavor Violation
(LFV) (especially
coherent muon-to-electron conversion in the field of a nucleus) is the
``best bet" for discovering signatures of new physics using low-energy
muons. The MECO experiment \cite{MECO}  proposed at BNL
offers, through a novel detector concept, very high sensitivity and some
four orders of magnitude improvement over the current limits from
PSI \cite{SINDRUM}.  At a future high muon flux facility, such as the Neutrino Factory, this could be  improved further by 1-2 orders of magnitude.

The search for $\mu\to e \gamma$ is also of great interest. The MEGA
experiment recently set an upper limit $B(\mu^+\to
e^+\gamma)<1.2\times10^{-11}$~\cite{MEGA}. Ways to extend sensitivity
to the $10^{-14}$ level have not only been discussed~\cite{Cooper97}
but also have lead to an active proposal at PSI \cite{Mori_99}. The
experiment aims for an improvement of three orders of magnitude over MEGA
which had systematics limitations.  The $\mu$-to-$e$-conversion approach
has the additional virtue of sensitivity to new physics that
does not couple to the photon.

An observation of a non-zero Electric Dipole Moment (EDM) of the muon, 
$d_{\mu}$, could prove equally exciting; This has generated a Letter of Intent~\cite{EDMLOI} to observe $d_\mu$, which  proposes to use the
the large electric fields associated with 
relativistic particles in a magnetic storage ring.  
As CP violation enters in the quark sector
starting with the second generation, the muon is a particularly
valuable probe in this regard, 
despite the already low limits for electrons. Moreover,
there exist some models in which the electric dipole moment 
scales stronger than linearly\cite{Ellis_01}.

It is worth noting that for  searches of rare muon decays 
and for $d_{\mu}$ that
the standard model predictions are orders of magnitude below
the presently established limits. Any observation which can be shown
to be not an artefact of the experimental method or due to background
would therefore be a direct sign of new physics.

There are  three experiments going on currently to improve the
muon lifetime $\tau_\mu$ \cite{tau_mu}. Note that the 
Fermi coupling constant $G_F$ is derived from a measurement of $\tau_\mu$. The
efforts are therefore worthwhile whenever experimental conditions allow
substantial improvement. One should however be aware that a comparison
with theory in this channel is presently dominated by 
theoretical uncertainties.

In the case of precision measurements ($\tau_\mu$, $a_\mu$, etc.),
new-physics effects appear  as small corrections arising from
the virtual exchange of new massive particles in loop diagrams. In
contrast, LFV and EDMs are forbidden in the standard model, thus their
observation at any level would constitute evidence for new physics. 
%%%%%%%%%%%%%%%%%%%%%%%%%%%%%%%%%%%%%
\newlength\slideback
\newenvironment{subtable}{\begin{tabular}[c]{c|c|c||c}&&&\\[\slideback]}{\\[\slideback]&&&\end{tabular}}
\newcommand\stcol[2]{\parbox[c]{#1}{\centering #2}}
\newcommand\stcola[1]{\stcol{0.85in}{#1}}
\newcommand\stcolb[1]{\stcol{0.85in}{#1}}
\newcommand\stcolc[1]{\stcol{1.7in}{#1}}
\newcommand\stcold[1]{\stcol{0.9in}{#1}}
\newcommand\mtcol[2]{\parbox[c]{#1}{\centering #2}}
\newcommand\mtcola[1]{\mtcol{0.73in}{#1}}
\newcommand\mtcolb[1]{\mtcol{1.54in}{#1}}
\begin{table*}[htbp]
  \centering
  \makeatletter
  \preprintsty@sw{\scriptsize}{\squeezetable}\makeatother
  \setlength\slideback{2pt}
  \addtolength\slideback{-\baselineskip}
% Delete next line
  \resizebox{\textwidth}{!}{
  \begin{tabular}{|c|c||c|}
    \hline
    \mtcola{Type of Experiment}&
    \mtcolb{Physics Issues}&
    \begin{subtable}
      \stcola{Possible experiments}&
      \stcolb{Previously established accuracy}&
      \stcolc{Present activities\\(proposed accuracy)}&
      \stcold{Projected for SMS @ CERN}
    \end{subtable}\\
    \hline\hline
    \mtcola{``Classical'' rare \& forbidden decays}&
    \mtcolb{Lepton number violation; searches for new physics: SUSY, L-R
      Symmetry, R-parity violation,\ldots}&
    \begin{subtable}
      \stcola{$\mu^-N \rightarrow e^-N$}&
      \stcolb{$6.1\cdot 10^{-13}$}&
      \stcolc{PSI, proposed BNL ($5 \cdot 10^{-17}$)}&
      \stcold{$<10^{-18}$}\\
      \stcola{$\mu\rightarrow e\gamma$}&
      \stcolb{$1.2\cdot10^{-11}$}&
      \stcolc{Proposed PSI ($1 \cdot 10^{-14}$)}&
      \stcold{$ < 10^{-15}$}\\
      \stcola{$\mu \rightarrow eee$}&
      \stcolb{$1.0 \cdot 10^{-12}$}&
      \stcolc{completed 1985 PSI}&
      \stcold{$ < 10^{-16}$}\\
      \stcola{$\mu^+e^- \rightarrow \mu^-e^+$}&
      \stcolb{$8.1 \cdot 10^{-11}$}&
      \stcolc{completed 1999 PSI}&
      \stcold{$ < 10^{-13}$}
    \end{subtable}\\
    \hline
    \mtcola{Muon Decays}&
    \mtcolb{$G_F$; searches for new physics; Michel parameters}&
    \begin{subtable}
      \stcola{$\tau_{\mu}$}&
      \stcolb{$18\cdot 10^{-6}$}&
      \stcolc{PSI (2x), RAL         ($1 \cdot 10^{-6}$)}&
      \stcold{$ < 10^{-7}$}\\
      \stcola{non $(V-A)$}&
      \stcolb{typ. few $10^{-3}$}&
      PSI, TRIUMF           ($1 \cdot 10^{-3}$)&
      $ < 10^{-4}$
    \end{subtable}\\
    \hline
    \mtcola{Muon Moments}&
    \mtcolb{Standard model tests; new physics; CPT tests
      T- resp. CP-violation in 2nd lepton generation}&
    \begin{subtable}
      &&&\\
      \stcola{$g_{\mu}-2$}&
      \stcolb{$1.3 \cdot 10^{-6} $}&
      \stcolc{BNL ($3.5\cdot10^{-7}$)}&
      \stcold{$ < 10^{-7}$}\\
      \stcola{$edm_{\mu}$}&
      \stcolb{$3.4 \cdot 10^{-19} e$ cm}&
      \stcolc{proposed BNL           ($10^{-24} e\,cm$)}&
      \stcold{$ < 5 \cdot 10^{-26} e$ cm}\\
      &&&
    \end{subtable}\\
    \hline
    \mtcola{Muonium Spectroscopy}&
    \mtcolb{Fundamental constants, $\mu_{\mu}$, $m_{\mu}$, $\alpha$;
      weak interactions; muon charge}&
    \begin{subtable}
      &&&\\[-0.5\baselineskip]
      \stcola{$M_{HFS}$}&
      \stcolb{$12 \cdot 10^{-9}$}&
      \stcolc{completed 1999 LAMPF}&
      \stcold{$ 5 \cdot 10^{-9}$}\\
      \stcola{$M_{1s2s}$}&
      \stcolb{$1 \cdot 10^{-9}$}&
      \stcolc{completed 2000 RAL}&
      \stcold{$ < 10^{-11}$}\\[-0.5\baselineskip]
      &&&
    \end{subtable}\\
    \hline
    \mtcola{Muonic Atoms}&
    \mtcolb{Nuclear charge radii;\\weak interactions}&
    \begin{subtable}
      \stcola{$\mu^-$ atoms}&
      \stcolb{depends}&
      \stcolc{PSI, possible CERN\\($<r_p>$ to $10^{-3}$)}&
      \stcold{new nuclear structure}
    \end{subtable}\\
    \hline
    \mtcola{Condensed Matter}&
    \mtcolb{surfaces, catalysis, bio sciences\ldots}&
    \begin{subtable}
      &&&\\[-0.5\baselineskip]
      \stcola{surface $\mu$SR}&
      \stcolb{n/a}&
      \stcolc{PSI, RAL (n/a)}&
      \stcold{high rate}\\[-0.5\baselineskip]
      &&&
    \end{subtable}\\
    \hline
  \end{tabular}
% Delete next line
  }
  
  \caption{
    Experiments which could beneficially take advantage of the intense
    future stopped muon source. The numbers were worked out for
    scenarios at a future Stopped Muon Source (SMS) of a neutrino
    factory at CERN \cite{Aysto_01}. They are based on a muon flux of
    $10^{21}$ particles per annum in which beam will be available for
    $10^7$ s. Typical beam requirements are given in
    Table~\ref{tab:LE_beams}.
  }
  \label{tab:LEexpts}
\end{table*}

%%%%%%%%%%%%%%%%%%%%%%%%%%%%%%%%%%%%% 
\begin{table}[tbh]
\caption[]{
Beam requirements for new muon experiments.  We show  the needed
muonic charge $q_{\mu}$ and the minimum of the total muon number
$\int I_{\mu}dt$ above which significant progress can be expected in
the physical interpretation of the experiments. Measurements which
require pulsed beams are sensitive to the muon suppression $I_0/I_{m}$
between pulses of length $\delta T$ and separation $\Delta T$.  Most
experiments require energies up to 4 MeV corresponding to 29 MeV/c
momentum. Thin targets, respectively storage ring acceptances, demand
rather small momentum spreads $\Delta p_{\mu}/p_{\mu}$
\cite{Aysto_01}.   
\label{tab:LE_beams} 
} 
\begin{center}
\begin{tabular}{|c|c|c|c|c|c|c|c|}
\hline
&&&&&&&\\
Experiment & $q_{\mu}$ &$\int I_{\mu}dt$&$I_0/I_{\mu}$&$\delta
T$&$\Delta T$&$E_{\mu}$&$\Delta p_{\mu}/p_{\mu}$\\ & & & & [ns] & [ns]
& [MeV] & [\%] \\
\hline
$\mu^-N \rightarrow e^-N$ &-- &$10^{19}$&$<10^{-9}$&$\leq 100$&$\geq
1000$ &$<20$ &1...5 \\ $\mu \rightarrow e \gamma$ &+ &$10^{16}$& n/a
&continuous &continuous &1...4 &1...5 \\ $\mu \rightarrow eee$ &+
&$10^{15}$& n/a &continuous &continuous &1...4 &1...5 \\ $\mu^+e^-
\rightarrow \mu^-e^+$&+ &$10^{16}$&$<10^{-4}$&$<1000$s &$\geq 20000$
&1...4 &1...2 \\
\hline
$\tau_{\mu}$ &+ &$10^{13}$&$<10^{-4}$&$<100 $ &$\geq 20000$ &4 &1...10
\\ $non (V-A)$ &$\pm$&$10^{13}$&$ n/a $ &continuous &continuous &4
&1...5 \\
\hline
$g_{\mu}-2$ &$\pm$&$10^{15}$&$<10^{-7}$&$\leq 50 $ &$\geq 10^6$ &3100
&$10^{-4}$ \\ $d_{\mu}$ &$\pm$&$10^{16}$&$<10^{-6}$&$\leq 50 $
&$\geq 10^6 $ &$\leq$1000&$\leq 10^{-5}$\\
\hline
$M_{HFS}$ &+ &$10^{15}$&$<10^{-4}$&$\leq 1000$ &$\geq 20000$ &4 &1...3
\\ $M_{1s2s}$ &+ &$10^{14}$&$<10^{-3}$&$\leq 500 $ &$\geq 10^6$ &1...4
&1...2 \\
\hline
$\mu^- atoms$ &-- &$10^{14}$&$<10^{-3}$&$\leq 500 $&$\geq 20000$
&1...4 &1...5 \\
\hline
$condensed$ $matter$ &$\pm$&$10^{14}$&$<10^{-3}$&$< 50 $ &$\geq 20000$
&1...4 &1...5 \\ $(incl.$$bio$ $ sciences)$ &&&&&&&\\
\hline
\end{tabular}
\end{center}
\end{table}
The current status and prospects for advances in these areas are
shown in Table~\ref{tab:LEexpts}, which lists present efforts in the
field and possible improvements at a Neutrino Factory or Muon
Collider facility. The beam parameters necessary for the expected
improvements are listed in Table~\ref{tab:LE_beams}.

It is worth recalling that LFV as a manifestation of neutrino mixing
is suppressed as $(\delta m^2)^2/m_W^4$ and is thus entirely
negligible. However, a variety of new-physics scenarios predict
observable effects.  Table~\ref{tab:newmuphys} lists some examples of
limits on new physics that would be implied by nonobservation of
$\mu$-to-$e$ conversion ($\mu^-N\to e^-N$) at the $10^{-16}$
level~\cite{Marciano97}.

\begin{table}
\caption[New physics probed by $\mu\rightarrow e$ experiments]
{Some examples of new physics probed by the nonobservation of
$\mu\rightarrow e$ conversion at the $10^{-16}$ level
(from~\protect\cite{Marciano97}).\label{tab:newmuphys}}
\begin{center}
\begin{tabular}{|lc|}
\hline
New Physics & Limit \\
\hline
Heavy neutrino mixing & $|V_{\mu N}^*V_{e N}|^2<10^{-12}$\\ Induced
$Z\mu e$ coupling & $g_{Z_{\mu e}}<10^{-8}$\\ Induced $H\mu e$
coupling & $g_{H_{\mu e}}<4\times10^{-8}$\\ Compositeness &
$\Lambda_c>3,000\,$TeV\\
\hline
\end{tabular}
\end{center}
\end{table}

The muon magnetic anomaly (muon g-2 value \cite{Farley_90}) has been
measured recently at the Brookhaven National Laboratory (BNL) with 0.7
ppm accuracy \cite{BNLg-2}.  At present, no definite statement can
be made whether this result agrees or disagrees with standard theory, which is sensitive to electroweak corrections.
The theory has recently come under severe scrutiny and 
in particlar an error was found in the calculation of 
hadronic light by light scattering
\cite{Knecht_02}. The theoretical calculations are being improved upon, 
and with more data, there is a good chance that this might eventually lead to
evidence for beyond the standard model effects\cite{Czarnecki_01}. 
The final goal of the experimental precision is
0.35 ppm for the current set of experiments.  
This value could be improved  by an order of
magnitude at a Neutrino Factory, 
provided cold muons of energy 3.1 GeV  are  made available.  
This could  further spur more accurate theoretical calculations 
that improve upon contributions from 
hadronic vacuum polarization and hadronic light
by light scattering~\cite{Marciano_2001}.  
In addition, the  muon g-2 experiments at CERN have provided the best
test of CPT invariance at a level of $2\cdot10^{-22}$ which is  more
than three orders of magnitude better than the mostly quoted result ${\rm
K}^0-\overline{{\rm K}^0}$ mass difference
\cite{Kostelecki_00}. 
A $g-2$ measurement at the Neutrino Factory front end that uses muons 
of both charges would lead to  further improvement in these CPT limits.

Precision studies of atomic electrons have provided notable tests of
QED ({ e.g,} the Lamb shift in hydrogen) and could in principle be
used to search for new physics were it not for nuclear corrections.
Studies of muonium ($\mu^+e^-$) are free of such corrections since it
is a purely leptonic system. Muonic atoms  can also yield new
information complementary to that obtained from electronic atoms. A
number of possibilities have been enumerated by Kawall {\it et
al.}~\cite{Kawall97}, Jungmann \cite{Jungmann_01} and
Molzon~\cite{Molzon97}.

By making measurements on the muonium system, for instance, one can
produce precise measurements of the fundamental constants and also do
sensitive searches for new physics.
The muonium ground state hyperfine structure has been measured to 12
ppb~\cite{Liu_99} and currently furnishes the most sensitive test of
the relativistic two-body bound state in QED~\cite{Jungmann_01}. The
precision could be further improved significantly with increased
statistics.  The theoretical error is 120~ppb. The uncertainty arising
from the muon mass is five times larger than that from calculations.
If one assumes the theory to be correct, the muon-electron mass ratio
can be extracted to 27~ppb. A precise value for the electromagnetic
fine structure constant $\alpha$ can be extracted.  Its good agreement
with the number extracted from the electron magnetic anomaly must be
viewed as the best test of the internal consistency of QED, as one case
involves bound state QED and the other that of free particles. The
Zeeman effect of the muonium hyperfine structure allows the best
direct measurement of the muon magnetic moment, respectively its mass,
to 120~ppb, improved by higher-precision measurements in muonium and
muon spin resonance. These are  also areas in which the Neutrino Factory front
end could contribute.  Laser spectroscopy of the muonium 1s-2s
transition
\cite{Meyer_00} has resulted in a
precise value of the muon mass as well as  the testing of the muon-electron 
charge
ratio to about $2\cdot 10^{-9}$. This is by far the best
test of charge equality in the first two generations.

The search for muonium-antimuonium conversion had been proposed by
Pontecorvo three years before the systemwas first produced by
Hughes {\it et al.}~\cite{Hughes_60}. Several 
new-physics models allow violation of lepton family number by two
units. The current limit is $R_g \equiv G_C / G_F<
0.0030$~\cite{Willmann_99}, where $G_C$ is the new-physics coupling
constant,
and $G_F$ the Fermi coupling constant. 
This sets a lower limit of $2.6 \,$TeV$/c^2$ (90\% C.L.) on the mass
of a grand-unified dileptonic gauge boson and also strongly disfavours
models with heavy lepton seeded radiative mass
generation~\cite{Willmann_99}. The search for muonium-antimuonium
conversion has by far the strongest gain in sensitivity of all rare
muon decay experiments \cite{Jungmann_01}.

The high intensity proton machine needed for the Neutrino Factory
can also find use as a new generation isotope facility which would
have much higher rates compared to the present ISOLDE
facility at CERN. Nucleids yet not studied  could be produced at quantities
which allow precision investigations of their properties
\cite{Aysto_01}.  The  measurements of muonic spectra can yield
most precise values for the charge radii of nuclei as well as other
ground state properties such as moments and even B(E2) transition
strengths for even-even nuclei. An improved understanding of nuclear
structure can be expected which may be of significance for
interpreting various neutrino experiments, rare decays involving
nuclei, and nuclear capture. An urgent need exists for accurate
charge and neutron radii of Francium and Radium isotopes which are of
interest for atomic parity violation research and $EDM$ searches in
atoms and nuclei.

Muonic x-ray experiments generally promise higher accuracy for most of
these quantities compared to electron scattering, particularly because
the precision of electron scattering data depends on the location of
the minimum of the cross section where rates are naturally low.  In
principle, for chains of isotopes charge radii can be inferred from
isotope shift measurements with laser spectroscopy. However, this
gives only relative information. For absolute values, calibration is
necessary and has been obtained in the past for stable nuclei from
muonic spectra.  In general, two not too distant nuclei are needed
for a good calibration.

The envisaged experimental approaches include i) the
technique pioneered by Nagamine and Strasser \cite{strasser_02}, which
involves cold films for keeping radioactive atoms and as a host
material in which muon transfer takes place; ii) merging beams if
radioactive ions and of muons; and iii) trapping of exotic isotopes in
a Penning trap which is combined with a cyclotron trap.  Large
formation rates can be expected from a setup containing a Penning
trap
\cite{Penning_trap},
the magnetic field of which serves also as a cyclotron muon trap
\cite{Simons}. 
For muon energies in the range of electron binding energies the muon
capture cross sections grow to atomic values, efficient atom
production results at the rate of approximately 50~Hz. 
It should be noted that antiprotonic atoms could be produced similarly
~\cite{Hayano_2001} 
and promise measurements of neutron distributions in nuclei.
\subsection[Physics Potential of a Higgs Factory Muon Collider]%
{Physics potential of a Low energy Muon Collider 
operating as a Higgs Factory}
Muon colliders~\cite{bargersnow,clinehanson} have a number of unique
features that make them attractive candidates for future
accelerators~\cite{INTRO:ref5}.  The most important and fundamental of
these derive from the large mass of the muon in comparison to that of
the electron.The synchrotron radiation loss in a circular accelerator
goes as the inverse fourth power of the mass and is two billion times
less for a muon than for an electron. Direct $s$ channel coupling to the
higgs boson goes as the mass squared and is 40,000 greater for the
muon than for the electron.  This leads to: a)~the possibility of
extremely narrow beam energy spreads, especially at beam energies
below $100\gev$; b)~the possibility of accelerators with very high
energy; c)~the possiblity of employing storage rings at high energy;
d)~the possibility of using decays of accelerated muons to provide a
high luminosity source of neutrinos as discussed in
Section~\ref{neuf}; e)~increased potential for probing physics in
which couplings increase with mass (as does the SM $\hsm f\bar f$
coupling)
.

The relatively large mass of the muon compared to the mass of the electron
means that the coupling of Higgs bosons to $\mu^+\mu^-$ is very 
much larger than to $e^+e^-$, implying much larger $s$-channel Higgs
production rates at a muon collider as compared to an electron collider.
For Higgs bosons with a very small (MeV-scale) width, 
such as a light SM Higgs boson,
production rates in the $s$-channel 
are further enhanced by the 
muon collider's ability to achieve beam energy spreads
comparable to the tiny Higgs width. 
In addition, there is little beamstrahlung, 
and the beam energy can be tuned to one part
in a million through continuous spin-rotation measurements~\cite{Raja:1998ip}.
Due to these important qualitative differences
between the two types of machines, only muon colliders can be 
advocated as potential $s$-channel
Higgs factories capable of determining the mass and decay width
of a Higgs boson to very high precision~\cite{Barger:1997jm,Barger:1995hr}.
High rates of Higgs production at $\epem$ colliders rely on
substantial $VV$ Higgs coupling for the
$Z+$Higgs (Higgstrahlung) or $WW\to$Higgs ($WW$ fusion) reactions.
In contrast, a $\mupmum$ collider can provide a factory for producing
a Higgs boson with little or no $VV$ coupling so long as it
has SM-like (or enhanced) $\mupmum$ couplings.

Of course, there is a tradeoff between small beam energy spread,
$\delta E/E=R$, and luminosity. Current estimates for yearly
integrated luminosities (using
$\call=1\times 10^{32}\rm\,cm^{-2}\ s^{-1}$ as implying $ L=1\fbi/{\rm yr}$) are:
$\lyear\gsim 0.1,0.22,1 \fbi$ at $\rts\sim 100\gev$
for beam energy resolutions of $R=0.003\%,0.01\%,0.1\%$, respectively;
$\lyear\sim 2,6,10 \fbi$ at $\rts\sim 200,350,400\gev$, respectively, for 
$R\sim 0.1\%$.
Despite this, studies show that for small Higgs width the $s$-channel
production rate (and statistical significance over background) is maximized
by choosing $R$ to be such that $\srts\lsim \gamhtot$. In particular,
in the SM context for $\mhsm\sim 110\gev$ this corresponds to $R\sim 0.003\%$.
 
If the $\mh\sim 115\gev$ LEP signal is real, or if the 
interpretation of the precision
electroweak data as an indication of a light Higgs boson (with
substantial $VV$ coupling) is valid,
then both $\epem$ and $\mupmum$ colliders will be valuable.
In this scenario the Higgs boson would have been discovered at a previous 
higher energy collider (even possibly a muon collider
running at high energy), and then the Higgs factory
would be built with a center-of-mass energy 
precisely tuned to the Higgs boson mass.
The most likely scenario is that the Higgs boson 
is discovered at the LHC via gluon fusion
($gg\to H$) or perhaps 
earlier at the Tevatron via associated production 
($q\bar{q}\to WH, t\overline{t}H$), and its mass is determined to an 
accuracy of about 100~MeV. If a linear collider has also observed the Higgs
via the Higgs-strahlung process ($e^+e^-\to ZH$), one might know the Higgs 
boson mass to better than 50~MeV with an integrated luminosity of 
$500$~fb$^{-1}$.
The muon collider would be optimized to run at $\sqrt{s}\approx m_H$, and this
center-of-mass energy would be varied over a narrow range
so as to scan over the Higgs resonance (see Fig.~\ref{mhsmscan} below). 
\subsubsection{Higgs Production}
The production of a Higgs boson (generically denoted $\h$)
in the $s$-channel with interesting rates is  
a unique feature of a muon collider~\cite{Barger:1997jm,Barger:1995hr}. 
The resonance cross section is
\begin{equation}
\sigma_h(\sqrt s) = {4\pi \Gamma(h\to\mu\bar\mu) \, \Gamma(h\to X)\over
\left( s - m_h^2\right)^2 + m_h^2 \left(\Gamma_{\rm tot}^h \right)^2}\,.
\label{rawsigform}
\end{equation}
In practice, however, there is a Gaussian spread ($\srts$) to
the center-of-mass energy and one must compute the
effective $s$-channel Higgs cross section after convolution 
assuming some given central value of $\rts$:
\begin{eqnarray}
\bar\sigma_h(\sqrt s) & =& {1\over \sqrt{2\pi}\,\srts} \; \int
\sigma_h  
(\sqrt{\what s}) \; \exp\left[ -\left( \sqrt{\what s} - \sqrt s\right)^2
\over  
2\sigma_{\sqrt s}^2 \right] d \sqrt{\what s}\\
&&\stackrel{\rts=\mh}{\simeq} {4\pi\over m_h^2} \; {\br(h\to\mu\bar\mu)
\,
\br(h\to X) \over \left[ 1 + {8\over\pi} \left(\srts\over\gamhtot 
\right)^2 \right]^{1/2}} \ .
%
%\bar\sigma_h(\sqrt s) & =& {1\over \sqrt{2\pi}\,\srts} \; \int \sigma_h  
%(\sqrt{\what s}) \; \exp\left[ -\left( \sqrt{\what s} - \sqrt s\right)^2 \over  
%2\sigma_{\sqrt s}^2 \right] d \sqrt{\what s}\\
%\stackrel{\rts=\mh}{\simeq} {4\pi\over m_h^2} \; {\br(h\to\mu\bar\mu) \,
%\br(h\to X) \over \left[ 1 + {8\over\pi} \left(\srts\over\gamhtot 
%\right)^2 \right]^{1/2}} \,.
\label{sigform}
\end{eqnarray}
\begin{figure}[tbh!]
\centering\leavevmode
\centerline{\includegraphics[width=4.0in]{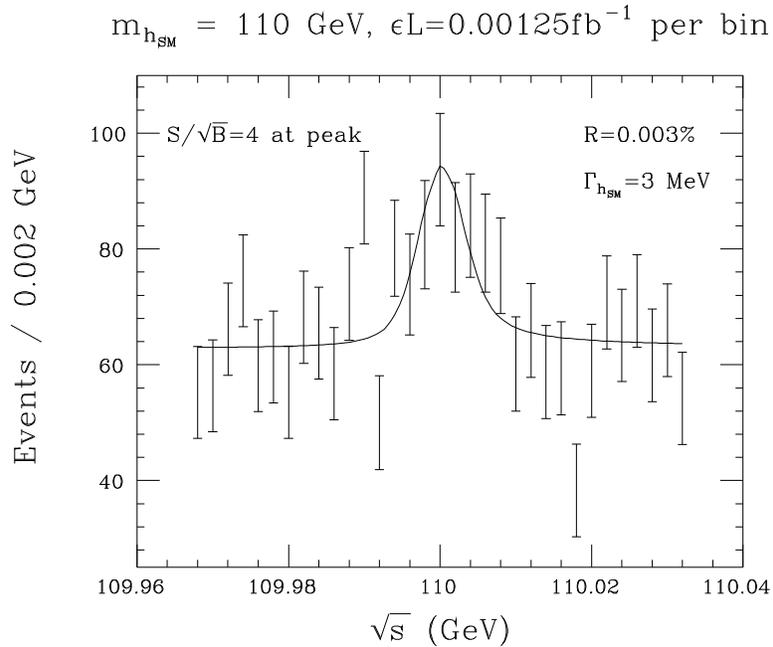}}
\caption[Scan of the Higgs resonance using a muon collider]{
Number of events and statistical errors in the $b\overline{b}$
final state as a function
of $\protect\rts$ in the vicinity of $\mhsm=110\gev$,
assuming $R=0.003\%$,
and $\epsilon L=0.00125$~fb$^{-1}$ at each data point.
%The precise theoretical prediction is given by the solid line.
%The dotted (dashed) curve is the theoretical prediction
%if $\Gamma _{tot}$ is decreased (increased) by 10\%, {\it keeping
%the $\Gamma(h\to\mu^+\mu^-)$ and $\Gamma(h\to b\overline{b})$
%partial widths fixed at the predicted SM value.}
\label{mhsmscan}}
\end{figure}
It is convenient to express $\srts$ in 
terms of the root-mean-square (rms) Gaussian spread
of the energy of an individual beam, $R$: 
\begin{equation}
\srts = (2{\rm~MeV}) \left( R\over 0.003\%\right) \left(\sqrt s\over  
100\rm~GeV\right) \,.
\end{equation}
From Eq.~(\ref{rawsigform}), it is apparent that a
resolution $\srts \lsim \gamhtot$ is needed to be
sensitive to the Higgs width. Further, Eq.~(\ref{sigform}) implies that
$\bar\sigma_h\propto 1/\srts$ for $\srts>\gamhtot$ {\it and}
that large event rates are only possible if $\gamhtot$ is not so large
that $\br(\h\to \mu\bar\mu)$ is extremely suppressed.
The width of a light SM-like Higgs is very small ({ e.g}, a few MeV
for $\mhsm\sim 110\gev$), implying the need for $R$
values as small as $\sim 0.003\%$ for studying a light SM-like $\h$.
Figure~\ref{mhsmscan} illustrates the result for the SM Higgs boson 
of an initial centering scan over $\rts$ values
in the vicinity of $\mhsm=110\gev$.
This figure dramatizes: a)~that the beam energy spread must be very small
because of the very small $\gamhsmtot$ (when $\mhsm$ is small
enough that the $WW^\star$ decay
mode is highly suppressed); b)~that we require
the very accurate {\it in situ} determination 
of the beam energy to one part in a million through the spin 
precession of the muon noted earlier in order to perform the scan
and then center on $\rts=\mhsm$ with a high degree of stability.
If the $\h$ has SM-like couplings to $WW$, its width will
grow rapidly for $\mh>2m_W$ and its $s$-channel production cross
section will be severely suppressed by the resulting 
decrease of $\br(\h\to\mu\mu)$. 
More generally, any $\h$ with SM-like or larger $\h\mu\mu$ coupling
will retain a large $s$-channel production rate when 
$\mh>2m_W$ only if the $\h WW$ coupling becomes 
strongly suppressed relative to the $\hsm WW$ coupling.

The general theoretical prediction within supersymmetric models is that the 
lightest supersymmetric Higgs boson $\hl$ will
be very similar to the $\hsm$ when the other Higgs bosons are
heavy.  This `decoupling limit' is very likely to arise if the
masses of the supersymmetric particles are large (since the Higgs
masses and the superparticle masses are typically similar in
size for most boundary condition choices).
Thus, $\hl$ rates will be very similar to $\hsm$ rates.
In contrast, the heavier Higgs bosons in a typical supersymmetric model
decouple from $VV$ at large mass  and remain reasonably
narrow. As a result, their $s$-channel production rates remain large.

For a SM-like $\h$, at $\sqrt s = \mh \approx 115$~GeV
and $R=0.003\%$, the $b\bar b$ rates are
\vspace{-.05in}
\begin{eqnarray}
\rm signal &\approx& 10^4\rm\ events\times L(fb^{-1})\,,\\
\rm background &\approx& 10^4\rm\ events\times L(fb^{-1})\,.
\end{eqnarray}

\subsubsection{What the Muon Collider Adds to LHC and LC Data}
An assessment of the need for a Higgs factory requires that one detail the 
unique capabilities of a muon collider versus the other possible future 
accelerators as well as comparing the abilities of all the machines to 
measure the same Higgs properties. 
Muon colliders, and a Higgs factory in particular,
would only become operational after the LHC physics program is well-developed 
and, quite possibly, after a linear collider program is mature as well. So one
important question is the following: if
a SM-like Higgs boson and, possibly, important
physics beyond the Standard Model have been discovered at the LHC and perhaps 
studied at a linear collider, what new information could a Higgs factory 
provide?
The $s$-channel production process allows one to determine the mass, 
total width, and the cross sections
$\overline \sig_h(\mupmum\to\h\to X)$ 
for several final states $X$ 
to very high precision. The Higgs mass, total width and the cross sections 
can be used to constrain the parameters of the Higgs sector. 
For example, in the MSSM their precise values will
constrain the Higgs sector parameters
$\mha$ and $\tanb$ (where $\tanb$ is 
the ratio of the two vacuum expectation values (vevs) of the 
two Higgs doublets of the MSSM). The main question is whether these
constraints will be a valuable addition to LHC and LC constraints.
The expectations for the luminosity available at linear colliders has risen 
steadily. The most recent studies assume an integrated luminosity of some
$500$~fb$^{-1}$ corresponding to 1--2 years of running at a 
few$\times100$~fb$^{-1}$ 
per year. This luminosity results in the production of greater than $10^4$
Higgs bosons per year through the Bjorken Higgs-strahlung process, 
$e^+e^-\to Z\h$, provided the Higgs boson is kinematically accessible. This is 
comparable or even better than can be achieved with the current machine
parameters for a muon collider operating at the Higgs resonance; in fact, 
recent studies have described high-luminosity linear colliders as ``Higgs
factories,'' though for the purposes of this report, we will reserve this term
for muon colliders operating at the $s$-channel Higgs resonance. 
A linear collider with such high luminosity can certainly perform quite 
accurate measurements of certain Higgs parameters, such as the Higgs mass, 
couplings to gauge bosons and  couplings to heavy quarks
~\cite{Battaglia:2000jb}.
Precise measurements of the couplings of the Higgs boson to the Standard 
Model particles is an important test of the mass generation mechanism.
In the Standard Model with one Higgs doublet, this coupling is proportional 
to the particle mass. In the more general case there can be mixing angles
present in the couplings. Precision measurements of the couplings can 
distinguish the Standard Model Higgs boson from that from a more general model
and can constrain the parameters of a more general Higgs sector.

\begin{table*}[h!]
\begin{center}
\caption[Comparison of a Higgs factory muon collider with LHC and LC]
{Achievable relative
uncertainties for a SM-like $\mh=110$~GeV for measuring the
Higgs boson mass and total width
for the LHC, LC (500~fb$^{-1}$), and the muon collider (0.2~fb$^{-1}$). 
}\label{unc-table}
\protect\protect
\begin{tabular}{|cccc|}
\hline
\ & LHC & LC & $\mu^+\mu^-$\\
\hline
$\mh$ & $9\times 10^{-4}$ & $3\times 10^{-4}$ & $1-3\times 10^{-6}$ \\
$\gamhtot$ & $>0.3$ & 0.17 & 0.2 \\
\hline
\end{tabular}
\end{center}
\end{table*}
The accuracies possible at different colliders
for measuring $\mh$ and $\gamhtot$ of
a SM-like $\h$ with $\mh\sim 110\gev$ are given in Table~\ref{unc-table}.
Once the mass is determined to about 1~MeV at the LHC and/or LC, 
the muon collider would employ a
three-point fine scan~\cite{Barger:1997jm} near the resonance peak.
Since all the couplings of the Standard Model are known, $\gamhsmtot$
is known. Therefore a precise determination of 
$\gamhtot$ is an important test of the Standard Model, and any deviation
would be evidence for a nonstandard Higgs sector. 
For a SM Higgs boson with a mass sufficiently below the $WW^\star$ 
threshold, the Higgs total width is very small (of order several MeV), and the 
only process where it can be measured {\it directly} is in the $s$-channel
at a muon collider. Indirect determinations at the LC can have
higher accuracy once $\mh$ is large enough that the $WW^\star$ mode
rates can be accurately measured, requiring $\mh>120\gev$.
This is because at the LC the total width must be determined 
indirectly by measuring a partial width and a branching fraction, and then 
computing the total width,
\begin{eqnarray}
&&\Gamma _{tot}={{\Gamma(h\to X)}\over {BR(h\to X)}}\;,
\end{eqnarray} 
for some final state $X$. For a Higgs boson so light that the $WW^\star$ decay
mode is not useful,  the total width measurement would probably require
use of the $\gamma \gamma $ decays~\cite{Gunion:1996cn}. This would require
information from a photon collider as well as the LC
and a small error is not possible.
The muon collider can measure the total width of the Higgs boson directly,
a very valuable input for precision tests of the Higgs sector.

To summarize,
if a Higgs is discovered at the LHC or possibly earlier at the Fermilab 
Tevatron, attention will turn to determining  whether this Higgs has the 
properties expected of the Standard Model Higgs. If the Higgs is discovered
at the LHC, it is quite possible that supersymmetric states will be 
discovered concurrently. The next goal for a linear collider or a muon collider
will be to better measure the Higgs boson properties to determine if 
everything is consistent within a supersymmetric framework or consistent
with the Standard Model.
A Higgs factory of even modest luminosity can provide uniquely
powerful constraints on the parameter space of the supersymmetric
model via its very precise measurement of the light Higgs mass, the
highly accurate determination of the total rate for $\mupmum\to\hl\to
b\bar b$ (which has almost zero theoretical systematic uncertainty
due to its insensitivity to the unknown $m_b$ value) and the
moderately accurate determination of the $\hl$'s total width.  In
addition, by combining muon collider data with LC data, a completely
model-independent and very precise determination of the
$h^0\mu^+\mu^-$ coupling is possible. This will provide another strong
discriminator between the SM and the MSSM.  Further, the
$h^0\mu^+\mu^-$ coupling can be compared to the muon collider and LC
determinations of the $h^0\tau^+\tau^-$ coupling for a precision test
of the expected universality of the fermion mass generation mechanism.

\subsection{Physics Potential of a High Energy Muon Collider}
Once one learns to cool muons, it becomes possible to build muon colliders with
energies of $\approx$ 3 TeV in the center of mass that fit on an
existing laboratory site~\cite{INTRO:ref5,rajawitherell}. At
intermediate energies, it becomes possible to measure the W mass
\cite{bbgh-wtt,bergerw} and the top quark mass~\cite{bbgh-wtt,bergertop}
with high
accuracy, by scanning the thresholds of these particles and making use
of the excellent energy resolution of the beams. We consider 
further here the ability of a higher energy muon collider to scan higher-lying
Higgs like objects such as the H$^0$ and the A$^0$ in the MSSM
that may be degenerate with each other.

\subsubsection{Heavy Higgs Bosons}
As discussed in the previous section, precision measurements of the
light Higgs boson properties might make it possible to not only
distinguish a supersymmetric boson from a Standard Model one, but also
pinpoint a range of allowed masses for the heavier Higgs bosons.  This
becomes more difficult in the decoupling limit where the differences
between a supersymmetric and Standard Model Higgs are
smaller. Nevertheless with sufficiently precise measurements of the
Higgs branching fractions, it is possible that the heavy Higgs boson
masses can be inferred.  A muon collider light-Higgs factory might be
essential in this process.
In the context of the MSSM, $\mha$ can probably be restricted to
within $50\gev$ or better if $\mha<500\gev$.
This includes the $250-500\gev$
range of heavy Higgs boson masses for which discovery is not possible 
via $\hh\ha$ pair production 
at a $\rts=500\gev$ LC. Further, the $\ha$ and $\hh$
cannot be detected in this mass range at either the LHC or LC 
in  $b\bar b \hh,b\bar b\ha$ production
for a wedge of moderate $\tanb$ 
values. (For large enough 
values of $\tanb$ the heavy Higgs bosons are expected to be observable
in $b\bar b \ha,b\bar b \hh$ production
at the LHC via their $\tau ^+\tau ^-$ decays and also at the LC.)
A muon collider can fill some, perhaps all of this moderate $\tanb$ wedge.
If $\tanb$ is large, the $\mupmum \hh$ and $\mupmum\ha$ couplings (proportional
to $\tanb$ times a SM-like value) are enhanced,
thereby leading to enhanced production rates in $\mupmum$ collisions.
The most efficient procedure is to operate the muon collider
at maximum energy and produce the $\hh$ and $\ha$ (often as overlapping
resonances) 
via the radiative return mechanism. By looking for a peak
in the $b\bar b$ final state, the $\hh$ and $\ha$ 
can be discovered and, once discovered, the machine $\rts$
can be set to $\mha$ or $\mhh$ and factory-like precision studies pursued.
Note that the $\ha$ and $\hh$ are typically broad enough that $R=0.1\%$
would be adequate to maximize their $s$-channel production rates.
In particular, $\Gamma\sim 30$~MeV
if the $t\overline{t}$ decay channel is not open, and $\Gamma\sim 3$~GeV if it
is. Since $R=0.1\%$ is sufficient, much higher luminosity
($L\sim 2-10~{\rm fb}^{-1}
/{\rm yr}$) would be possible as compared to that 
for $R=0.01\%-0.003\%$ required for studying the $\hl$.
 
In short, for these moderate $\tanb$--$\mha\gsim 250\gev$
scenarios that are particularly difficult for both
the LHC and the LC, the muon collider would be the only 
place that these extra Higgs bosons can be discovered and their properties 
measured very precisely.  

\begin{figure}[tbh!]
\centering\leavevmode
\centerline{\includegraphics[width=4.0in]{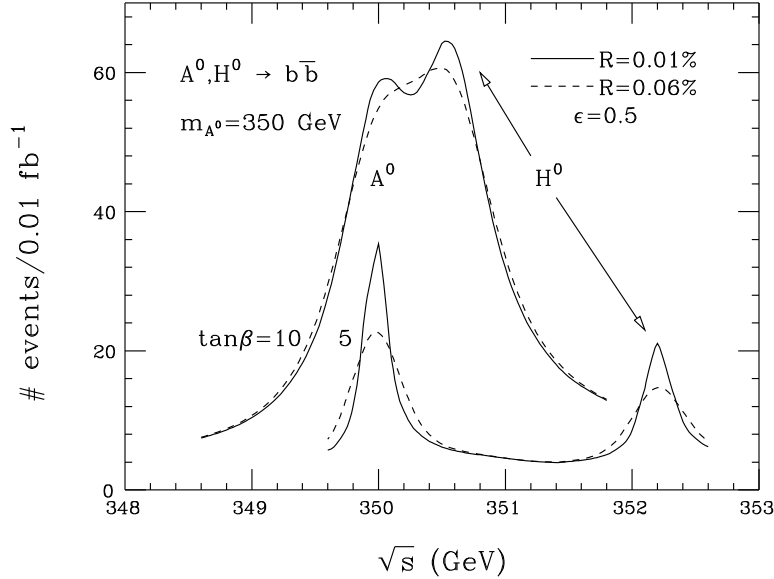}}
\caption[Separation of $A$ and $H$ signals for $\tan\beta=5$ and $10$]
{Separation of $A$ and $H$ signals for $\tan\beta=5$ and $10$. From  
Ref.~\cite{Barger:1997jm}. \label{H0-A0-sep}}
\end{figure}

In the MSSM, the heavy Higgs bosons are largely degenerate, especially in the 
decoupling limit where they are heavy. Large values of $\tan \beta$ heighten
this degeneracy.
A muon collider with sufficient energy resolution might be
the only possible means for separating out these states.
Examples showing the $H$ and $A$ resonances for $\tan \beta =5$ and $10$
are shown in Fig.~\ref{H0-A0-sep}. For the larger value of 
$\tan \beta$ the resonances are clearly overlapping. For the better energy 
resolution of $R=0.01\%$, the two distinct resonance peaks are still 
visible, but become smeared out for $R=0.06\%$. 

Once muon colliders of these intermediate energies can be built,
higher energies such as 3--4~TeV in the center of mass become feasible.
Muon colliders with these energies will be complementary to hadron
colliders of the SSC class and above. The background radiation from
neutrinos from the muon decay becomes a problem at $\approx$~3~TeV in
the CoM~\cite{kingnu}. 
Ideas for ameliorating this problem have been discussed and include
optical stochastic cooling to reduce the number of muons needed for a
given luminosity, elimination of straight sections via wigglers or
undulators, or special sites for the collider such that the neutrinos
break ground in uninhabited areas.

%\input neufact 
% Mike's new version, received Jan.9
\section{ Neutrino Factory}

\label{neufact}

In this Section we describe the various components of a Neutrino Factory,
based on the most recent Feasibility Study (Study~II)~\cite{EPP:studyii}
that was carried out jointly by BNL and the MC. We also describe the stages
that could be constructed incrementally to provide a productive physics
program that evolves eventually into a full-fledged Neutrino Factory.
Details of the design described here are based on the specific scenario of
sending a neutrino beam from Brookhaven to a detector in Carlsbad, New
Mexico. More generally, however, the design exemplifies a Neutrino Factory
for which our two Feasibility Studies demonstrated technical feasibility
(provided the challenging component specifications are met), established a
cost baseline, and established the expected range of physics performance. As
noted earlier, this design typifies a Neutrino Factory that could fit
comfortably on the site of an existing laboratory, such as BNL or FNAL.

A list of the main ingredients of a Neutrino Factory is given below: 

\begin{itemize}
\item  \textbf{Proton Driver:} Provides 1--4 MW of protons on target from an
upgraded AGS; a new booster at Fermilab would perform equivalently.

\item  \textbf{Target and Capture:} A high-power target immersed in a 20~T
superconducting solenoidal field to capture pions produced in proton-nucleus
interactions.

\item  \textbf{Decay and Phase Rotation:} Three induction linacs, with
internal superconducting solenoidal focusing to contain the muons from pion
decays, that provide nearly non-distorting phase rotation; a
``mini-cooling'' absorber section is included after the first induction
linac to reduce the beam emittance and lower the beam energy to match the
downstream cooling channel acceptance.

\item  \textbf{Bunching and Cooling:} A solenoidal focusing channel, with
high-gradient rf cavities and liquid hydrogen absorbers, that bunches the
250~MeV/c muons into 201.25~MHz rf buckets and cools their transverse
normalized rms emittance from 12 mm$\cdot $rad to 2.7 mm$\cdot $rad.

\item  \textbf{Acceleration:} A superconducting linac with solenoidal
focusing to raise the muon beam energy to 2.48 GeV, followed by a four-pass
superconducting RLA to provide a 20 GeV muon beam; a second RLA could
optionally be added to reach 50 GeV, if the physics requires this.

\item  \textbf{Storage Ring:} A compact racetrack-shaped superconducting
storage ring in which $\approx $35\% of the stored muons decay toward a
detector located about 3000 km from the ring.
\end{itemize}

\subsection{Proton Driver}

\begin{figure}[tbh]
\begin{center}
\includegraphics[width=5.5in]{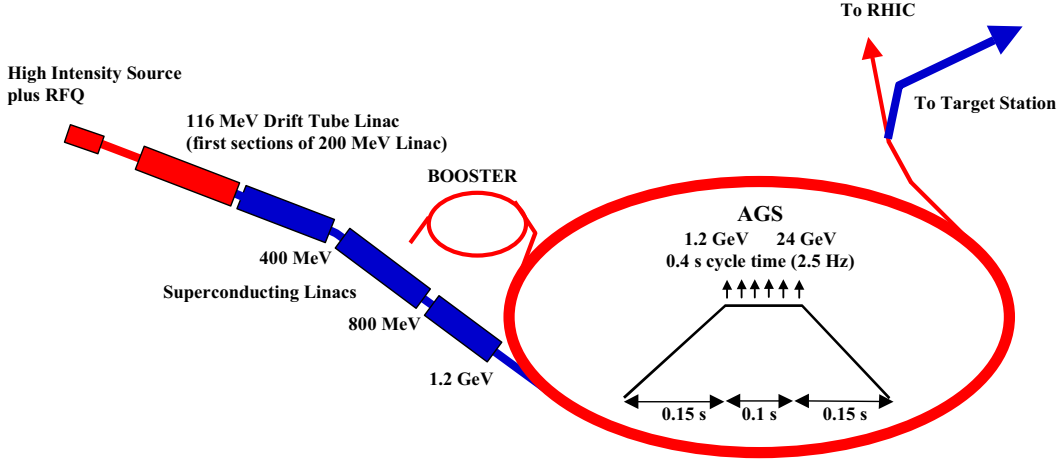}
\end{center}
\caption{(Color)AGS proton driver layout.}
\label{Proton:bnl}
\end{figure}

The proton driver considered in Study~II is an upgrade of the BNL
Alternating Gradient Synchrotron (AGS) and uses most of the existing
components and facilities; parameters are listed in Table~\ref{Proton:tb1}.
To serve as the proton driver for a Neutrino Factory, the existing booster
is replaced by a 1.2~GeV superconducting proton linac. The modified layout
is shown in Fig.~\ref{Proton:bnl}. 
\begin{figure}[tbh]
\begin{center}
\includegraphics[width=5.5in]{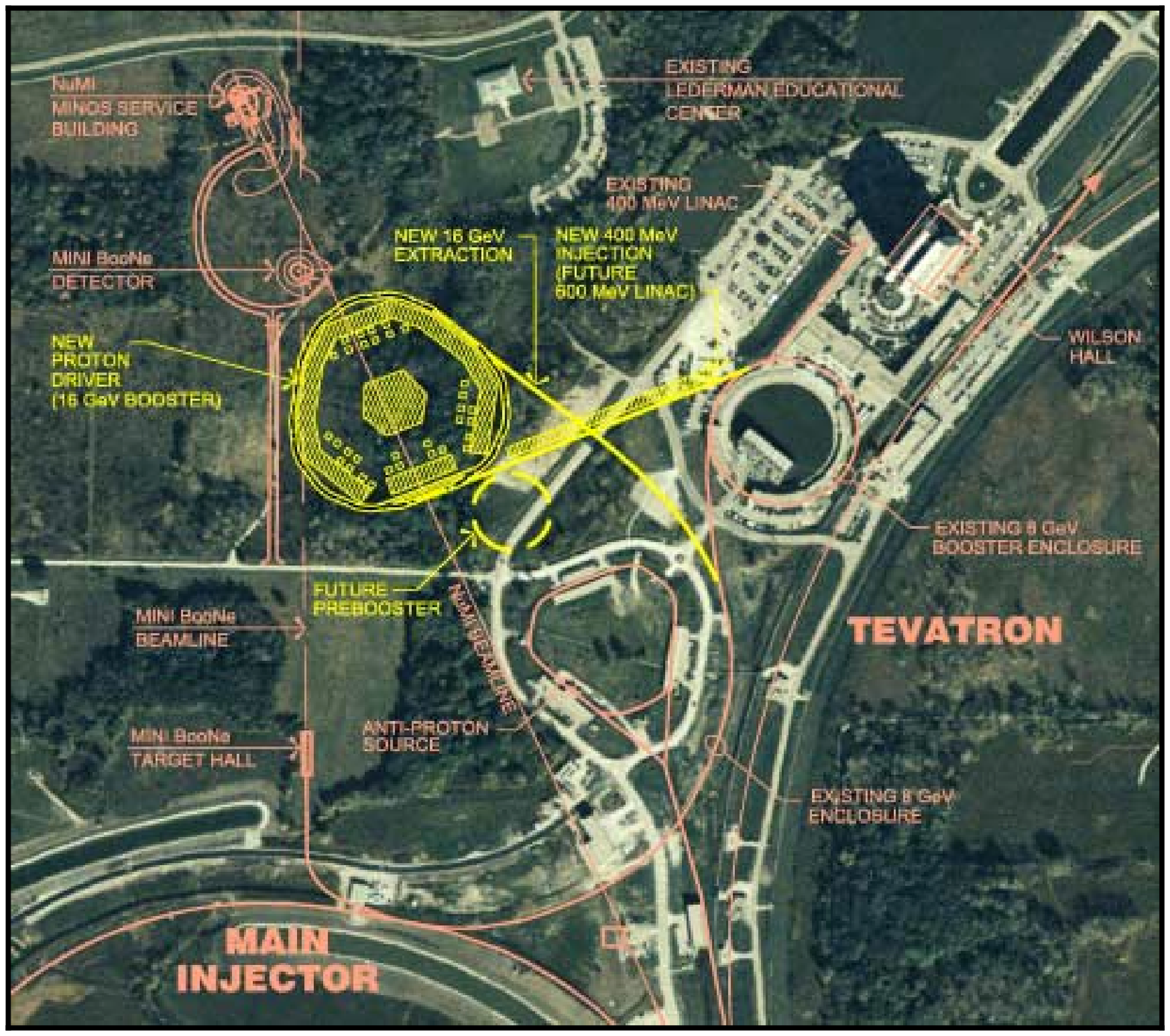}
\end{center}
\caption{(Color)FNAL proton driver layout from Ref. \protect\cite{FNALbooster}.}
\label{Proton:fnal}
\end{figure}
The AGS repetition rate is increased from 0.5 Hz to 2.5 Hz by adding power
supplies to permit ramping the ring more quickly. No new technology is
required for this---the existing supplies are replicated and the magnets are
split into six sectors rather than the two used presently. The total proton
charge (10$^{14}$ ppp in six bunches) is only 40\% higher than the current
performance of the AGS. However, due to the required short bunches, there is
a large increase in peak current and concomitant need for an improved vacuum
chamber; this is included in the upgrade. The six bunches are extracted
separately, spaced by 20 ms, so that the target, induction linacs, and rf
systems that follow need only deal with single bunches at an instantaneous
repetition rate of 50 Hz (average rate of 15 Hz). The average proton beam
power is 1 MW. A possible future upgrade to 2 $\times $10$^{14}$ ppp and 5
Hz could give an average beam power of 4 MW. At this higher intensity, a
superconducting bunch compressor ring would be needed to maintain the rms
bunch length at 3 ns.

If the facility were built at Fermilab, the proton driver would be a newly
constructed 16~GeV rapid cycling booster synchrotron~\cite{FNALbooster}. The
planned facility layout is shown in Fig.~\ref{Proton:fnal}. The initial beam
power would be 1.2 MW, and a future upgrade to 4 MW would be possible. The
Fermilab design parameters are included in Table~\ref{Proton:tb1}. A less
ambitious and more cost-effective 8~GeV proton driver option has also been
considered for Fermilab \cite{FNALbooster}; this too might be the basis for
a proton driver design.

\begin{table}[tbh]
\caption{Proton driver parameters for BNL and FNAL designs.}
\label{Proton:tb1}
\begin{center}
\begin{tabular}{|l|c|c|}
\hline
& BNL & FNAL \\ \cline{2-3}
Total beam power (MW) & 1 & 1.2 \\ 
Beam energy (GeV) & 24 & 16 \\ 
Average beam current ($\mu $A) & 42 & 72 \\ 
Cycle time (ms) & 400 & 67 \\ 
Number of protons per fill & $1\times 10^{14}$ & $3\times 10^{13}$ \\ 
Average circulating current (A) & 6 & 2 \\ 
No. of bunches per fill & 6 & 18 \\ 
No. of protons per bunch & $1.7\times 10^{13}$ & $1.7\times 10^{12}$ \\ 
Time between extracted bunches (ms) & 20 & 0.13 \\ 
Bunch length at extraction, rms (ns) & 3 & 1 \\ \hline
\end{tabular}
\end{center}
\end{table}

\subsection{Target and Capture}

\begin{figure}[tbh]
\begin{center}
\includegraphics*[width=4in]{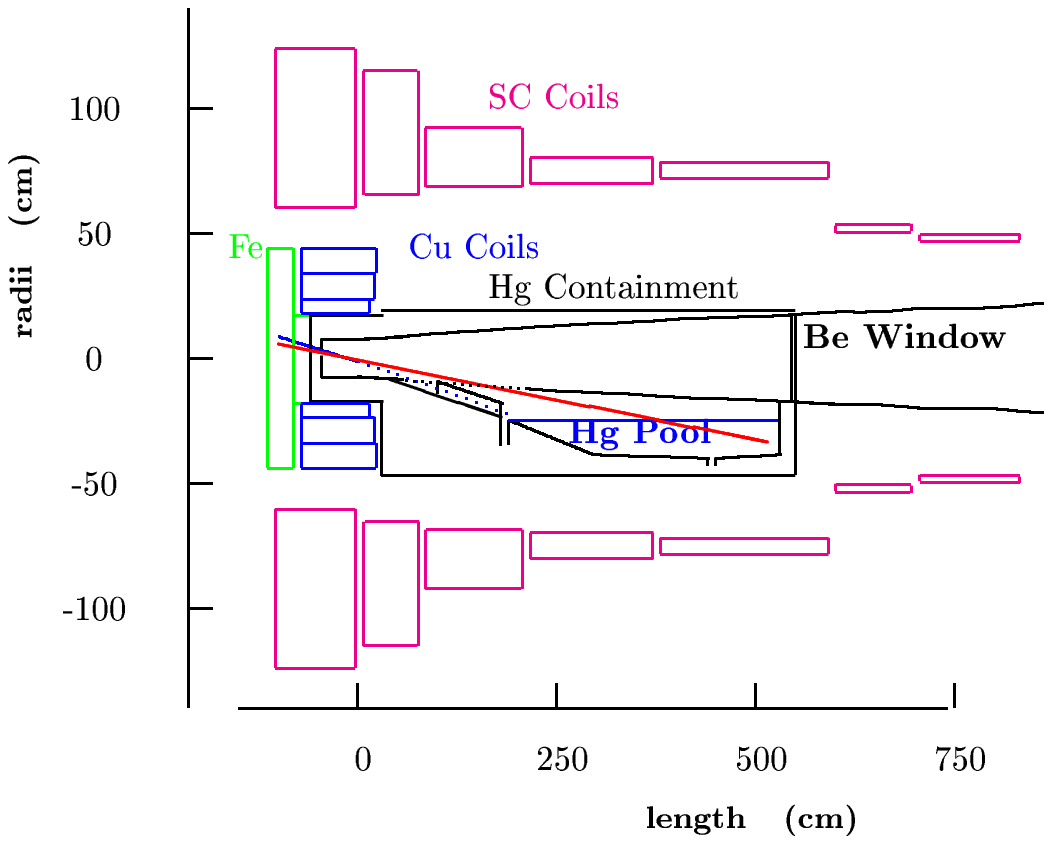}
\end{center}
\caption[Target, capture solenoids and mercury containment ]{(Color)Target, capture
solenoids and mercury containment.}
\label{tgtc}
\end{figure}

A mercury jet target is chosen to give a high yield of pions per MW of
incident proton power. The 1~cm diameter jet is continuous, and is tilted
with respect to the beam axis. The target layout is shown in Fig.~\ref{tgtc}.  
We assume that the thermal shock from the interacting proton bunch fully
disperses the mercury, so the jet must have a velocity of 20--30 m/s to be
replaced before the next bunch. Calculations of pion yields that reflect the
detailed magnetic geometry of the target area have been performed with the
MARS code~\cite{MARSstudyii}. To avoid mechanical fatigue problems, a
mercury pool serves as the beam dump. This pool is part of the overall
target---its mercury is circulated through the mercury jet nozzle after
passing through a heat exchanger.

Pions emerging from the target are captured and focused down the decay
channel by a solenoidal field that is 20 T at the target center, and
tapers down, over 18 m, to a periodic (0.5~m) superconducting solenoid
channel ($B_{z}=1.25\text{ T}$) that continues through the phase
rotation to the start of bunching.  Note that the longitudinal
direction of the fields in this channel do not change sign from cell
to cell as they do in the cooling channel.  The 20~T solenoid, with a
resistive magnet insert and superconducting outer coil, is similar in
character to the higher field (up to 45 T), but smaller bore, magnets
existing at several laboratories \cite {ITERmag}. The magnet insert is
made with hollow copper conductor having ceramic insulation to
withstand radiation. MARS~\cite{MARSstudyii} simulations of radiation
levels show that, with the shielding provided, both the copper and
superconducting magnets could have a lifetime greater than 15 years at
1 MW.

In Study~I, the target was a solid carbon rod. At high beam energies, this
implementation has a lower pion yield than the mercury jet, and is expected
to be more limited in its ability to handle the proton beam power, but
should simplify the target handling issues that must be dealt with. At lower
beam energies, say 6 GeV, the yield difference between C and Hg essentially
disappears, so a carbon target would be a competitive option with a lower
energy driver. Present indications \cite{Ref:ORNLtgt} are that a
carbon-carbon composite target can be tailored to tolerate even a 4 MW
proton beam power---a very encouraging result. Other alternative approaches,
including a rotating Inconel band target, and a granular Ta target are also
under consideration, as discussed in Study~II \cite{EPP:studyii}. Clearly
there are several target options that could be used for the initial facility.

\subsection{Phase Rotation}

The function of the phase rotation section in a neutrino factory is to
reduce the energy spread of the collected muon beam to a manageable
level, allowing reasonable throughput in the subsequent system
components. The following description refers specifically to the
properties of the U.S. Feasibility Study 2 for a neutrino factory. The
initial pions are produced in the mercury target with a very wide
range of momenta. The momentum spectrum peaks around 250~MeV/$c$, but
there is a tail of high energy pions that extends well beyond 1~GeV.
The pions are spread in time over about 3 ns, given by the pulse
duration of the proton driver. After the 18 m long tapered collection
solenoid and an 18~m long drift section, where the beam is focused by
1.25~T solenoids, most of the low energy pions have decayed into
muons. At this point the muon energy spectrum also extends over an
approximately 1~GeV range and the time spectrum extends over
approximately 50~ns. However, there is a strong correlation between
the muon energy and time that can be used for ``phase rotation''.

\begin{table}
\caption{Properties of the induction linacs used in Feasibility Study 2.}
\label{induc}
\begin{tabular}{lcccc}
Induction Linac & & 1 & 2 & 3 \\\hline\hline
Length & m & 100 & 80 & 100 \\
Peak gradient & MV/m & 1.5 & -1.5 & 1.0 \\
Pulse FWHM & ns & 250 & 100 & 380 \\
Pulse start offset & ns & 55 & 0 & 55 \\
\end{tabular}
\end{table}
In the phase rotation process an electric field is applied at
appropriate times to decelerate the leading high energy muons and to
accelerate the trailing low energy ones. Since the bunch train
required by a neutrino factory can be very long, it is possible to
minimize the energy spread using induction linacs. The induction linac
consists of a simple non-resonant structure, where the drive voltage is
applied to an axially symmetric gap that encloses a toroidal
ferromagnetic material. The change in flux in the magnetic core
induces an axial electric field that provides particle acceleration.
The induction linac is typically a low gradient structure that can
provide acceleration fields of varying shapes and time durations from
tens of nanoseconds to several microseconds. Some properties of the
induction linacs are given in Table~\ref{induc}.
 
\begin{figure}[tbh]
  \centering
  \includegraphics*[width=4in]{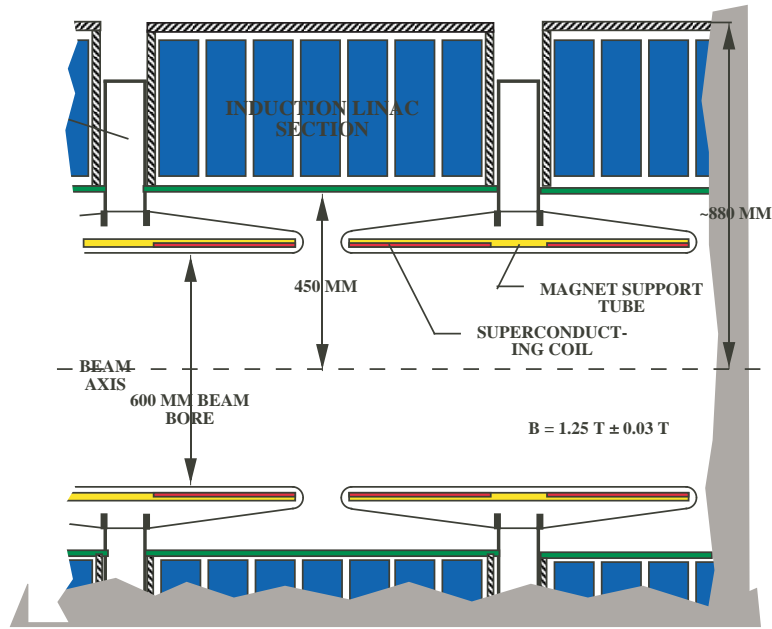}
  \caption[Induction cell and mini-cooling solenoid]{(Color)Cross section of the
    induction cell and transport solenoids.}
  \label{CandPR:fg1}
\end{figure}
Three induction linacs are used in a system that reduces distortion in
the phase-rotated bunch, and permits all induction units to operate
with unipolar pulses. The induction units are similar to those being
built for the DARHT project \cite{daarht}. The 1.25 T beam transport solenoids
are placed inside the induction cores in order to avoid saturating the
core material, as shown in Fig.~\ref{CandPR:fg1}.

Between the first and second induction linacs two liquid hydrogen
absorbers (each 1.7~m long and 30~cm radius) are used to (1) provide
some initial cooling of the transverse emittance of the muon beam and
(2) lower the average momentum of the beam to match better the
downstream cooling channel acceptance. This process is referred to as
``mini-cooling''. The direction of the solenoid magnetic field is
reversed between the two absorbers. The presence of material in the
beam path destroys the conservation of canonical angular momentum that
occurs when a particle enters and leaves a solenoid in vacuum. The
build-up of this angular momentum would eventually lead to emittance
growth. However, this growth can be minimized by periodically
reversing the direction of the field.

\begin{figure}[t]
  \centering
  \includegraphics[width=\textwidth]{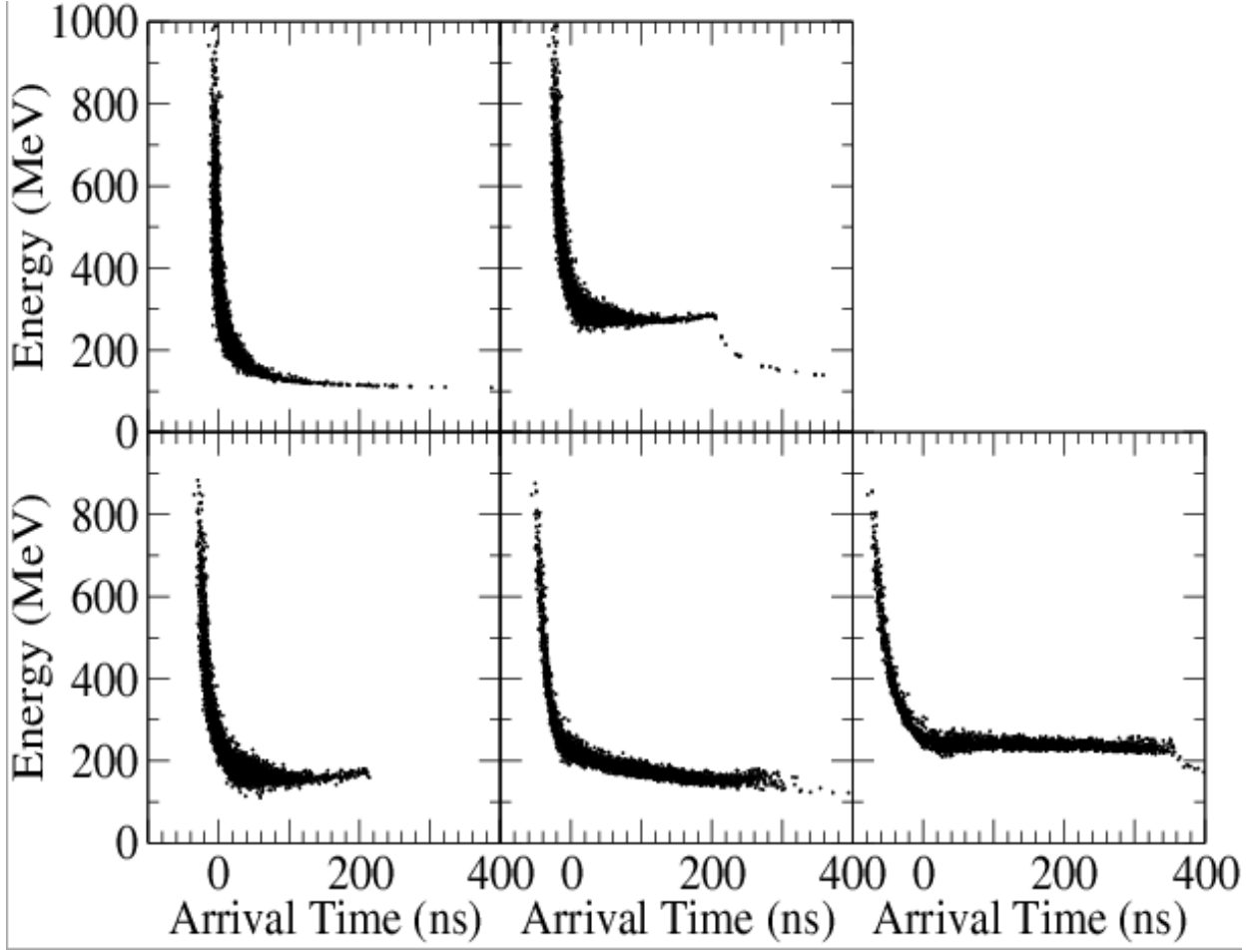}
  \caption{Evolution of the beam distribution in the phase rotation section.
    The graphs show the distribution before the phase rotation, after
    the first induction linac (top row, left to right), after
    mini-cooling, and after the second and third induction linacs
    (bottom row).}
  \label{fig:phaserot}
\end{figure}
The beam at the end of the phase rotation section has an average
momentum of about 250 MeV/c and an rms fractional energy spread of
$\approx$4.4\%.  Figure~\ref{fig:phaserot} shows the evolution of the
beam distribution in the phase rotation section.

\subsection{Buncher}

\begin{figure}[t]
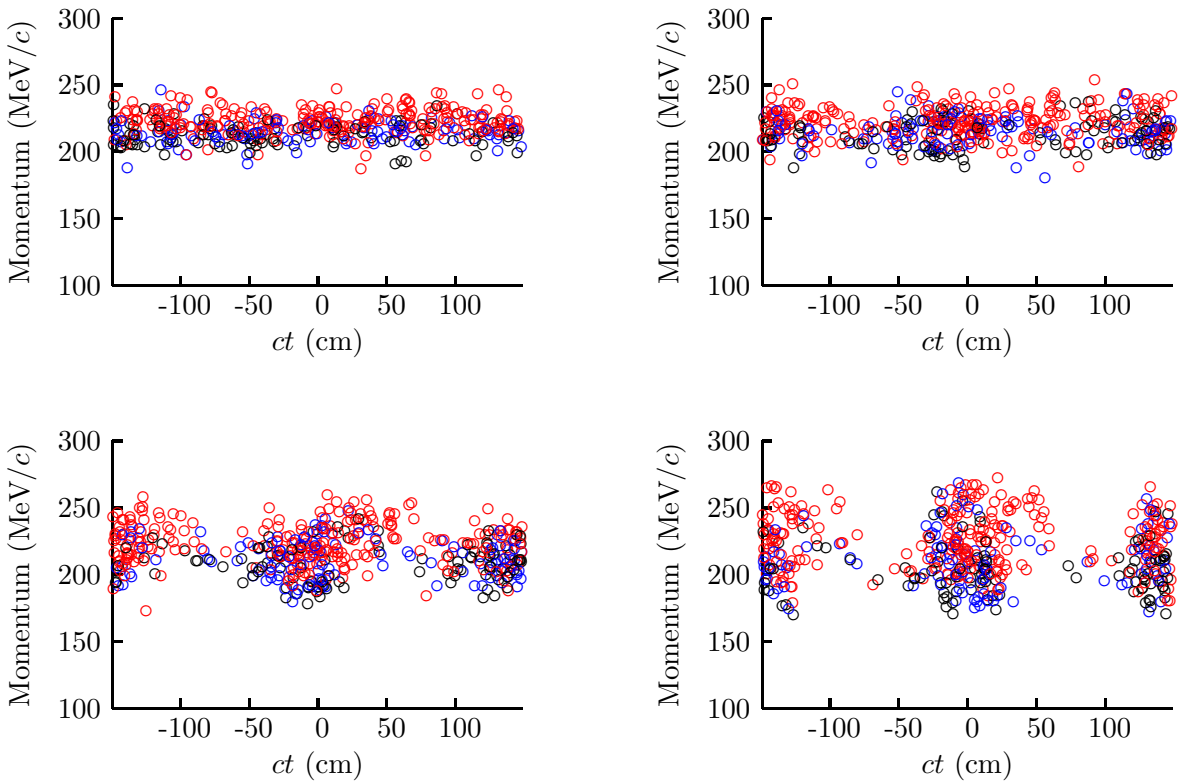

  \centering
  \setlength\unitlength{0.2in}


  \caption{(Color)Evolution of beam in buncher.
    Plots are at the beginning of the buncher (top left), and at the ends
    of the three bunching stages (top right, bottom left, and bottom
    right, in that order).}
  \label{fig:buncher}
\end{figure}
The long beam pulse (400 ns) after the phase rotation is then bunched
at 201.25 MHz prior to cooling and acceleration at that frequency. The
bunching is done in a lattice identical to that at the start of the
cooling channel, and is preceded by a matching section from the 1.25~T
solenoids into this lattice. The bunching has three stages, each
consisting of rf (with increasing acceleration) followed by drifts
(with decreasing length). In the first two rf sections,
second-harmonic 402.5~MHz rf is used together with the 201.25 MHz
primary frequency to improve the capture efficiency. The 402.5~MHz
cavities are designed to fit into the bore of the focusing solenoids,
in the location corresponding to that of the liquid hydrogen absorber
in the downstream cooling channel.  Their aperture radius for the
402.5~MHz cavities is 20~cm at the IRIS, while that of the 201.25~MHz
cavities is 25~cm.  The gradients on axis in the cavities are 6.4~MV/m
for the 402.5~MHz cavities, and range from 6 to 8~MV/m for the
201.25~MHz cavities.  The resulting bunches fill the 201.25~MHz
stationary rf bucket.  Figure~\ref{fig:buncher} shows the evolution of
the longitudinal distribution in the buncher.

\subsection{Cooling}

The transverse emittance of the muon beam after phase rotation and
bunching must be reduced in order to fit into the downstream
accelerators and storage ring. Ionization cooling is currently the
only feasible option for cooling the beam within the muon lifetime. In
ionization cooling the transverse and longitudinal momenta are lowered
in the absorbers, but only the longitudinal momentum is restored by
the rf. The following description refers specifically to the
properties of the U.S.\ Feasibility Study 2 for a neutrino factory.
Transverse emittance cooling is achieved using cooling cells that
(1) lower the beam energy by 7-12~MeV in liquid hydrogen absorbers, (2)
use 201~MHz rf cavities to restore the lost energy, and (3) use 3--5~T
solenoids to strongly focus the beam at the absorbers. At the end of
the cooling channel the rms normalized transverse emittance is reduced
to about 2.5~mm~rad.
 
\begin{figure}[hbt!]
\centering
\includegraphics[width=4in]{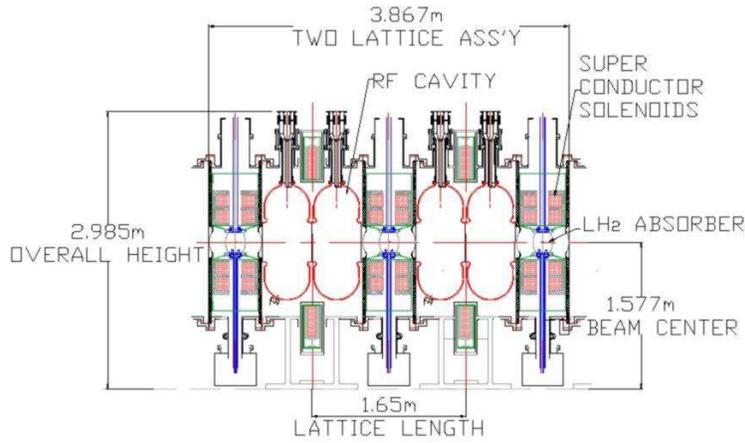}
\caption{(Color)Two cells of the 1.65 m cooling lattice.}
\label{RF:fg18.Q}
\end{figure}
Each cell of the lattice contains three solenoids. The direction of
the solenoidal field reverses in alternate cells in order to prevent
the build-up of canonical angular momentum, as mentioned above in the
discussion of mini-cooling. In analogy with the FODO lattice this
focusing arrangement is referred to as a (S)FOFO lattice.  Multiple
Coulomb scattering together with the focusing strength determine the
asymptotic limit on the transverse emittance that the cooling channel
can reach. The focusing strength in the channel is tapered so that
the angular spread of the beam at the absorber locations remains
large compared to the characteristic spread from scattering. This is
achieved by keeping the focusing strength inversely proportional to
the emittance, \textit{i.e.}, increasing it as the emittance is
reduced. The solenoidal field profile was chosen to maximize the
momentum acceptance ($\pm$22\%) through the channel. To maintain the
tapering of the focusing it was eventually necessary to reduce the
cell length from 2.75~m in the initial portion of the channel to 1.65
m in the final portion. A layout of the shorter cooling cells is
shown in Fig.~\ref{RF:fg18.Q}.

\begin{figure}[tbh]
\centering
\includegraphics*[width=100mm]{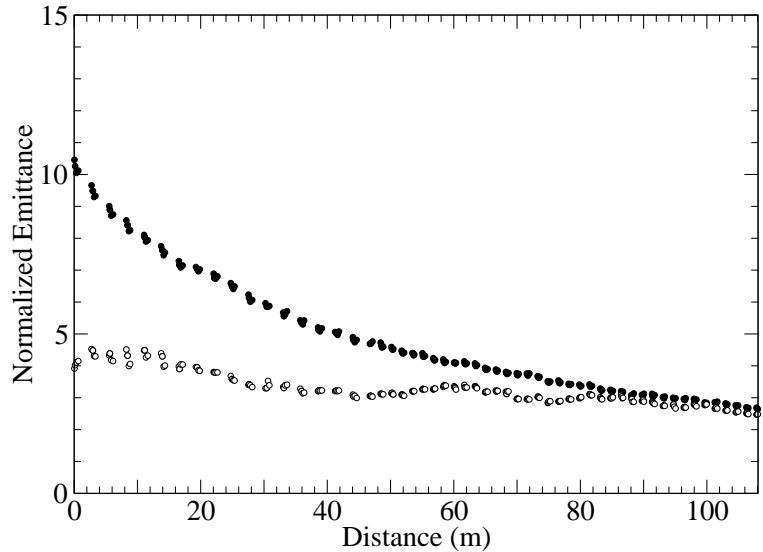}
\caption[The transverse and longitudinal emittances]{The transverse
  (filled circles, in mm) and longitudinal (open circles, in cm)
  emittances, as a function of the distance down the cooling channel.}
\label{EmittCool}
\end{figure}
\begin{figure}[tbh]
\centering
\includegraphics*[width=100mm]{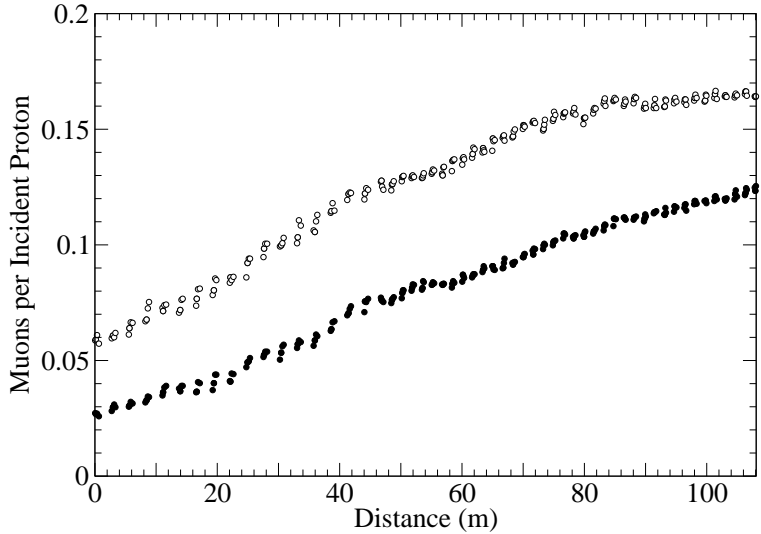}
\caption[$\protect\mu /p$ yield ratio for the two transverse emittance
cuts]{Muons per incident proton in the cooling channel that would
  fall within a normalized transverse acceptance of 15~mm (open circles) or
  9.75~mm (filled circles).}
\label{YieldCool}
\end{figure}
Figure~\ref{EmittCool} shows a simulation of cooling in this channel.
The transverse emittance decreases steadily along the length of the
channel. This type of channel only cools transversely, so the
longitudinal emittance increases until the rf bucket is full and then
remains fairly constant as particles are lost from the bucket. A
useful figure of merit for cooling at a neutrino factory is the
increase in the number of muons that fit within the acceptance of the
downstream accelerators. This is shown in Fig.~\ref{YieldCool}. At
each axial position the number of muons is shown that fall within two
acceptances appropriate to a downstream accelerator. Both acceptances
require the muon longitudinal phase space be less than 150~mm. The
density of particles within a normalized transverse acceptance, for
example, steadily increases by a factor of about 3 over the channel
length, clearly showing the results of cooling. The saturation of the
yield determined the chosen channel length of 108~m.

\subsection{Acceleration}

\begin{figure}[tbh]
  \centering
  \includegraphics[width=\textwidth]{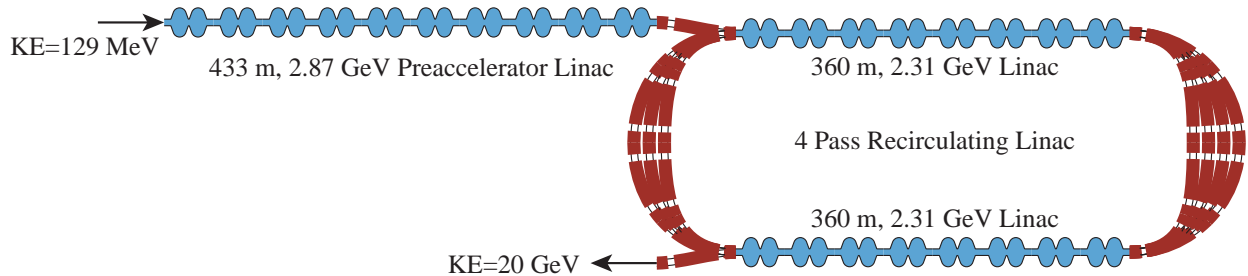}
  \caption{(Color)Accelerating system layout.}
  \label{fig:acc:layout}
\end{figure}
The layout of the acceleration system is shown in
Fig.~\ref{fig:acc:layout}, and its parameters are listed in
Table~\ref{tab:acc:parm}.  The acceleration system consists of a
preaccelerator linac followed by a four-pass recirculating linac.  The
recirculating linac allows a reduction in the amount of rf required
for acceleration by passing the beam through linacs multiple times.
The linacs are connected by arcs, and a separate are is used for each
pass.  At low energies, however, the large emittance of the beam would
require a much shorter cell length and larger aperture than is
desirable and needed at higher energies.  This, combined with
difficulties in injecting the large emittance and energy spread beam
into the recirculating accelerator, and the loss of efficiency due to
the phase slip at low energies, lead to the necessity for a linac
that precedes the recirculating linac.

\begin{figure}[tbh]
\centering\includegraphics[angle=270,width=4.0in]{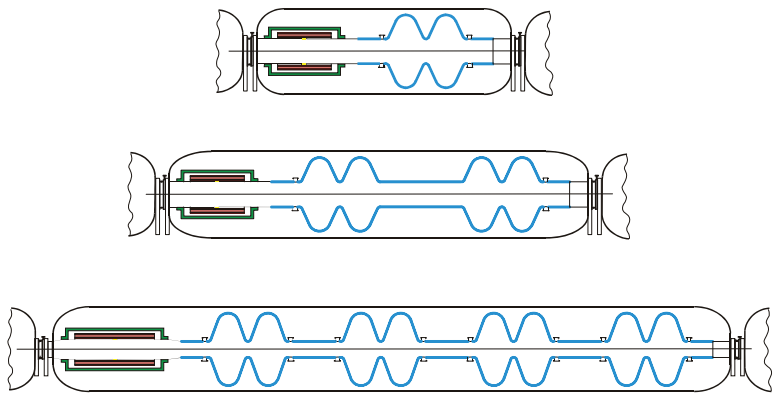}
\caption[Layouts of cryomodules.]{(Color)Layouts of short (top), intermediate
(middle) and long (bottom) cryomodules. Blue lines are the SC walls of the
cavities. Solenoid coils are indicated in red.}
\label{fig:acc:cryomod}
\end{figure}

A 20~m SFOFO matching section, using normal conducting rf systems, matches
the beam optics to the requirements of a 2.87 GeV superconducting rf linac
with solenoidal focusing.  The linac is in three parts. The first part has a
single 2-cell rf cavity unit per period. The second part, as a longer period
becomes possible, has two 2-cell cavity units per period. The last section,
with still longer period, accommodates four 2-cell rf cavity units per
period. See Tables~\ref{tab:acc:linac} and ~\ref{tab:acc:cav} for 
details of the rf cryostructures and cavities. 
Figure~\ref{fig:acc:cryomod} shows the three cryomodule types that
make up the linac.

\begin{table}[tbh]
\caption{Main parameters of the muon accelerator.}
\label{tab:acc:parm}
\begin{center}
\begin{tabular}{|l|c|}
\hline
Injection momentum (MeV/c)/Kinetic energy (MeV) & 210/129.4 \\ 
Final energy (GeV) & 20 \\ 
Initial normalized transverse acceptance (mm~rad) & 15 \\ 
\quad rms normalized transverse emittance (mm~rad) & 2.4 \\ 
Initial longitudinal acceptance, $\Delta pL_{b}/m_{\mu }$ (mm) & 170 \\ 
\quad momentum spread, $\Delta p/p$ & $\pm 0.21$ \\ 
\quad bunch length, $L_{b}$ (mm) & $\pm 407$ \\ 
\quad rms energy spread & 0.084 \\ 
\quad rms bunch length (mm) & 163 \\ 
Number of bunches per pulse & 67 \\ 
Number of particles per bunch\textbf{/}per pulse & $4.4\times 10^{10}$ 
\textbf{/}$3\times 10^{12}$ \\ 
Bunch frequency\textbf{/}accelerating frequency (MHz) & 201.25\textbf{/}
201.25 \\ 
Average beam power (kW) & 150 \\ \hline
\end{tabular}
\end{center}
\end{table}

\begin{figure}[!tbh]
\centering\includegraphics[angle=270,width=4.0in]{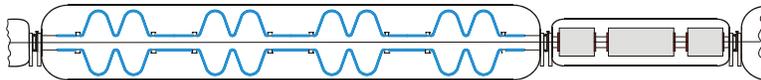}
\caption{(Color)Layout of an RLA linac period.}
\label{fig:acc:rlalinac}
\end{figure}

\begin{table}[tbh]
  \caption{Parameters for three types of linac cryomodules.}
  \label{tab:acc:linac}
  \centering
  \begin{tabular}{|l|c|c|c|}
    \hline
    Cavities per period&1&2&4\\
    Period length (m)&5&8&13\\
    Number of periods&11&16&19\\
    Cavity type&A&A&B\\
    Solenoid full aperture (cm)&46&46&36\\
    Solenoid length (m)&1&1&1.5\\
    Maximum solenoid field (T)&2.1&2.1&4.2\\
    \hline
  \end{tabular}
\end{table}
\begin{table}[tbh]
  \caption{Parameters for superconducting cavities.}
  \label{tab:acc:cav}
  \centering
  \begin{tabular}{|l|c|c|}
    \hline
    &A&B\\
    \hline
    Frequency (MHz)&201.25&201.25\\
    Cells per cavity&2&2\\
    Aperture diameter (cm)&46&30\\
    Energy gain (MV)&22.5&25.5\\
    rf pulse length (ms)&3&3\\
    Input Power (kW)&980&1016\\
    Peak surface field (MV/m)&23.1&24.3\\
    $Q_0$&$6\times10^9$&$6\times10^9$\\
    \hline
  \end{tabular}
\end{table}
This linac is followed by a single four-pass recirculating linear
accelerator (RLA) that raises the energy from 2.5 GeV to 20 GeV. The
RLA uses the same layout of four 2-cell superconducting rf cavity
structures as the long cryomodules in the linac, but utilizes
quadrupole triplet focusing, as indicated in
Fig.~\ref{fig:acc:rlalinac}.

The arcs have an average radius of 62 m, and are all in the same
horizontal plane. They also utilize triplet focusing.  There are
around 120 arc cells, with 2.15~m dipoles, and triplet quadrupoles
which are very similar to those in the linacs.  The required full
quadrupole apertures vary from 20~cm to 12~cm, and the dipole gaps
vary from 14~cm to 9~cm.  All magnet pole tip fields are under 2~T,
except in the switchyard where they are as high as 2.3~T in some
cases (and the magnet apertures rise to 21~cm).

The 4.5~K-equivalent cryogenic load for the entire acceleration system is 
27.9~kW. In Study~I, where the final beam energy was chosen to be 50 GeV, a second
RLA is needed. This second RLA is similar to the RLA just described, but
considerably larger.

\subsection{Storage Ring}

After acceleration in the RLA, the muons are injected into the
upward-going straight section of a racetrack-shaped storage ring with
a circumference of 358 m. Parameters of the ring are summarized in
Table~\ref{SRING:tb}.  High-field superconducting arc magnets are used
to minimize the arc length.  Minimizing the arc length for a given
length of straight maximizes the fraction of the circumference
contained in the straight section, thereby maximizing the fraction of
neutrinos (around 35\% in our case) decaying toward the detector.

Furthermore, the beta functions in the downward-going straight (which
is pointed toward the detector) are made large to reduce the angular
divergence of the beam.  This ensures that the angular divergence of
the beam is dominated by the calculable relativistic angular
divergence of the decay neutrinos.  The goal of this is not only to
make the angular divergence of the neutrino beam as small as possible
and therefore maximize the flux, but it also reduces the experimental
uncertainty associated with an uncertainty in the flux that would come
from an uncertainty in the angular divergence of the muon beam.

\begin{figure}[t]
  \centering
  \includegraphics[width=\textwidth]{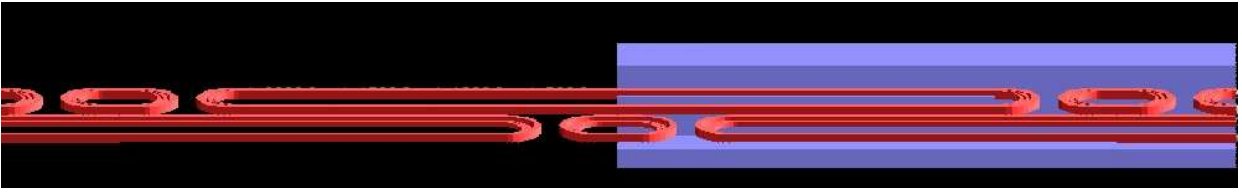}
  \caption{(Color)Three-dimensional view of the storage ring magnets.}
  \label{fig:sr:mag}
\end{figure}
\begin{figure}[tbh]  %% figure 21
\centering
\includegraphics[width=4.0in]{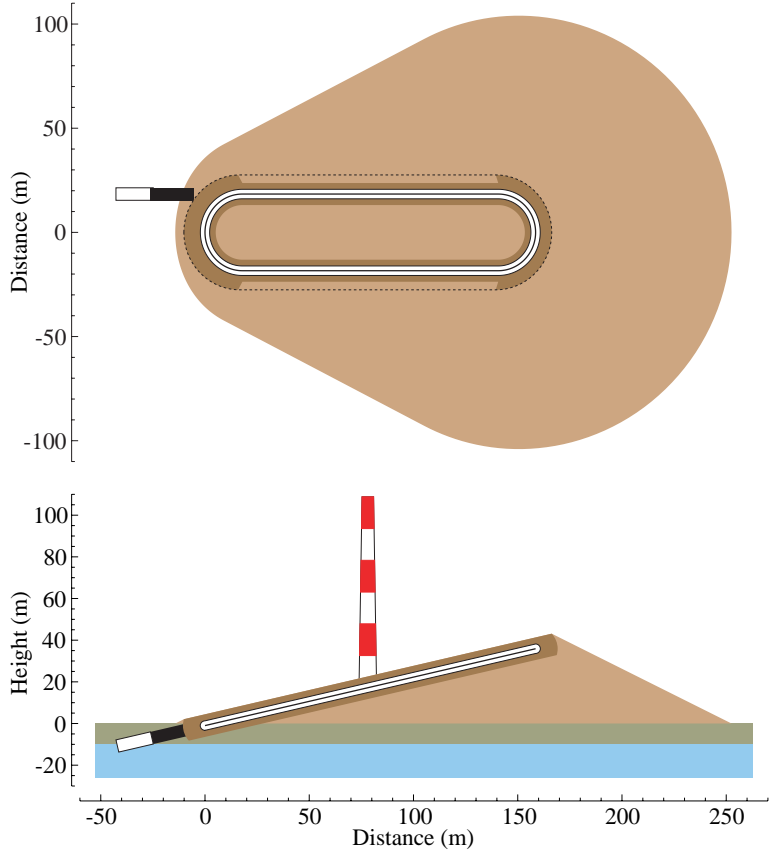}
\caption[Top view and cross section through ring and berm]{(Color)Top view and
cross section through 20~GeV ring and berm. The existing 110~m tower, drawn
to scale, gives a sense of the height of the ring on the BNL landscape.}
\label{EPP:fgsection}
\end{figure}

All muons are allowed to decay; the maximum heat load from their decay
electrons is 42 kW (126 W/m). This load is too high to be dissipated
in the superconducting coils. For Study~II, a magnet design
(see Fig.~\ref{fig:sr:mag}) has been
chosen that allows the majority of these electrons to exit between
separate upper and lower cryostats, and be dissipated in a dump at
room temperature. To maintain the vertical cryostat separation in
focusing elements, skew quadrupoles are employed in place of standard
quadrupoles.  The result is a skew FODO lattice, giving
diagonal oscillations, as opposed to the horizontal and vertical
oscillations of the usual upright FODO lattice.  In order to maximize
the average bending field, Nb$_{3}$Sn pancake coils are employed.  One
coil of the bending magnet is extended and used as one half of the
previous (or following) skew quadrupole to minimize unused space.
Figure~\ref {EPP:fgsection} shows a layout of the ring as it would be
located at BNL.  (The existing 110~m~high BNL smokestack is shown for
scale.) For site-specific reasons, the ring is kept above the local
water table and is placed on a roughly 30~m~high berm. This
requirement places a premium on a compact storage ring.

The beam is allowed to debunch in the storage ring.  In one muon
lifetime (0.42~ms), a bunch with the full energy spread
($\pm\text{2.2\%}$) will have its total length increase by 0.51~$\mu$s
(the storage ring is 1.19~$\mu$s long, and the bunch train starts out
0.33~$\mu$s long).  If one wishes to avoid the increase in the bunch
train length, one could install rf cavities, but the voltage required
to avoid energy spread increase for the momentum compaction in this
ring is prohibitive:  a better solution would be a ring re-designed to
have very low momentum compaction.

\begin{table}[tb]
\caption{Muon storage ring parameters.}
\label{SRING:tb}
\begin{center}
\begin{tabular}{|l|l|}
\hline
Energy (GeV) & 20 \\ 
Circumference (m) & 358.18 \\ 
Normalized transverse acceptance (mm~rad) & 30 \\ 
Energy acceptance (\%) & 2.2 \\
Momentum compaction&0.028\\
\hline
\multicolumn{2}{c}{\underline{Arc}} \\ 
\hline
Length (m) & 53.09 \\ 
No. cells per arc & 10 \\ 
Cell length (m) & 5.3 \\ 
Phase advance ($\deg $) & 60 \\ 
Dipole length (m) & 1.89 \\ 
Dipole field (T) & 6.93 \\ 
Skew quadrupole length (m) & 0.76 \\ 
Skew quadrupole gradient (T/m) & 35 \\ 
$\beta _{\text{max}}$ (m) & 8.6 \\
\hline 
\multicolumn{2}{c}{\underline{Production Straight}} \\ 
\hline
Length (m) & 126 \\ 
$\beta _{\text{max}}$ (m) & 200 \\ \hline
\end{tabular}
\end{center}
\end{table}

For Study~I, a conventional superconducting ring was utilized to store the
50 GeV muon beam. The heat load from muon decay products in this scenario is
managed by having a liner inside the magnet bore to absorb the decay
products. This approach is likewise available for BNL, provided some care is
taken to keep the ring compact; acceptable lattice solutions have been found
for this option as well.

\subsection{Overall Machine Summary}

\begin{figure}[t]
  \centering\includegraphics[width=100mm]{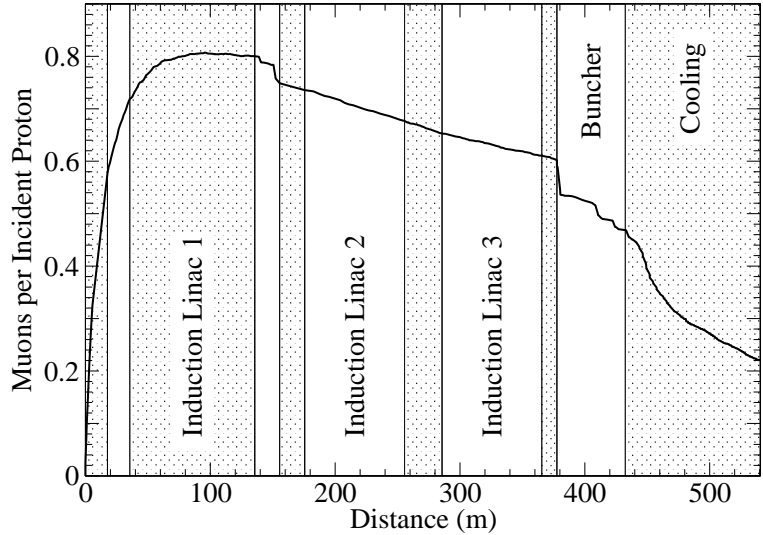}
  \caption{Muons per incident proton in the Study~II neutrino factory
    front end.}
  \label{fig:all:muonp}
\end{figure}
\begin{table}[t]
  \caption{Muon survival and losses in several sections of the Study~II
    neutrino factory.  $\mu/p$ is the number of muons per proton at the
    end of that section,
    and ``Loss'' is the loss in that section.}
  \label{tab:all:muloss}
  \begin{tabular}{|l|r|r|}
    \hline
    Section&$\mu$/p&Loss\\
    \hline
    Phase rotation&0.60&\multicolumn{1}{c|}{---}\\
    Buncher&0.47&22\%\\
    Cooling&0.22&53\%\\
    Accelerator aperture&0.16&26\%\\
    Preaccelerator Linac&0.15&10\%\\
    Recirculating Linac&0.13&10\%\\
    \hline
  \end{tabular}
\end{table}
Figure~\ref{fig:all:muonp} shows the muons per incident proton in the
``front end'' (everything before the acceleration) of the Study~II
neutrino factory.
Table~\ref{tab:all:muloss} gives the values at the ends of several
sections and the losses in those sections.  These significant losses
are a necessary cost of making a low-emittance beam that can be
accelerated and injected into a storage ring.

\begin{figure}[tbh]  %% figure 22
\begin{center}
\includegraphics[angle=90,width=0.9\linewidth]{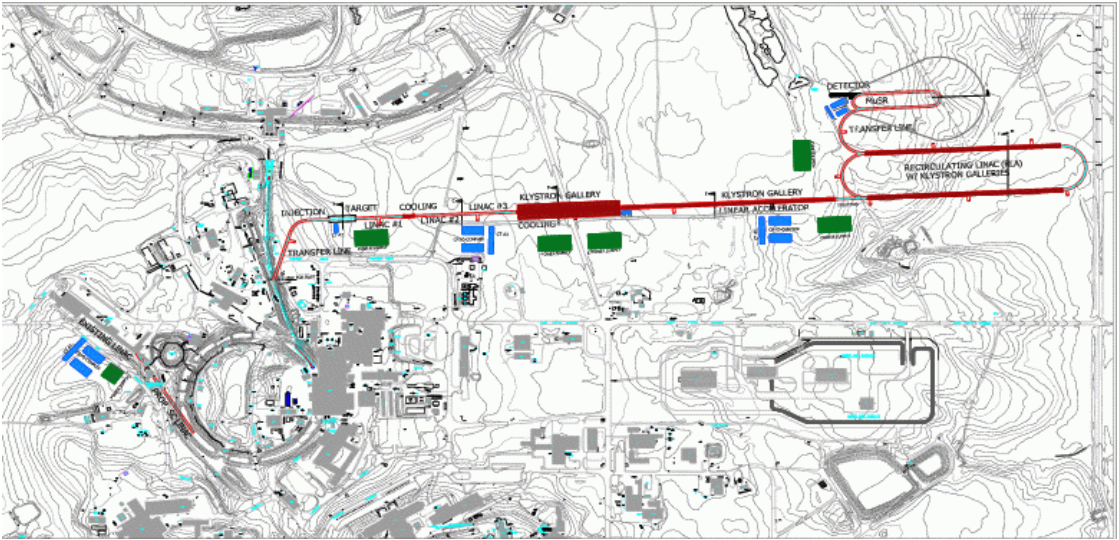}
\end{center}
\caption[(Color)Schematic of a Neutrino Factory at Brookhaven]{Schematic of a
20~GeV Neutrino Factory at BNL.}
\label{bnlsite}
\end{figure}

\begin{figure}[tbh]  %% figure 23
\begin{center}
\includegraphics[width=4in,angle=270]{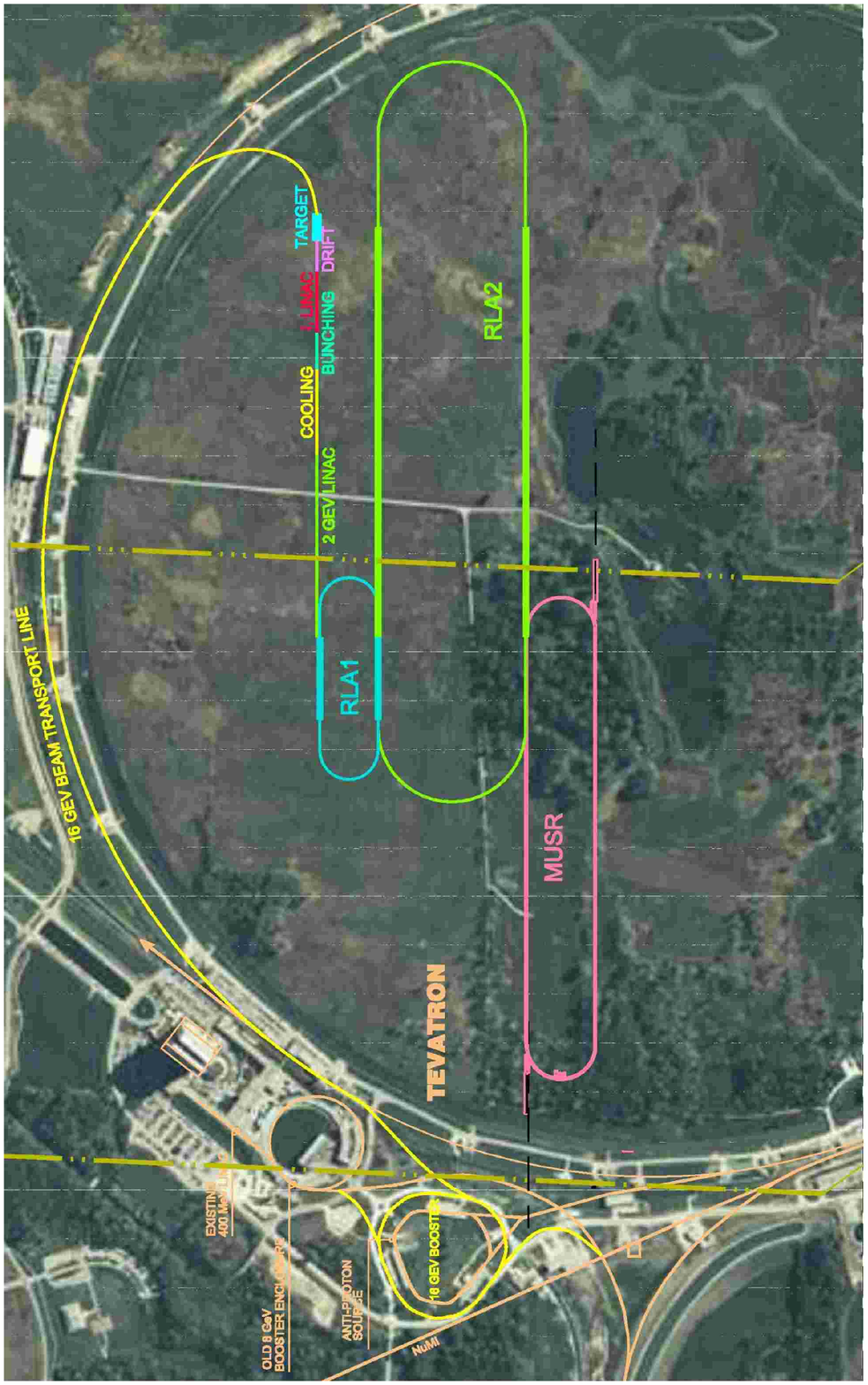}
\end{center}
\caption[Schematic of a Neutrino Factory at Fermilab]{(Color)Schematic of a 50~GeV
Neutrino Factory at Fermilab.}
\label{fnalsite}
\end{figure}

An overall layout of the Neutrino Factory on the BNL site is shown in Fig.~\ref{bnlsite}.  
Figure~\ref{fnalsite} shows the equivalent picture for a facility on the
Fermilab site. In this latter case, the layout includes the additional RLA
and longer storage ring needed to reach 50 GeV. Clearly the footprint of a
Neutrino Factory is reasonably small, and such a machine would fit easily on
the site of either BNL or Fermilab. 

\subsection{Detector}

The Neutrino Factory, plus its long-baseline detector, will have a physics
program that is a logical continuation of current and near-future neutrino
oscillation experiments in the U.S., Japan, and Europe. Moreover, detector
facilities located in experimental areas near the neutrino source will have
access to integrated neutrino intensities $10^{4}$--$10^{5}$ times larger
than previously available ($10^{20}$ neutrinos per year compared with $10^{15}$--$10^{16}$).

The detector site taken for Study~II is the Waste Isolation Pilot Plant
(WIPP) in Carlsbad, New Mexico. The WIPP site is approximately 2900~km from
BNL. Space is potentially available for a large underground physics facility
at depths of 740--1100~m, and discussions are under way between DOE and the
UNO project~\cite{DET:uno} on the possible development of such a facility.

\subsubsection{Far Detector}

Specifications for the long-baseline Neutrino Factory detector are rather
typical for an accelerator-based neutrino experiment. However, the need to
maintain a high neutrino rate at these long distances requires detectors
3--10 times more massive than those in current neutrino experiments.
Clearly, the rate of detected neutrinos depends on two factors---the source
intensity and the detector size. In the final design of a Neutrino Factory,
these two factors would be optimized together.

\begin{figure}[tbh]
\begin{center}
\includegraphics[width=3.5in]{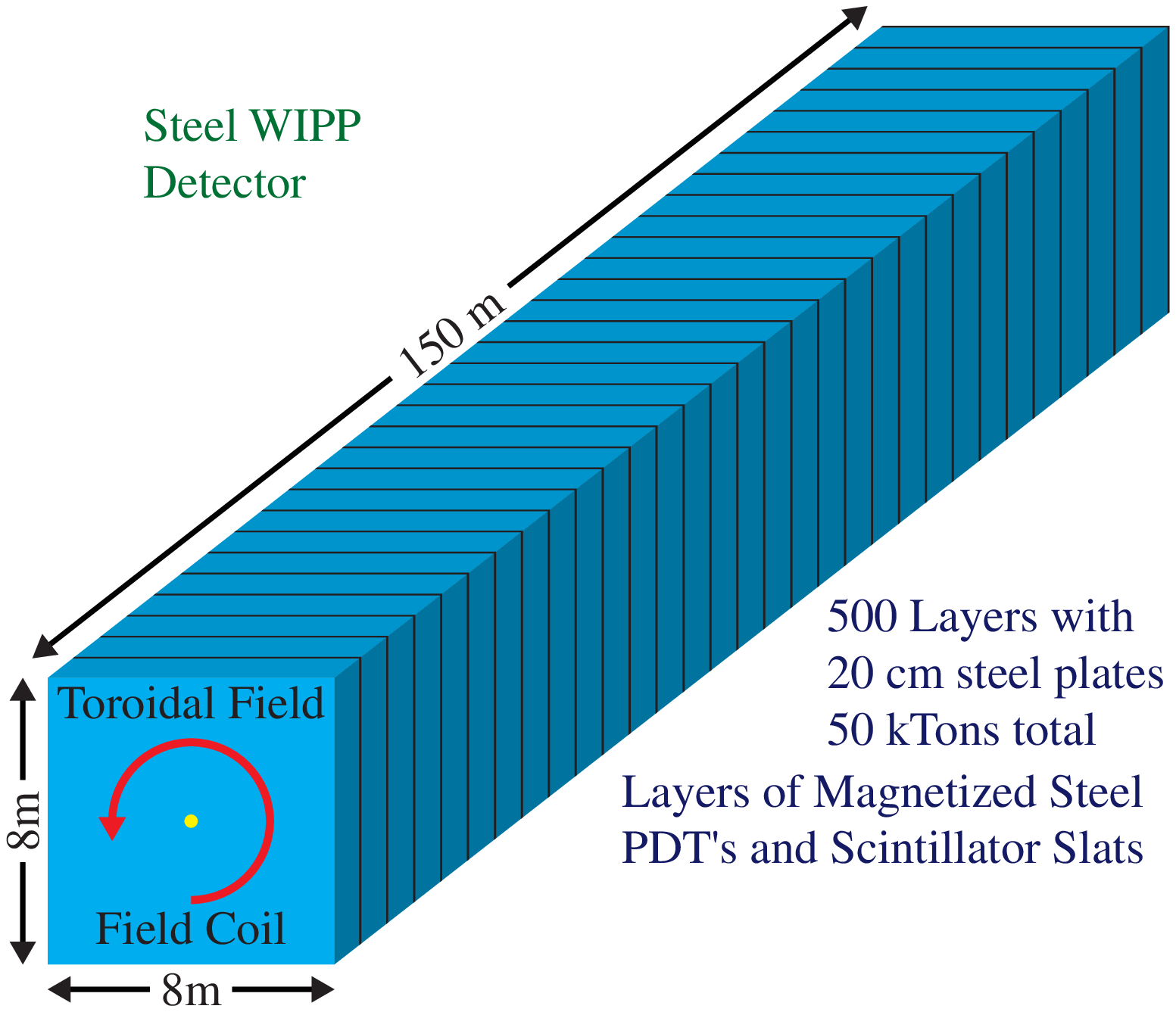}
\end{center}
\caption[A possible 50 kton Steel/Scintillator/PDT detector at WIPP]{(Color)A
possible 50~kton steel-scintillator-PDT detector at WIPP.}
\label{fg:steelwipp}
\end{figure}

Two options are considered for the WIPP site: a 50 kton
steel--scintillator--proportional-drift-tube (PDT) detector or a
water-Cherenkov detector. The PDT detector would resemble MINOS. Figure~\ref
{fg:steelwipp} shows a 50~kton detector with dimensions $8~\text{m}\times 8~%
\text{m}\times 150$~m. A detector of this size would record up to $4\times
10^{4}$ $\nu _{\mu }$ events per year. 

\begin{figure}[tbh]
\begin{center}
\includegraphics[width=3.0in]{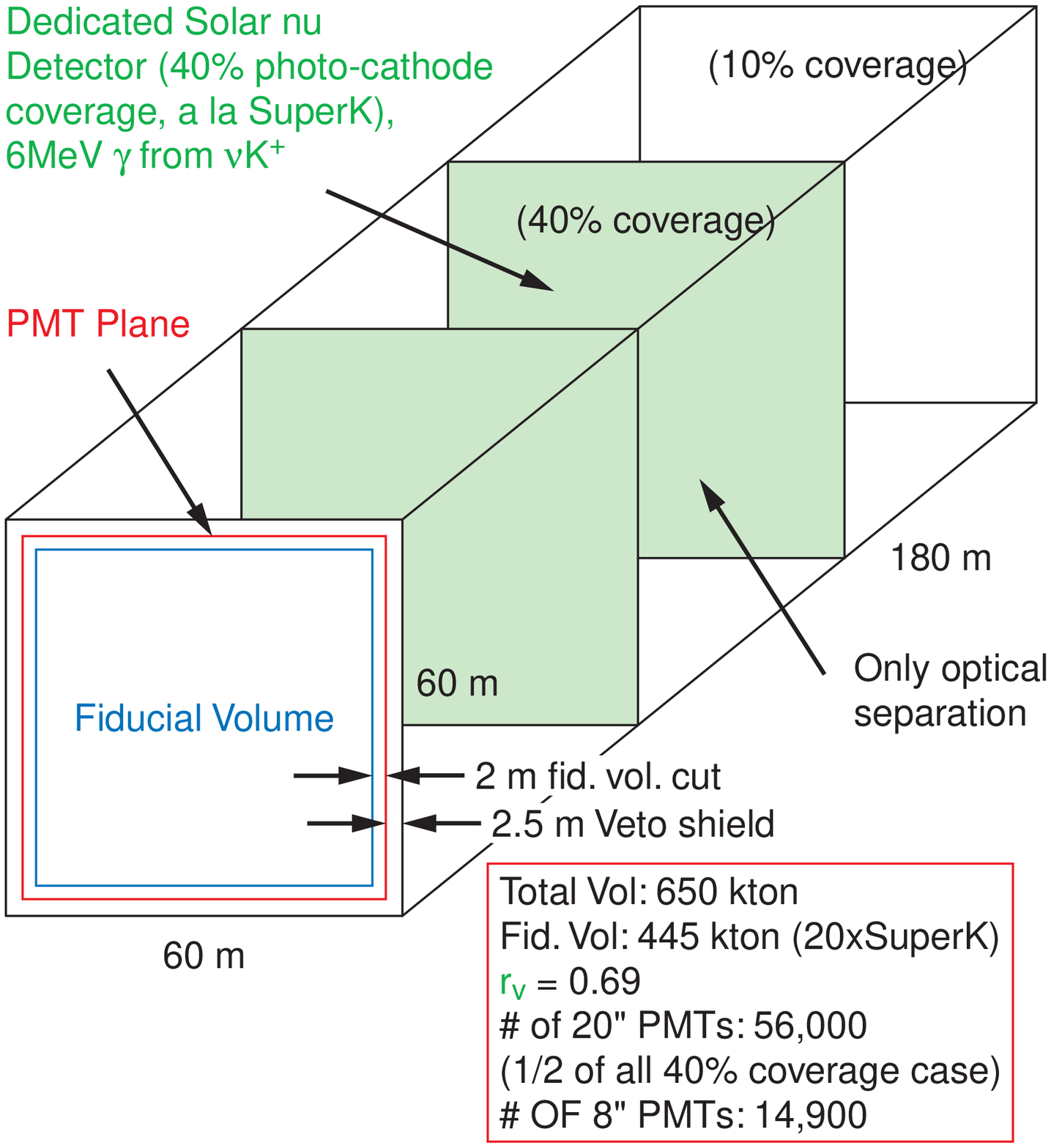}
\end{center}
\caption[Block schematic of the UNO detector]{(Color)Block schematic of the UNO
detector, including initial design parameters.}
\label{fg:unodet}
\end{figure}

A large water-Cherenkov detector would be similar to SuperKamiokande, but
with either a magnetized water volume or toroids separating smaller water
tanks. The detector could be the UNO detector~\cite{DET:uno}, currently
proposed to study both proton decay and cosmic neutrinos. UNO would be a
650~kton water-Cherenkov detector segmented into a minimum of three tanks
(see Fig.~\ref{fg:unodet}). It would have an active fiducial mass of
440~kton and would record up to $3\,\times \,10^{5}$ $\nu _{\mu }$ events
per year from the Neutrino Factory beam. 

Another possibility for a Neutrino Factory detector is a massive
liquid-argon magnetized detector~\cite{landd} that would also attempt to
detect proton decay, as well as solar and supernova neutrinos.

\subsubsection{Near Detector}

As noted, detector facilities located on-site at the Neutrino Factory would
have access to unprecedented intensities of pure neutrino beams. This would
enable standard neutrino physics studies, such as ${\sin }^{2}\theta _{W}$,
structure functions, neutrino cross sections, nuclear shadowing and pQCD to
be performed with much higher precision than previously obtainable. In
addition to its primary physics program, the near detector can also provide
a precise flux calibration for the far detector, though this may not be
critical given the ability to monitor the storage ring beam intensity
independently.

A compact liquid-argon time projection chamber (TPC; similar to the ICARUS
detector~\cite{ICARUS}),
cylindrically shaped with a radius of 0.5 m and a length of 1~m, would have
an active mass of $10^{3}$ kg and a neutrino event rate \textsl{O}(10~Hz).
The TPC could be combined with a downstream magnetic spectrometer for muon
and hadron momentum measurements. At these neutrino intensities, it is even
possible to have an experiment with a relatively thin Pb target (1~$L_{rad}$), followed by a standard fixed-target spectrometer containing tracking
chambers, time-of-flight, and calorimetry, with an event rate ${\cal O}$(1~Hz).

\subsection{Staging Options}

\label{StagingOps}

It seems quite possible---perhaps even likely---that the Neutrino Factory
would be built in stages, both for programmatic and for cost reasons. Here
we outline a possible staging concept that provides good physics
opportunities at each stage. The staging scenario we consider is not unique,
nor is it necessarily optimized. Depending on the results of our technical
studies and the results of continued searches for the Higgs boson, it is
hoped that the Neutrino Factory is really the penultimate stage, to be
followed later by a Muon Collider (Higgs Factory). We assume this
possibility in the staging discussion that follows. Because the physics
program would be different at different stages, it is impractical at this
time to consider detector details.

\subsubsection{Stage 1}

In the first stage, we envision a Proton Driver and a Target Facility to
create superbeams. The Driver could begin with a 1 MW beam level (Stage 1)
or could be designed from the outset to reach 4 MW (Stage 1a). (Since the
cost differential between 1 and 4 MW is not expected to be large, we do not
consider any intermediate options here.) It is assumed, as was the case for
both Study~I and Study~II, that the Target Facility is built from the outset
to accommodate a 4 MW beam. Based on the Study~II results, a 1 MW beam would
provide about $1.2\times 10^{14}$ $\mu $/s ($1.2\times 10^{21}$ $\mu $/year)
and a 4 MW beam about $5\times 10^{14}$ $\mu $/s ($5\times 10^{21}$ $\mu$/year) into a solenoid channel.

In addition to the neutrino program, this stage will also benefit $\pi $, $K$, and $\overline{p}$ programs, as discussed in~\cite{proton-physics,low-pbar}.

\subsubsection{Stage 2}

In Stage 2, we envision a muon beam that has been phase rotated (to give a
reasonably low momentum spread) and transversely cooled. In the nomenclature of
Study~II, this stage takes us to the end of the cooling channel. Thus, we
have access to a muon beam with a central momentum of about 200 MeV/c, a
transverse (normalized) emittance of 2.7~mm~rad and an rms energy spread of
about 4.5\%. The intensity of the beam would be about $4\times 10^{13}$ $\mu 
$/s ($4\times 10^{20}$ $\mu $/year) at 1 MW, or $1.7\times 10^{14}$ $\mu $/s
($1.7\times 10^{21}$ $\mu $/year) at 4 MW. If more intensity were needed,
and if less cooling could be tolerated, the length of the cooling channel
could be reduced. As an example, stopping at the end of Lattice 1 instead of
the end of Lattice 2 in the Study~II cooling channel would result in an
increase of transverse emittance by roughly a factor of two. This is an 
appropriate stage to mount an experiment to search for a non-zero 
muon electric dipole moment.

\subsubsection{Stage 3}

In Stage 3, we envision using the Pre-acceleration Linac to raise the beam
energy to roughly 2.5 GeV. At this juncture, it may be appropriate to
consider a small storage ring, comparable to the $g-2$ ring at BNL, to be
used for the next round of muon $g-2$ experiments.

\subsubsection{Stage 4}

At Stage 4, we envision having a complete Neutrino Factory operating with a
20~GeV storage ring. This stage includes the RLA and the storage ring. If it
were necessary to provide a 50 GeV muon beam as Stage 4a, an additional RLA
and a larger storage ring would be needed.

\subsubsection{Stage 5}

In Stage 5, we could envision an entry-level Muon Collider operating as a
Higgs Factory. Because the initial muon beam must be prepared as a single
bunch of each charge, an additional ring for the proton driver to coalesce
proton bunches into a single pulse is anticipated. The cooling will have to
be significantly augmented. First, a much lower transverse emittance is
needed, and second, it will be necessary to provide longitudinal cooling
(emittance exchange) to maintain a reasonable transmission of the muons. The
additional cooling will permit going to smaller solenoids and higher
frequency rf systems (402.5 or perhaps 805 MHz), which should provide more
cost-effective cooling. Next, we will need considerably more acceleration,
though with smaller energy acceptance and aperture requirements than at
present. Lastly, we will need a very low $\beta ^{\ast }$ lattice for the
collider ring, along with mitigation of the potentially copious background
levels near the interaction point. In this case the detector is, in effect,
part of the collider ring, and its design must be an integral part of the
overall ring design.

\section{Muon Colliders} 
\label{higgsfact}

\begin{figure}[bth!]
\includegraphics[width=6in,height=4.75in]{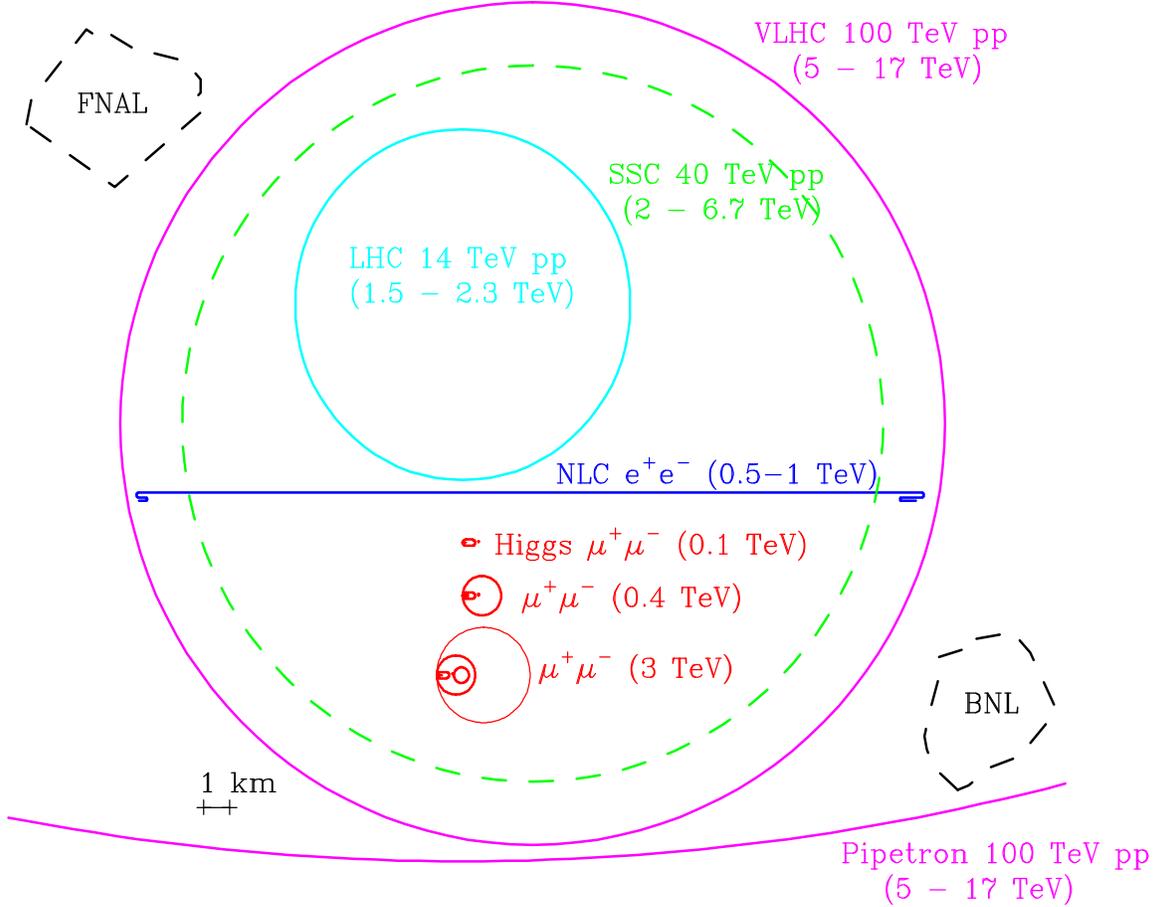}
%\centerline{\epsfig{file=machine_comparison_new.ps,height=4.75in,width=6.in}}
\vspace{0.5cm}
\caption[Sizes of various proposed high energy colliders]
{(Color)Comparative sizes of various proposed high energy colliders compared
with the FNAL and BNL sites. The energies in parentheses give for
lepton colliders their CoM energies and for hadron colliders the
approximate range of CoM energies attainable for hard parton-parton
collisions.}
\label{compare}
\end{figure}

The primary advantage of using  muons in a lepton collider arises from the fact that the muon is
$\approx$~200 times heavier than the electron. It is thus possible to 
accelerate  muons using circular accelerators that are
compact and fit on existing accelerator sites. See
Figure~\ref{compare} for a comparison of relative sizes of muon
colliders ranging from 500 GeV to 3 TeV center of mass energies with
respect to the LHC, SSC, and NLC. Once  the problem of
cooling a muon beam to sufficiently small emittances 
is solved and the beams  can be accelerated, higher energies are
much more easily obtained in a muon collider than in the linear
electron-positron collider.
Because the muon is unstable, it is necessary to cool and accelerate
the beam before a substantial number have decayed.  The number of
turns in a muon lifetime is independent of the muon momentum for a
given magnetic field, since both the circumference and the muon
lifetime in the laboratory frame scale with muon momentum.  With
typical bending magnetic fields($\approx$~5~Tesla) available with
today's technology, the muons last $\approx$~1000 turns before half of
them have decayed in the collider ring.

 Muon decay also gives rise to large numbers of electrons that can
affect the cryogenics of the magnets in the machine as well as 
pose serious background problems for detectors in the collision
region. The 1999 Status Report~\cite{INTRO:ref5} contains an excellent
summary of the problems (and possible solutions) one faces on the way to
a muon collider.

\begin{figure}[bth!]
\centerline{\includegraphics[width=0.6\linewidth]{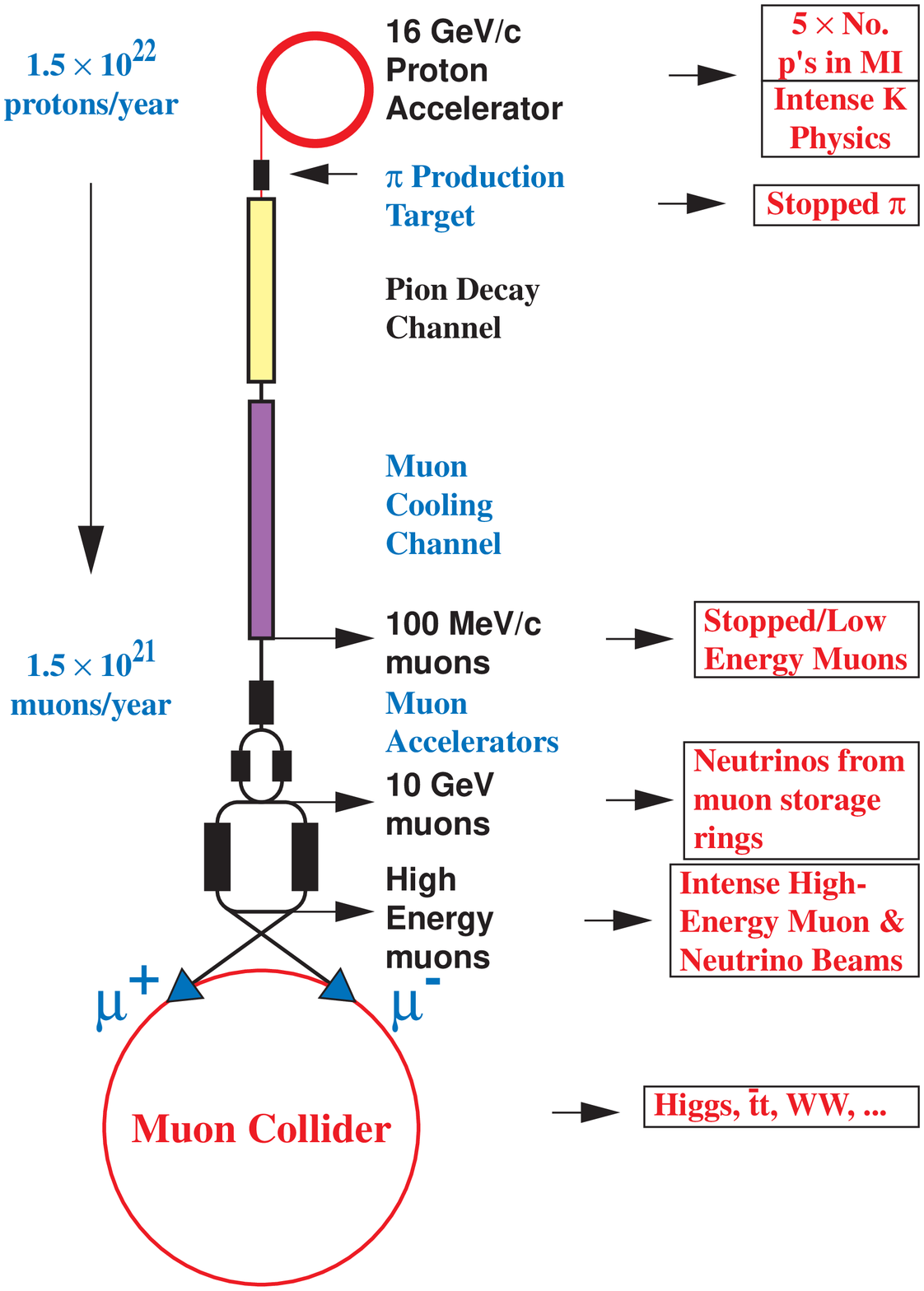}}
\vspace{0.5cm}
\caption[Schematic of a  muon collider]{(Color)Schematic of a  muon collider.}
\label{schematic}
\end{figure}

Figure~\ref{schematic} shows a schematic of such a muon collider,
along with a depiction of the possible physics that can be addressed
with each stage of the facility. Some of the ideas needed to obtain
longitudinal cooling necessary for the Muon Collider are discussed in
section~\ref{long-cool} and some of the parameters of the accelerator
system for higher energy colliders are discussed in section~\ref{high-acc}
below.

\subsection{Higgs Factory Requirements}
The emittance of the muon beam needs to be reduced by a factor of
$\approx$ 10$^6$ from production~\cite{INTRO:ref5} to the point of
collision for there to be significant luminosity for
experiments. Table~\ref{emitable} lists the transverse and
longitudinal emittances at the end of the decay channel,
Study-II~\cite{EPP:studyii} cooling channel and those needed for a
0.1~TeV Center of Mass Energy Muon Collider, also known as a Higgs Factory. 
It can be seen that one needs to cool by a factor of
$\approx$ 20 in the transverse dimension and $\approx$ 3 in the
longitudinal dimension from the Study-II emittances to achieve the
emittances necessary for a Higgs factory.
\begin{table*}[thb!]
\centering  
\caption[Emittances at the end of various machines. ]
{Transverse and longitudinal emittances at the end of the decay
channel, Study-II cooling channel and the Higgs factory cooling
channel.\label{emitable}}
\begin{tabular}{|c|c|c|}
\hline
Emittance at end of & Transverse emittance ($\pi$ mm) & Longitudinal
emittance ($\pi$ mm) \\
\hline
Decay Channel & 17 & 150 \\ Study-II cooler & 2.6 & 30 \\
Higgs Factory cooler & 0.14 & 9 \\
\hline
\end{tabular}
\end{table*}
This can be
achieved by ionization cooling similar to the scheme described in the
section~\ref{neufact}. The transverse emittance is reduced during
ionization cooling, since only the longitudinal energy loss is
replaced by rf acceleration. However, due to straggling, the
longitudinal energy spread of the beam increases, even if the average
longitudinal energy of the beam is kept constant. The longitudinal
emittance thus grows in a linear cooling channel.  In order to cool
longitudinally, one needs to create dispersion in the system and have
wedge absorbers at the point of maximum dispersion so that the faster
particles go through the thicker parts of the wedge. This results in a
reduction in longitudinal emittance accompanied by an increase in
transverse emittance and is thus called emittance exchange.

\begin{figure*}[tbh!]
\includegraphics[height=2.9in,width=5.7in]{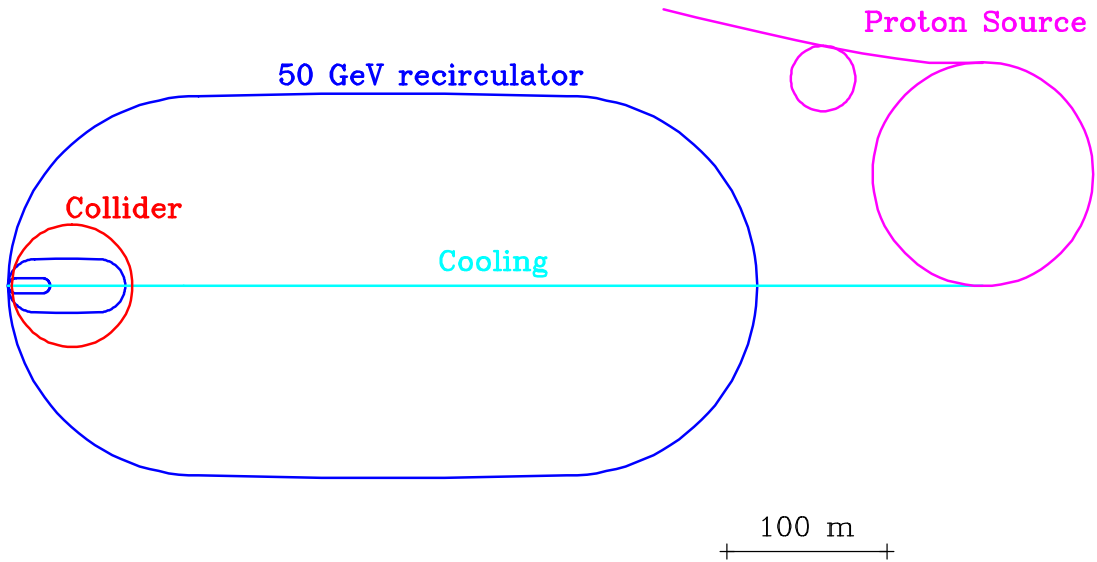}
\caption[Plan of a 0.1-TeV-CoM muon collider]
{(Color)Plan of a 0.1-TeV-CoM muon collider, also known as a Higgs Factory.}
\label{plan1}
\end{figure*}

The status report~\cite{INTRO:ref5} outlines the details of the
acceleration and collider ring for the 0.1~TeV Higgs factory, shown 
schematically in Fig.~\ref{plan1}. Table \ref{sum}
gives a summary of the parameters of various muon colliders including
three different modes of running the Higgs Collider that have varying
beam momentum spreads. Additional information about the Muon Collider
can be found in~\cite{gail,higgsreport}.

\begin{table*}[thb!]
\centering  
\caption[Baseline parameters for high- and low-energy muon colliders. ]
{Baseline parameters for high- and low-energy muon colliders.
Higgs/year assumes a cross section $\sigma=5\times 10^4$~fb; a Higgs
width $\Gamma=2.7$~MeV; 1~year = $10^7$~s.}
\label{sum}
\begin{tabular}{|l|c|c|c|c|c|}
\hline
\rr CoM energy~ (TeV)   &\rr 3 &\rr 0.4 &
\multicolumn{3}{c|}{\rr 0.1 }  \\
% & & & & & & \\
$p$ energy~(GeV) & 16 & 16 & \multicolumn{3}{c|}{16}\\ $p$'s/bunch &
$2.5\times 10^{13}$ & $2.5\times 10^{13}$ &
\multicolumn{3}{c|}{$5\times 10^{13}$  }  \\  
Bunches/fill   & 4  & 4  & \multicolumn{3}{c|}{2 } \\ 
Rep.~rate~(Hz) & 15 & 15 & \multicolumn{3}{c|}{15 } \\ 
$p$ power~(MW) & 4 & 4
&\multicolumn{3}{c|}{4} \\ $\mu$/bunch & $2\times 10^{12}$ &
$2\times10^{12}$ &\multicolumn{3}{c|}{$4\times 10^{12}$ } \\
\rr $\mu$ power~(MW)     & \rr 28 &\rr 4 & \multicolumn{3}{c|}{\rr 1 }  \\
\rr Wall power~(MW)    &  \rr  204 &\rr 120  & \multicolumn{3}{c|}{\rr
81 } \\ Collider circum.~(m) & 6000 & 1000 & \multicolumn{3}{c|}{350 }\\
Ave bending field~(T) & 5.2 & 4.7 &\multicolumn{3}{c|}{3 } \\
%Depth~ m          &  500 & 100 & \multicolumn{3}{c}{10 }  \\
\hline
\rr Rms ${\Delta p/p}$~\%          &\rr 0.16 &\rr 0.14 &\rr
0.12 &\rr 0.01&\rr 0.003 \\
\hline
6-D $\epsilon_{6,N}$~$(\pi \textrm{m})^3$&$1.7\times
10^{-10}$&$1.7\times 10^{-10}$&$1.7\times 10^{-10}$&$1.7\times
10^{-10}$&$1.7\times 10^{-10}$\\ Rms $\epsilon_n$~($\pi$ mm-mrad) & 50 &
50 & 85 & 195 & 290\\ $\beta^*$~(cm) & 0.3 & 2.6 & 4.1 & 9.4 & 14.1\\
$\sigma_z$~(cm) & 0.3 & 2.6 & 4.1 & 9.4 & 14.1 \\ $\sigma_r$spot~$(\mu$m)
& 3.2 & 26 & 86 & 196 & 294\\ $\sigma_{\theta}$ IP~(mrad) & 1.1 & 1.0 &
2.1 & 2.1 & 2.1\\ Tune shift &0.044 &0.044 & 0.051 &0.022 & 0.015\\
$n_{\rm turns}$ (effective) & 785 & 700 & 450 & 450 & 450 \\
\hline
\rr Luminosity~cm$^{-2}$s$^{-1}$&\rr $7\times 10^{34}$ & $10^{33}$ &\rr
$1.2\times 10^{32}$ &\rr $2.2\times 10^{31}$&\rr $10^{31}$ \\ & & & &
& \\ Higgs/year & & & $1.9\times 10^3$ & $4\times 10^3$ & $3.9\times
10^3$ \\
\hline
\end{tabular}
\end{table*}

\subsection{Longitudinal Cooling}
\label{long-cool}
At the time of writing of the status report~\cite{INTRO:ref5} 
there was no satisfactory solution for the  emittance exchange problem and
this remained a major stumbling block towards realizing a muon
collider. However, ring coolers have been found to hold significant
promise in cooling in 6-dimensional phase space. Another advantage of
ring coolers is that one can circulate the muons many turns, thereby
reusing the cooling channel elements.  
Several meetings on emittance exchange were held~\cite{eemeets} and a 
successful workshop~\cite{eework} was held in 2001, where we
explored in some depth several kinds of ring coolers.
These options differ primarily in the type of focusing used to 
contain the beam. We describe the current status of our
understanding of three types of ring coolers here.

\subsubsection{Solenoidal Ring Coolers}

\begin{figure}[tbh!]
\vspace{0mm}
\begin{minipage}[tbh!]{0.47\linewidth}
\includegraphics[width=\linewidth]{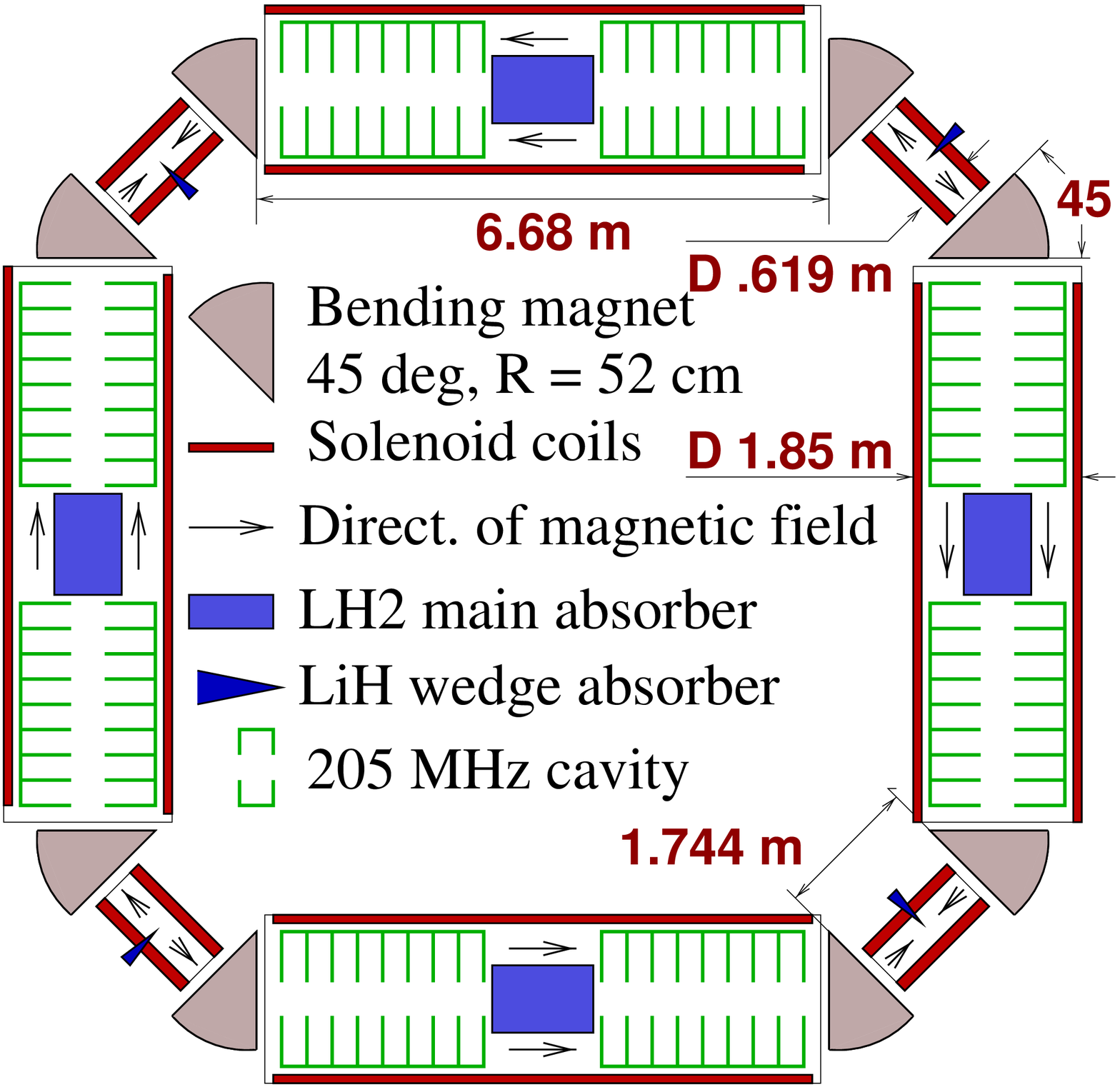}
%\centerline{\epsfig{file=02_ringA.eps,width=\linewidth}}
\end{minipage}
\begin{minipage}[tbh!]{0.47\linewidth}
\begin{tabular}{ll}
Circumference & 36.963 m \\ Nominal energy at short & \\  straight section & 250 MeV
\\ Bending field & 1.453 T \\ Norm. field gradient & 0.5 \\
Max. solenoid field & 5.155 T \\ rf frequency & 205.69 MHz \\
Accelerating gradient & 15 MeV/m \\ Main absorber length & 128 cm \\
LiH wedge absorber & 14 cm \\ Grad. of energy loss & 0.75 MeV/cm \\
\end{tabular}
\end{minipage}
\caption{(Color)Layout and parameters of the solenoid based ring cooler
\label{ring}}
\end{figure}

\begin{figure}[htb!] 
\includegraphics[width=2.5in]{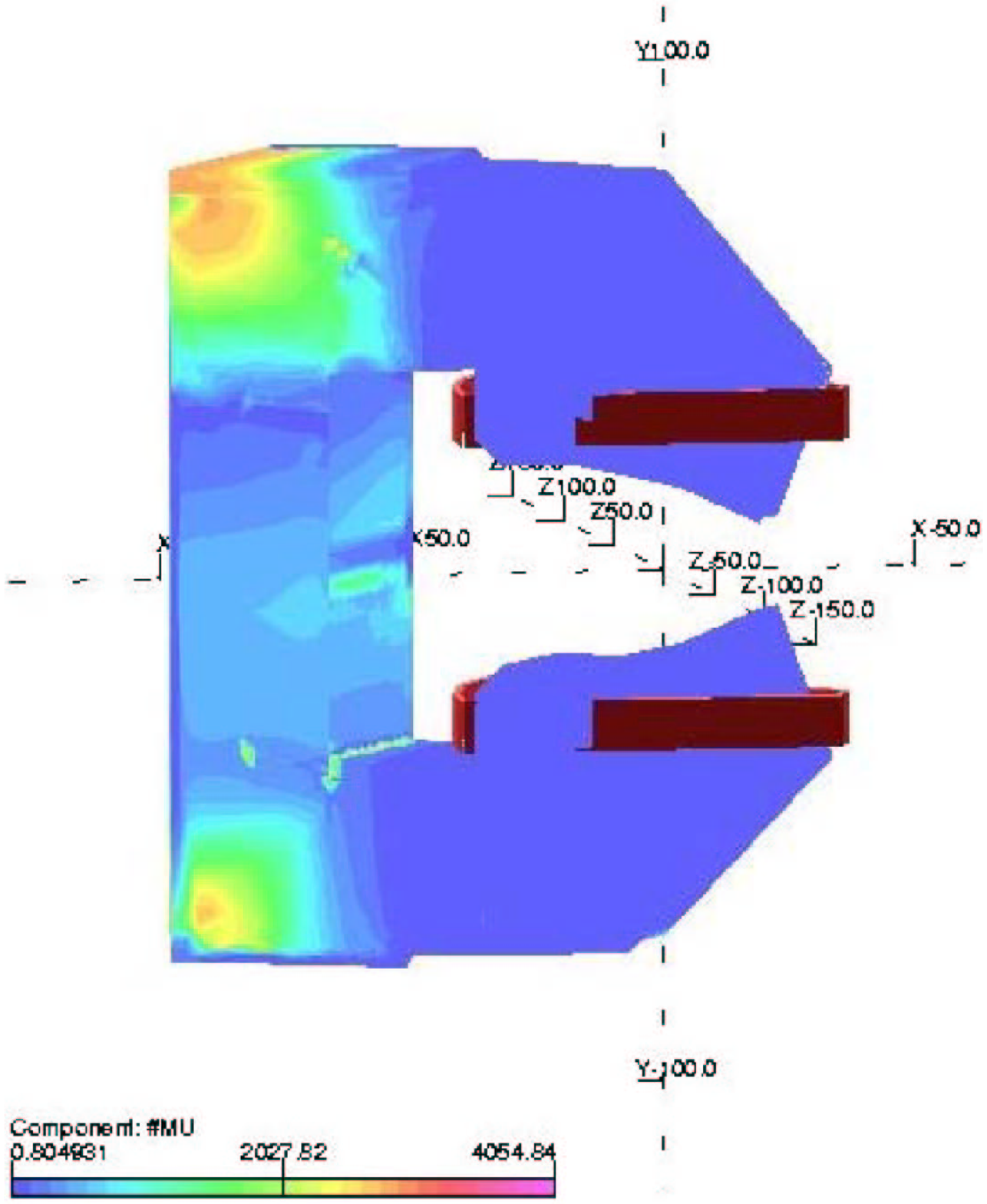}
\includegraphics[width=2.5in]{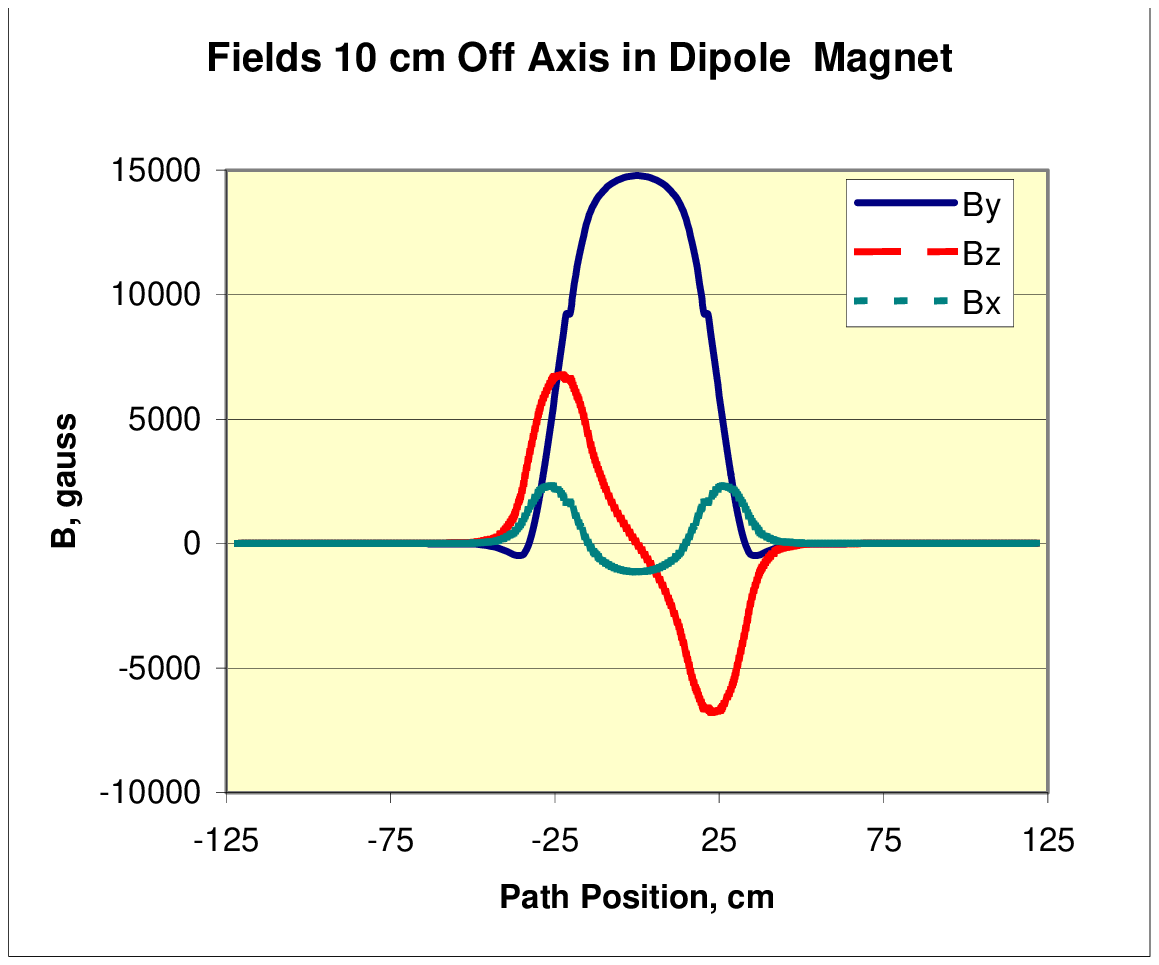}
\caption{(Color)Figure shows a computer model of a 52cm radius dipole with 
index $n=-\frac{1}{2}$ and the calculated field components radially  10cm off axis.
\label{dipole}}
\end{figure}

The basic design of the solenoidal ring cooler~\cite{balb1} is
presented in Figure~\ref{ring}.  Eight focusing dipole magnets with an
index $n=-\frac{1}{2}$ are used for bending and focusing of the
beam. Each of these dipoles bends the beam through 45 degrees with a
central orbit bending radius of 52 cm. We have done calculations to
show that such dipoles are buildable. Figure~\ref{dipole} shows a
configuration of such a dipole and the resulting magnetic field
components calculated using a 3D field calculation program.  
There are
4 long solenoids containing rf cavities and liquid hydrogen absorbers
for transverse cooling.  A magnetic field of 2.06~T at the end regions
of the solenoids provides the same transverse focusing as the bending
magnets.  The magnetic field adiabatically increases to 5.15 T towards
the center of the solenoid in order to produce a small $\beta$
function (25-30 cm) at the absorbers.  The short solenoids are
designed to create an appropriate dispersion function that is zero at
the long solenoids, which house the 200 MHz rf cavities.  Their field
is $\pm 2.06~T$ at the edges and $\pm 2.75$~T centrally.  A symmetric
field flip is required in the short solenoids to prevent the build up
of canonical angular momentum. This field flip causes the dispersion
in the long solenoids housing the rf cavities to be zero while
permitting a non-zero dispersion at the lithium hydride wedge
absorbers at the centers of the short solenoids which then produce
longitudinal cooling via emittance exchange.

%%%%%%%%%%%%%%%%%%%%%%%%%%%%%%%%%%%%%%%%%%%%%%%
%%%%%%%%%%%%%%%%%%%%%%%%%%%%%%%%%%%%%%%%%%%%%%%%
\begin{figure}[tbh!]
\begin{minipage}[t!]{0.47\linewidth}
\includegraphics[width=1.15\linewidth]{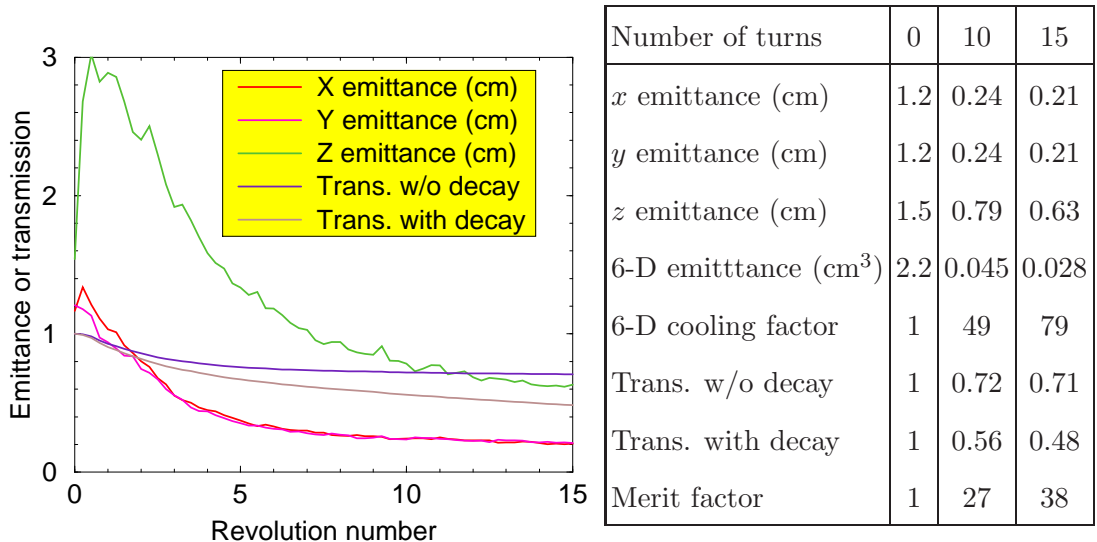}
%\centerline{\epsfig{file=10_evol.eps,width=1.15\linewidth}}
\end{minipage}
\begin{minipage}[t!]{0.47\linewidth}
\vspace{-5mm}
\begin{tabular}{|l|c|c|c|}
\hline
Number of turns & 0 & 10 & 15 \\
\hline
$x$ emittance (cm) & 1.2 & 0.24 & 0.21 \\ $y$ emittance (cm) & 1.2 & 0.24
& 0.21 \\ $z$ emittance (cm) & 1.5 & 0.79 & 0.63 \\ 6-D emitttance (cm$^3$) &
2.2 & 0.045 & 0.028 \\ 6-D cooling factor & 1 & 49 & 79 \\ Trans. w/o
decay & 1 & 0.72 & 0.71 \\ Trans. with decay & 1 & 0.56 & 0.48 \\
Merit factor & 1 & 27 & 38 \\
\hline
\end{tabular}
\end{minipage}
\vspace{-3mm}
\caption{(Color)Evolution of the beam emittance/transmission at the ring cooler.
\label{evol}}
\vspace{-3mm}
\end{figure}
%%%%%%%%%%%%%%%%%%%%%%%%%%%%%%%%%%%%%%%%%%%%%%%%
%

Evolution of the beam emittance and transmission is
shown in Figure~\ref{evol} as a function of the number of turns in the
ring.  In 15 turns, the transverse emittance decreases from 1.2
cm to 0.21 cm yielding a cooling factor of 5.7, the longitudinal
emittance decreases from 1.5 cm to 0.63 cm (cooling factor 2.4), and the 6-D
emittance decreases from 2.2 cm$^3$~to 0.028 cm$^3$, with an overall cooling
factor $\approx$~79.  The transmission is 0.71 without decay and 0.48
with decay. We define a merit factor for cooling that is 
the total transmission including decay times the 6-D cooling factor.
The  merit factor for this ring is then 38.  This implies that transverse
emittance at the ring cooler is about the same as at a linear SFOFO
cooling channel employed in Study-II~\cite{EPP:studyii}, whereas the
longitudinal emittance is noticeably less.

This cooler provides mainly transverse cooling and can be used as a
part of Neutrino Factory or a muon collider.  A cooler specially designed for
strong longitudinal cooling (``bunch compressor'') can also be created using
a similar scheme.  Such a compressor would be a part of a muon collider
to shorten muon bunches from 6-8 m (minimal length after $\pi -
\mu$~decay and phase rotation, see Ref.~\cite{INTRO:ref5}) to 0.6-0.8
m acceptable for further cooling by a 200 MHz channel.

Two options for the bunch compressor are considered in
Ref.~\cite{balb4}.  The first one is a two-step cooler where each step
is very similar to the ring cooler shown in Figure~\ref{ring}.  The main
difference is that  the primary goal in the first cooler is the
longitudinal bunching of the beam. This leads to a uniform magnetic
field in the long solenoids and lower frequency/voltage of the
accelerating rf system (15.6 MHz/4 MeV/m at the first stage vs.  62.5
MHz/8 MeV/m at the second one).  Another option is a 15 MHz octagonal
cooler composed of the same cells as in Figure~\ref{ring}, but with half the
bending magnet angle.  Decrease of longitudinal emittance
from 43 cm to 2.5-3 cm, as required for muon collider, is obtained in
both cases.

We are proceeding with a realistic simulation of this system using
Geant and ICOOL that employs realistic magnetic fields~\cite{kahn} produced by
field calculation programs.

After the two stage cooler, we still need a factor of $\approx$~30 in
transverse cooling, but we are within a factor of 4 in longitudinal
cooling relative to the Higgs factory goals. Lithium lens cooling, which with
its strong focusing will cool transversely further while degrading
longitudinally due to straggling, is a possibility and is being
investigated.

\subsubsection{RFOFO ring coolers}

The cooling lattice for the Neutrino Factory (see section~\ref{neufact}) 
employs a
configuration of fields known as an SFOFO lattice (super-FOFO) where
the axial magnetic field profile changes polarity in alternate cells
of the lattice. For the ring cooler design under consideration here,
we employ an RFOFO lattice (regular-FOFO) where the axial field
profile changes polarity in the middle of a cell. As a result all
cells in an RFOFO lattice are identical.

\begin{figure}[htb!] 
\includegraphics[width=\linewidth]{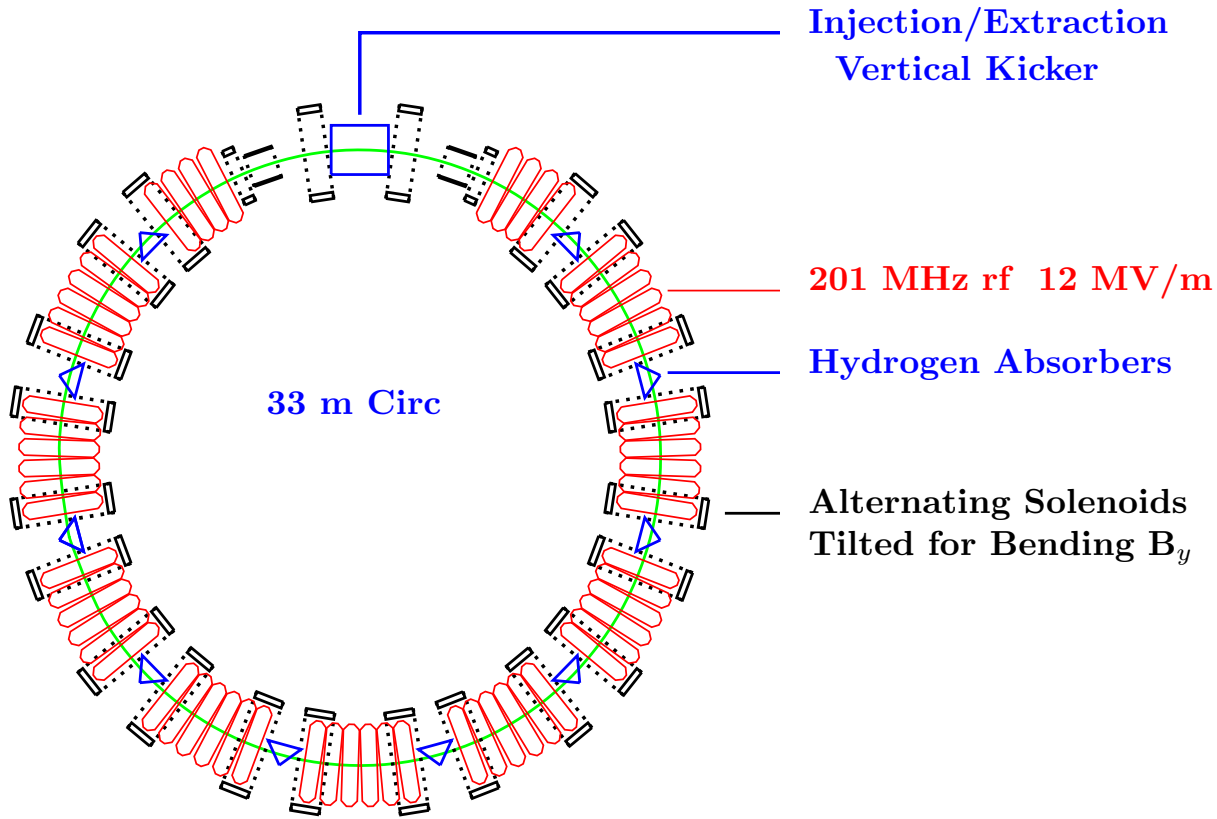} 
\caption{(Color)Layout of an RFOFO cooling ring. \label{rforing}}
\end{figure}

The ring cooler design employs a single cell for both transverse
cooling and emittance exchange. It uses solenoids for focusing, giving
large angular and momentum acceptances. The cell includes dispersion,
acceleration, and energy loss in a single thick hydrogen wedge. Figure
\ref{rforing} shows the layout of the cooling ring drawn to scale.  The
RFOFO lattice was chosen because, unlike in the SFOFO case used in
Study-II, all cells are strictly identical, and the presence of an
integer betatron resonance within the momentum acceptance is
eliminated.

\begin{figure}[tbh] 
\begin{center} 
a)
\includegraphics[width=3.5in] {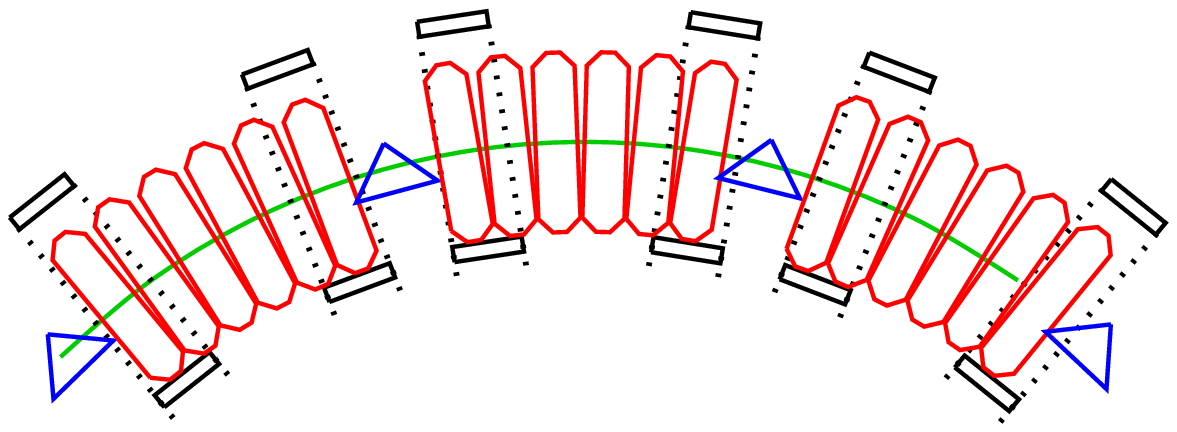}                              

b)
\includegraphics[width=3.5in] {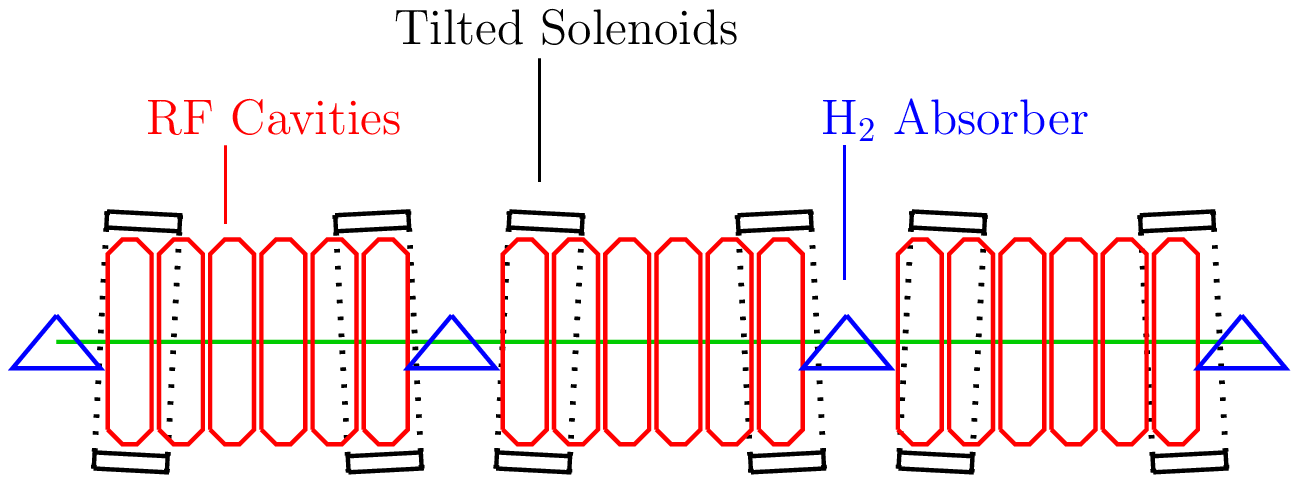}                              
\end{center}
\caption{(Color)Three cells of the RFOFO lattice; a) plan, b) side. \label{cells}} 
\end{figure}

The basic 33 m circumference ring is made up of 12 identical 2.75-m long cells. In the
figure, this symmetry is broken for injection and extraction, but the
magnetic fields in this insertion are nearly identical to those in the
rest of the ring.  Figure \ref{cells} shows a detailed view of three
cells of the lattice, in plan (a) and side (b) views.                    

The longitudinal field on-axis has an approximately sinusoidal
dependence on position. The beam axis is displaced laterally 
with respect to the coil centers (as shown in figure \ref{cells}a) to 
minimize horizontal fields that cause vertical beam deviations.
The lattice transmits
particles in the momentum band from 150 to 250 MeV/c. 
The minimum value of the beta
function at the central momentum is 40 cm.

Bending round the ring, and the required dispersion, are provided by applying an approximately 0.125 T vertical bending field generated by alternately tilting
the solenoids (as shown in figure \ref{cells}b).  There is no attempt to set the
field index $n$ (where $B\propto r^n$) to the 0.5 value;  
so the focusing is not identical in x and y.

It is found that the acceptance is reduced as the bending field is
increased. We thus use a wedge with maximum possible angle (giving
zero thickness on one side), and the lowest bending field consistent
with adequate emittance exchange.  The dispersion at the absorber is
approximately 8 cm in a direction 30 degrees from the $y$ axis, The dispersion at the
center of the rf is of the opposite sign, and also mostly in the $y$
direction. 

The liquid-hydrogen wedge has a central thickness of 28.6 cm and a
total wedge angle of 100 degrees and is rotated 30 degrees from the
vertical.  The rf
cavities are at a frequency of 201.25 MHz and have an accelerating 
gradient of 12 MV/m. 

\begin{figure}[tbh] 
\includegraphics[height=3.6in]{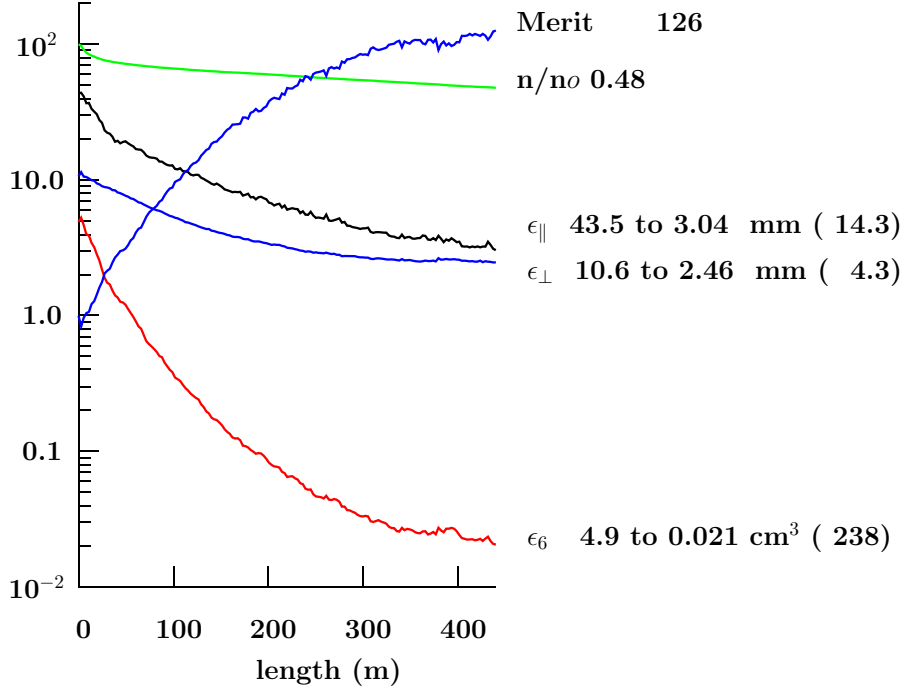}
%\vskip.5in 
\caption{(Color)Transmission,
normalized transverse emittance, normalized longitudinal emittance,
normalized 6-dimensional emittance, and the merit factor, as a
function of distance.  \label{all}}
\end{figure}

The ICOOL~\cite{icool} simulation (with results shown in shown Figure~\ref{all})
used fields calculated from the specified coils, and thus 
neglects no end field effects. But in this simulation,
no absorber, or rf, windows, are included; nor did it include
the injection/extraction insertion.
The rf was represented by the fields in perfect pillbox
cavities. The input tracks were taken from a Study-II~\cite{EPP:studyii} simulation, using
distributions from just upstream of the  transverse cooling system. 
The use of Study-II
simulated distributions is intended to allow a more realistic estimate
of the ring's performance.  No attempt was made to match this beam to the ring
dispersion or slight differences in the transverse beta functions.

Figure~\ref{all} shows the simulated transmission, transverse emittance,
 longitudinal emittance, 6-dimensional emittance, and a merit
factor $M$ vs. length in the ring. $M$ is given by:

$$M~=~{\epsilon_6(initial) \over \epsilon_6(final)}~\times~{\rm
Transmission}$$

Initially, the $x$ emittance falls more rapidly than the $y$, 
because it is the $y$ emittance that is exchanged with the
longitudinal. But the Larmor rotations soon mix the $x$ and $y$
emittances bringing them to a common value.

 After a distance of 400 m ($\approx$~12 turns), the 6-dimensional
emittance has fallen by a factor of 238, with a transmission of 48 \%
(66\% without decay). The merit factor is 136. The same factor for the
Study-II cooling lattice, also without windows, is 13. Studies with realistic
windows and the injection/extraction insertion added, show lower merit 
factors, but always far better than the Study-II example.

The design of the injection/ejection channels and kickers will be challenging, 
and this ring could not be used, as is, to replace the Study-II cooling
channel because the bunch train in that case is too long to fit in
the ring. Both problems would be removed in a helical cooling channel. 
The merit factor 
for such a channel could be even better than that of the ring
because it would be possible to ``taper'' the optics, as a function 
of distance down the channel, and thus
lower the final equilibrium emittance.

\subsubsection{Quadrupole Ring Coolers}

\begin{figure}[tbh!]
\includegraphics[height=6cm,width=8cm]{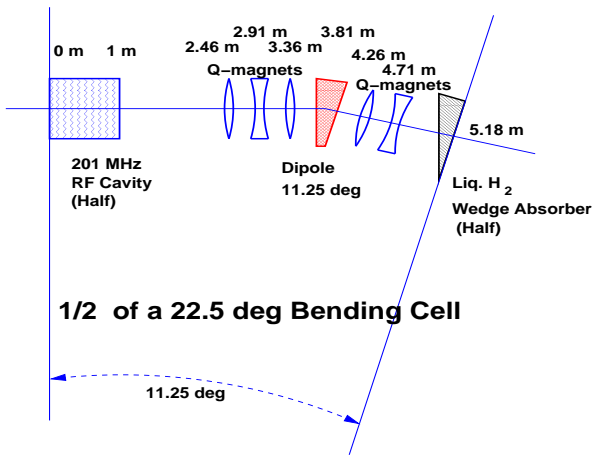}
\caption{\label{fig:half}(Color) Schematic diagram of half of a 22.5$^\circ$
bending cell. A wedge absorber is located in the middle of the cell.}
\end{figure}

\begin{figure}[tbh!]
\includegraphics[height=5cm,width=8cm]{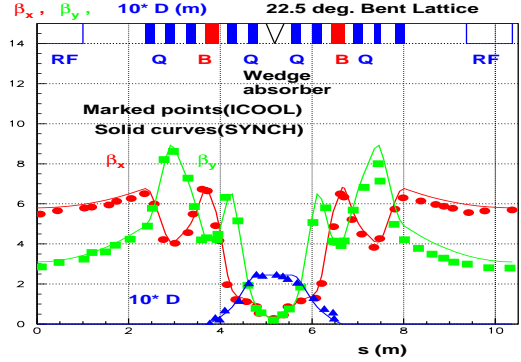}
\caption{\label{fig:cell}(Color)The $\beta_{x}$, $\beta_{y}$, and $D$(dispersion)
in a 22.5$^\circ$ bending cell. SYNCH calculations(solid curves) and
beam parameters from an ICOOL simulation(marked points) are compared.}
\end{figure}

\begin{figure}[tbh!]
\includegraphics[width=8cm]{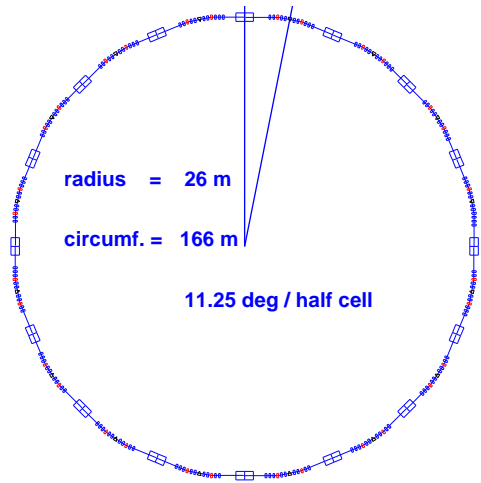}
\caption{\label{fig:ringucla}(Color) Top view of a sixteen cell muon cooling ring.}
\end{figure}

Alternative ring designs have been explored which are based on storage
rings which consist of conventional quadrupole and dipole magnetic elements
instead of solenoids~\cite{ucla}.  The strategy has been to utilize the SYNCH
storage ring design code~\cite{synch} to develop linear lattice solutions
and then transfer the lattice parameters into the ray-tracing tracking 
code ICOOL~\cite{icool} in which absorbers and energy recovery with rf cavities
can be included for full simulation.  An example of such a ring is shown
in Figure~\ref{fig:half} where the elements of a half cell for a 22.5$^\circ$
bending cell are depicted schematically.  The correspondence between the beam
envelope beta and dispersion functions resulting from a simulation with the 
ICOOL tracking code and the SYNCH calculated values are shown in 
Figure~\ref{fig:cell}.   The full sixteen cell ring is shown in 
Figure~\ref{fig:ringucla}.


\begin{figure}[tbh!]
\includegraphics[height=5cm,width=8cm]{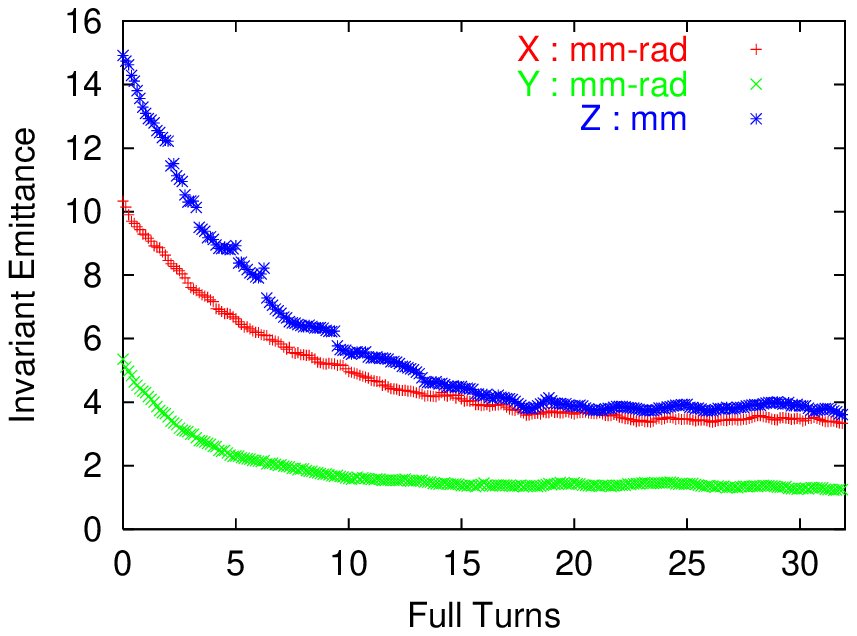}
\caption{\label{fig:figaa} (Color)The evolution of x, y, z normalized emittances
in 32 full turns.}
\vskip0.5cm
\includegraphics[height=5cm,width=8cm]{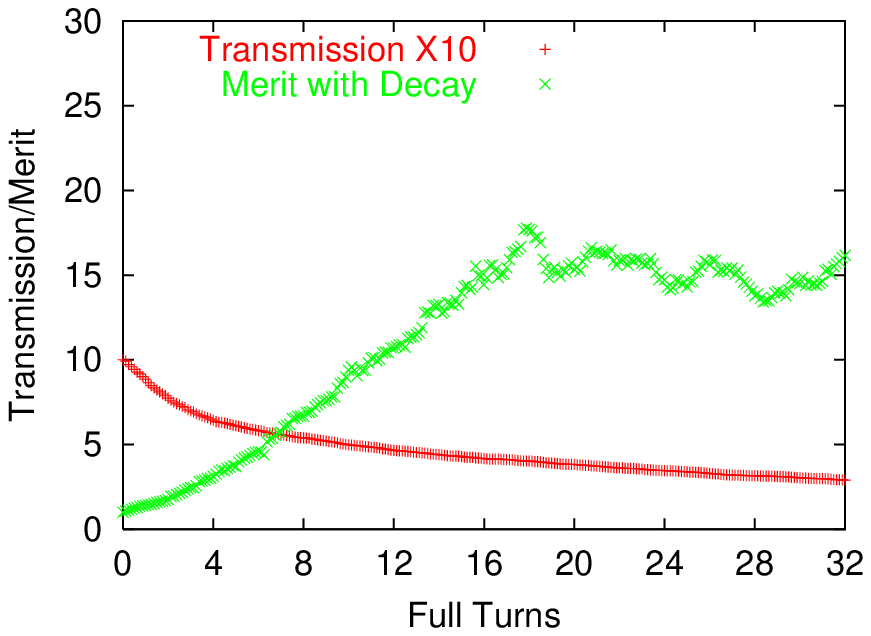}
\caption{\label{fig:figbb}(Color) The transmission and the figure of merit factor
as a function of the arc length in 32 full turns.}
 
\end{figure}

In general, we find that the performance of the rings, as measured in emittance
reduction along with particle transmission and decay losses, improves when
more compact lattice designs are considered.  In Figure~\ref{fig:figaa}, the
variation of the normalized emittances as a function of ring turns
is shown for an eight cell ring.  A reduction of normalized emittance
is observed for all three dimensions.  This particular
ring has a total circumference of 30.9~m.  Each half cell contains
one 22.5$^\circ$ combined function dipole proceeded and followed by
a single horizontally focusing quadrupole.
The average muon beam momentum is 250~MeV/c and liquid hydrogen
absorbers with wedge opening angles of 40$^\circ$ are used. For each cell, the
central beam orbit traverses 24 cm of absorber.  The energy loss in the wedge
absorbers is compensated with 201~MHz rf cavities with peak on-axis
gradients of 16~MV/m.

The muon transmission efficiency and total merit factor (muon survival
rate times the ratio of initial to final 6-dimensional emittance) is shown
in Figure~\ref{fig:figbb} as a function of ring turns.
The merit factor reaches 18, while the muon transmission efficiency
including decay losses after eighteen full turns is 40\%.

Rings in which the focusing function is handled exclusively by the
dipole elements have also been explored.  In this case, the natural
focusing power of the dipole is utilized for horizontal focusing while
the entrance and exit dipole edge angles are adjusted to provide the
required vertical focusing.  Several examples of such lattices have
been examined.  Because these lattices are more compact than lattices
which also include quadrupoles, the performance of these rings tend to
be better.  As an example, performances with merit factors on the
order of 100 have been observed with rings based on four cell lattice
with eight 45$^\circ$ dipoles.  For one case, the entrance and exit
angles dipole faces are 7.4$^\circ$ and 21.8$^\circ$ relative to the
normal of the beam trajectory.  The ring for this example has a
circumference of only 9.8 m and the design of injection/ejection cells
may prove to be very challenging.

\subsubsection{Injection into Ring Coolers}
The most serious technical problem facing the ring cooler approach is
the injection system which may require a very powerful kicker
magnet~\cite{balb5}. The energy stored in the injection kicker goes as
the square of the emittance of the beam and inversely as the
circumference of the ring. A promising injection scheme that does not
use kicker magnets, but instead uses absorbers to degrade the beam
energy and rf phase manipulations has been proposed~\cite{balb6} and
is being studied.
\subsection{Higher Energy Muon colliders}
\label{high-acc}
Once the required cooling  has been achieved to make the first
muon collider feasible, acceleration to higher energies becomes possible.
Colliders with 4 TeV center of mass energy have been
studied~\cite{INTRO:ref5} and Table \ref{dntable} lists the parameters
for such a collider.  The muons are accelerated initially by a linear
accelerator followed by a series of recirculating linear accelerators
(RLA's) followed by rapid cycling synchrotrons (RCS's).  The radiation
from the neutrinos from the muon decay begins to become a problem at
CoM energies of 3 TeV~\cite{kingnu}. 

 There have been preliminary attempts to study colliders of
even higher energy, starting at 10~TeV all the way up to 
100~TeV in the center of mass and
we include the references to these studies~\cite{kinghi} for the sake 
of completeness.

\begin{table*}[tbh!]
 \caption[Parameters of Acceleration for a 4~TeV Muon Collider] {Parameters
 of Acceleration for a 4~TeV Muon Collider.}
\label{dntable}
\begin{tabular}{|l|c|c|c|c|c|}
\hline
                       & Linac & RLA1 & RLA2 & RCS1 & RCS2 \\
\hline
E (GeV) & 0.1$\rightarrow$ 1.5 & 1.5 $\rightarrow$ 10 & 10
$\rightarrow$ 70 & 70 $\rightarrow$ 250 & 250 $\rightarrow$ 2000 \\
f$_{rf}$ (MHz) & 30 $\rightarrow$ 100 & 200 & 400 & 800 & 1300 \\
N$_{turns}$ & 1 & 9 & 11 & 33 & 45 \\ V$_{rf}$(GV/turn) & 1.5 & 1.0 &
6 & 6.5 & 42 \\ C$_{turn}$(km) & 0.3 & 0.16 & 1.1 & 2.0 & 11.5 \\ Beam transit
time (ms) & 0.0013 & 0.005 & 0.04 & 0.22 & 1.73 \\
$\sigma_{z,beam}$(cm) & 50 $\rightarrow$ 8 & 4 $\rightarrow$ 1.7 & 1.7
$\rightarrow$ 0.5 & 0.5 $\rightarrow$ 0.25 & 0.25 $\rightarrow$ 0.12
\\ $\sigma_{E,beam}$(GeV) & 0.005 $\rightarrow$ 0.033 & 0.067
$\rightarrow$ 0.16 & 0.16 $\rightarrow$ 0.58 & 0.58 $\rightarrow$ 1.14
& 1.14 $\rightarrow$ 2.3 \\ Loss (\%) & 5 & 7 & 6 & 7 & 10 \\
\hline
\end{tabular}
\end{table*}

\subsection{Muon Collider Detectors} 
\begin{figure}[bth!]
\centerline{\includegraphics[width=0.5\linewidth]{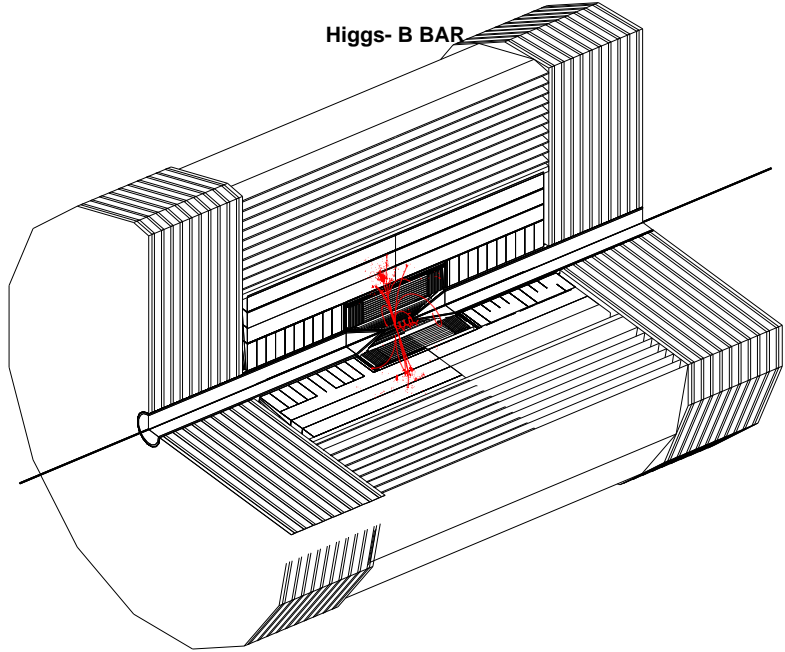}}
\caption[Strawman Geant detector for a muon collider]
{(Color)Cut view of a strawman detector in Geant for the Higgs factory with a
Higgs$\rightarrow b\bar b$ event superimposed. No backgrounds
shown. The tungsten cones on either side of the interaction region
mask out a 20~$\deg$ area.}
\label{geant}
\end{figure}

Figure~\ref{geant} shows a strawman muon collider detector for a Higgs
factory simulated in Geant. The background from muon decay sources has
been extensively studied~\cite{INTRO:ref5}. At the Higgs factory, the
main sources of background are from photons generated by the showering
of muon decay electrons. At the higher energy colliders, Bethe-Heitler
muons produced in electron showers become a problem.  Work was done to
optimize the shielding by using specially shaped tungsten
cones~\cite{INTRO:ref5} that reduce the backgrounds resulting from electomagnetic 
showers from entering the detector.  The occupancy levels due to background photons and neutrons in detectors
 were shown to
be similar to those predicted for the LHC experiments. It still needs
to be established whether pattern recognition is possible in the
presence of these backgrounds, especially the Bethe Heitler muons, which are a unique source of 
background to muon collider detectors..
%

%\input r_and_d 
% Mike's new version, received Jan.9

\section{R\&D Status}

\label{r_and_d}

A successful construction of a muon storage ring to provide a copious
source of neutrinos requires several novel approaches to be developed
and demonstrated; a high-luminosity Muon Collider requires an even
greater extension of the present state of accelerator design. Thus,
reaching the full facility performance in either case requires an
extensive R\&D program.

The required R\&D program has been identified for each of the major
neutrino factory systems. In particular, some critical components must
be prototyped and their performance verified. For example, the cooling
channel assumes a normal conducting rf (NCRF) cavity gradient of 17
MV/m at 201.25 MHz, and the acceleration section demands similar
performance from superconducting rf (SCRF) cavities at this
frequency. In both cases, the requirements are beyond the performance
reached to date for cavities in this frequency range. The ability of
the target to withstand a proton beam power of up to 4 MW must be
confirmed, and, if it remains the technology of choice, the ability of
an induction linac unit to coexist with its internal SC solenoid must
be verified. Finally, an ionization cooling experiment should be
undertaken to validate the implementation and performance of the
cooling channel, and to confirm that our simulations of the cooling
process are accurate.

Below we give an overview of the MC R\&D program goals and list the
specific questions we expect address. We also summarize briefly the
R\&D accomplishments to date

\subsection{R\&D Program Overview}

A Neutrino Factory comprises the following major systems: Proton Driver,
Target and (Pion) Capture Section, (Pion-to-Muon) Decay and Phase Rotation
Section, Bunching and Matching Section, Cooling Section, Acceleration
Section, and Storage Ring. These same categories apply to a Muon Collider,
with the exception that the Storage Ring is replaced by a Collider Ring
having a low-beta interaction point and a local detector. Parameters and requirements for the various systems are generally more severe in the case
of the Muon Collider, so a Neutrino Factory can properly be viewed as a scientifically productive first step toward the eventual goal of a collider.

The R\&D program we envision is designed to answer the key questions
needed to embark upon a Zeroth-order Design Report (ZDR). The ZDR will
examine the complete set of systems needed for a Neutrino Factory, and
show how the parts can be integrated into a coherent whole. Although a
fully engineered design with a detailed cost estimate is beyond the
scope of a ZDR, enough detailed work must been accomplished to ensure
that the critical items are technically feasible and that the proposed
facility could be successfully constructed and operated at its design
specifications. By the end of the full R\&D program, it is expected
that a formal Conceptual Design Report (CDR) for a Neutrino Factory
could be written. The CDR would document a complete and fully
engineered design for the facility, including a detailed bottom-up
cost estimate for all components. This document would form the basis
for a full technical, cost, and schedule review of the construction
proposal. Construction could then commence after obtaining approval.

The R\&D issues for each of the major systems must be addressed by a
mix of theoretical calculation, simulation modeling, and experimental
studies, as appropriate. A list of the key physics and technology
issues for each major system is given:-

\bigskip

\textbf{Proton Driver}

\begin{itemize}
\item  Production of intense, short proton bunches, e.g., with space-charge
compensation and/or high-gradient, low frequency rf systems
\end{itemize}

\textbf{Target and Capture Section}

\begin{itemize}
\item  Optimization of target material (low-\textit{Z} or high-\textit{Z})
and form (solid, moving band, liquid-metal jet)

\item  Design and performance of a high-field solenoid ($\approx $20 T) in a
very high radiation environment
\end{itemize}

\textbf{Decay and Phase Rotation Section}

\begin{itemize}
\item  Development of high-gradient induction linac modules having an
internal superconducting solenoid channel

\item  Examination of alternative approaches, e.g., based upon combined rf
phase rotation and bunching systems or fixed-field, alternating gradient
(FFAG) rings
\end{itemize}

\textbf{Bunching and Matching Section}

\begin{itemize}
\item  Design of efficient and cost-effective bunching system
\end{itemize}

\textbf{Cooling Section}

\begin{itemize}
\item  Development and testing of high-gradient normal conducting rf (NCRF)
cavities at a frequency near 200 MHz

\item  Development and testing of efficient high-power rf sources at a
frequency near 200 MHz

\item  Development and testing of liquid hydrogen absorbers for muon cooling

\item Development and testing of an alternative gaseous-absorber cooling-channel 
design incorporating pressurized, high-gradient rf cavities.

\item  Development and testing of candidate diagnostics to measure emittance
and optimize cooling channel performance

\item  Design of beamline and test setup (e.g., detectors) needed for
demonstration of transverse emittance cooling

\item  Development of full six-dimensional analytical theory to guide the
design of the cooling section
\end{itemize}

\textbf{Acceleration Section}

\begin{itemize}
\item  Optimization of acceleration techniques to increase the energy of a
muon beam (with a large momentum spread) from a few GeV to a few tens of GeV
(e.g., recirculating linacs, rapid cycling synchrotrons, FFAG rings) for a
Neutrino Factory, or even higher energy for a Muon Collider

\item  Development of high-gradient superconducting rf (SCRF) cavities at
frequencies near 200 MHz, along with efficient power sources (about 10 MW
peak) to drive them

\item  Design and testing of components (rf cavities, magnets, diagnostics)
that will operate in the muon-decay radiation environment
\end{itemize}

\textbf{Storage Ring}

\begin{itemize}
\item  Design of large-aperture, well-shielded superconducting magnets that
will operate in the muon-decay radiation environment
\end{itemize}

\textbf{Collider}

\begin{itemize}
\item  Cooling of 6D emittance (\textit{x}, \textit{p}$_{x}$, \textit{y}, 
\textit{p}$_{y}$, \textit{t}, \textit{E}) by up to a factor of $10^{5}-10^{6}
$

\item  Design of a collider ring with very low $\beta ^{\ast }$ (a few mm)
at the interaction point having sufficient dynamic aperture to maintain
luminosity for about 500 turns

\item  Study of muon beam dynamics at large longitudinal space-charge
parameter and at high beam-beam tune shift
\end{itemize}

\textbf{Detector}

\begin{itemize}
\item  A study of cost trade-offs between increasing the detector mass compared with increasing the beam intensity. 

\item Simulation studies to define acceptable approaches for a Muon Collider detector operating in a high-background environment
\end{itemize}

Most of these issues are being actively pursued as part of the ongoing MC R\&D program. In a few areas, notably the proton driver and
detector, the MC does not currently engage in R\&D activities, though
independent efforts are under way. Longer-term activities, related primarily to the Muon Collider, are supported and encouraged as resources permit.

\subsection{Recent R\&D Accomplishments}

\subsubsection{Targetry}

A primary effort of the Targetry experiment E951 at BNL has been to carry out
initial beam tests of both a solid carbon target and a mercury target. Both
of these tests have been made with a beam intensity of about $ 4\times10^{12}$ ppp, with encouraging results.

In the case of the solid carbon target, it was found that a carbon-carbon
composite having nearly zero coefficient of thermal expansion is largely
immune to beam-induced pressure waves. Sublimation losses are a concern. However, a carbon target in a helium atmosphere is expected to have negligible sublimation loss. If radiation damage is the limiting effect for a carbon target, the predicted lifetime would be about 12 weeks when bombarded with a 1 MW proton beam. Carbon targets would therefore seem viable for beam powers up to 1~MW, or perhaps a little higher.

For a mercury jet target, tests with about 2 $\times $\ 10$^{12}$ ppp showed
that the jet is not dispersed until long after the beam pulse has passed
through the target. Measurements of the velocity of droplets emanating from
the jet as it is hit with the proton beam pulse from the AGS ($\thickapprox $%
10 m/s for 25 J/g energy deposition) compare favorably with simulation
estimates. High-speed photographs indicate that the beam disruption at the
present intensity does not propagate back upstream toward the jet nozzle. If
this remains true at the higher intensity of $1.6 \times 10^{13}$ ppp,
it will ease mechanical design issues for the nozzle. The tests so far have used a jet with an initial mercury velocity of 2~m/sec. At a Neutrino Factory a 20~m/sec jet is envisioned. A prototype is under development.

\subsubsection{MUCOOL}

The primary effort in the muon ionization cooling (MUCOOL) R\&D has been to complete the Lab G high power rf test area at Fermilab and begin high-power tests of 805-MHz rf cavities. A test solenoid for the facility, capable of operating either in solenoid mode (its two independent
coils powered in the same polarity) or gradient mode (with the two coils
opposed), was commissioned up to its design field of 5~T.

An 805 MHz open-cell cavity has been tested in Lab G to look at gradient
limitations, magnetic field effects and compatibility of the rf cavities with
other systems.  We have measured the dark currents over a range covering
14 orders of magnitude, and accumulated data on the momentum spectrum, angular distribution, pulse shape, dependence on conditioning and dependence on magnetic fields~\cite{DarkCurrentnote}.
The dark currents seem to be described by the Fowler Nordheim field emission process, which results from very small emitter sources (sub-micron sizes) at very high local electric fields (5 - 8 GV/m).  This implies that the emitter fields are enhanced by large factors,  $\beta_{FN} = \sim 500$, over the accelerating field.  (At these electric fields the electrostatic stress
becomes comparable to the strength of hardened copper.)  We have shown how both
normal conditioning and nitrogen processing can reduce dark currents.  Our data
from the 805 MHz cavity has been compared with other data from NLC cavities,
superconducting TESLA cavities and 200 MHz proton linacs, showing that all
cavities seem to be affected by the same processes.

We have also looked at damage produced on irises and windows, primarily when the
system is run with the solenoid magnet on.  A number of effects are seen: copper
splatters on the inside of the thin Ti window, burn marks on the outside of the
window due to electron beamlets, and some craters, evidently produced by
breakdown on the irises.  The electron beamlets burned through the
windows twice.  We have measured the parameters of the beamlets produced from
individual emitters when the magnetic field is on, and we have seen ring beams,
presumably produced by E$\times$B 
drifts during the period when the electrons are
being accelerated.  The radius of the beamlets is found to be proportional to
E/B$^2$.

We are proceeding with an experimental program designed to minimize the dark
currents using surface treatment of the copper cavity.

A second cavity, a single-cell pillbox having foils to close the beam
iris, has been tuned to final frequency, shipped to Fermilab, and testing has begun at Lab G. This cavity will permit an assessment of the
behavior of the foils under rf heating and give indications about
multipactor effects. It will also be used to study the dark current effects
discussed above. An advantage of the pillbox cavity is that its windows can
be replaced with ones made from (or coated with) various materials and
cleaned or polished by various techniques.

Development of a prototype liquid hydrogen absorber is in progress. Several large diameter, thin (125 $\mu $m) aluminum windows have been successfully
fabricated by machining from solid disks. These have been pressure tested
with water and found to break at a pressure consistent with design
calculations. A new area, the MUCOOL\ Test Area (MTA), is being developed at
FNAL for testing the absorbers. The MTA, located at the end of the proton
linac, will be designed to eventually permit beam tests of components and
detectors with 400 MeV protons. It will also have access to 201-MHz
high-power rf amplifiers for testing of future full-sized 201-MHz cavities.

Initial plans for a cooling demonstration are well under way.
This topic is covered separately in more detail in Section~\ref{mice}.

A parallel cooling channel development effort based on the use
of gaseous hydrogen or helium energy-absorber has begun.  Muons
Inc.~\cite{muonsinc} 
has received a DOE Small Business Technology Transfer grant with
the Illinois Institute of Technology to develop cold,
pressurized high-gradient rf cavities for use in muon ionization
cooling.  These cavities will be filled with dense gas, which
suppresses high voltage breakdown by virtue of the Paschen
effect and also serves as the energy absorber.  A program of
development for this alternative approach to ionization cooling
is foreseen that starts with Lab G tests, evolves to an MTA
measurement program, and leads to the construction of a cooling
channel section suitable for tests in MICE.

\subsubsection{Feasibility Study-II}

The MC has participated  in a second Feasibility Study for a
Neutrino Factory, co-sponsored by BNL. The results of the study were 
encouraging (see Section III), indicating that a neutrino intensity of $1\times 10^{20}$ per 
Snowmass year per MW can be sent to a detector located
3000 km from the muon storage ring. It was also clearly demonstrated that a Neutrino Factory could be sited at either FNAL
or BNL. Hardware R\&D needed for such a facility was identified, and is a
major part of the program outlined here.

\subsubsection{Beam Simulations and Theory}

In addition to work on Study-II, our present effort has focused on
longitudinal dynamics~\cite{longdyn}. 
We are developing theoretical tools for understanding
the longitudinal aspects of cooling, with the goal of developing approaches to
6D cooling, i.e., ``emittance exchange.'' This is a crucial aspect for the
eventual development of a Muon Collider, and would also benefit a Neutrino
Factory. Improved designs for the phase rotation, bunching, and acceleration systems are also being explored, with an emphasis on preparing the way for a future design study in which the performance obtained by the study II design is maintained, but with a reduction in cost. 

\subsubsection{Other Component Development}

At present, the main effort in this area is aimed at development of a
high-gradient 201-MHz SCRF cavity. A test area of suitable
dimensions has been constructed at Cornell. In addition, a prototype cavity has
been fabricated for the Cornell group by our CERN colleagues. Mechanical
engineering studies of microphonics and Lorentz detuning issues are being
carried out.

\subsubsection{Collider R\&D}

Studies of possible hardware configurations to perform emittance exchange,
such as the compact ring proposed by Balbekov~\cite{balb1}, are now
getting under way. A ring cooler has the potential to cool in 6D phase space,
provided the beam can be injected into and extracted from it. 
A series of workshops have been held on the topics of emittance exchange and 
ring coolers that have helped further our understanding of both.

\section{International Muon Ionization Cooling Experiment}
\label{mice}

\subsection{Motivation}

Ionization cooling of minimum-ionizing muons is an important
ingredient in the performance of a Neutrino Factory. However, it has
not been demonstrated experimentally. We seek to carry out an
experimental demonstration of cooling in a muon beam. Towards this
goal, we have developed (in collaboration with a number of physicists
from Europe and Japan interested in neutrino factories) a conceptual
design for an International Muon Ionization Cooling Experiment
(MICE).A proposal for MICE  has recently  
been submitted to the Rutherford Appleton
Laboratory in England~\cite{mice-prop}. 

The aim of the proposed cooling experimental demonstration is
\begin{itemize} 
\item	to show that we can design, engineer and build a section of cooling
channel capable of giving the desired performance for a neutrino
factory;
\item	to place it in a beam and measure its performance, {\em i.e.},
experimentally validate our ability to simulate precisely the passage
of muons confined within a periodic lattice as they pass through
energy absorbers and rf cavities.
\end{itemize}
The experience gained from this experimental demonstration will
provide important input to the final design of a real cooling
channel. The successful operation of a section of a muon cooling
channel has been identified (most recently by the U.S. Muon Technical
Advisory Committee~\cite{MUTAC-report}) as a key step in demonstrating
the feasibility of a Neutrino Factory or Muon Collider.

\subsection{Principle of the experiment}

Fundamentally, in a muon cooling experiment one needs to measure,
before and after the cooling channel, the phase space distribution of 
a muon beam in six dimensions~\cite{cernmice}. Such a measurement must include
the incoming and outgoing beam intensities and must avoid biases due
to the decay of muons into electrons within the channel and due to
possible contamination of the incoming beam by non-muons~\cite{summers}. 
Two techniques have been considered: i) the multi-particle method, in
which emittance and number of particles in any given volume of phase
space are determined from the global properties of a bunch; and ii)
the single-particle method, in which the properties of each particle
are measured and a ``virtual bunch" formed off-line. The full
determination of the covariance matrix in six dimensions is a delicate
task in a multi-particle experiment, and the desired diagnostics would
have to be developed specifically for this purpose; moreover, a
high-intensity muon beam bunched at an appropriate frequency would
need to be designed and built. For these reasons, the single particle
method is preferred. The single-particle approach, typical of
particle-physics experiments, is one for which experimental methods
already exist and suitable beams are already available.

\begin{figure}
\centerline{{\includegraphics[width=\linewidth]{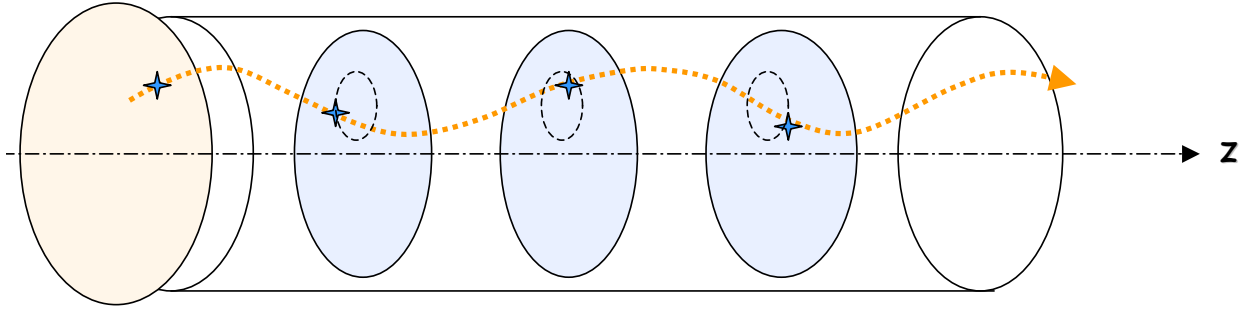}}}
\caption{(Color)Conceptual layout of MICE upstream spectrometer: 
following an initial time-of-flight (TOF) measurement, muons are 
tracked using detector planes located within a solenoidal magnetic field.
 Although in principle three $x,y$ measurements as shown suffice to 
determine the parameters of each muon's helical trajectory, in practice 
additional measurement redundancy will be employed; for example, a 
fourth measurement plane can be used to eliminate very-low-momentum
 muons that would execute multiple cycles of helical motion. 
A similar spectrometer (but with the time-of-flight measurement 
at the end) will be used downstream of the cooling apparatus.}
\label{fig:MICE-measurement}
\end{figure}

In the particle-by-particle approach, the properties of each particle
are measured in magnetic spectrometers before and after the cooling
channel (Figure~\ref{fig:MICE-measurement}). Each spectrometer measures,
at given $z$ positions, the coordinates $x, y$ of every incident
particle, as well as the time. Momentum and angles are reconstructed
by using more than one plane of measurement. For the experimental
errors not to affect the measurement of the emittance by a significant
factor, the rms resolution of the measurements must be smaller than
typically 1/10th of the rms equilibrium beam size in each of the six
dimensions~\cite{Blondel-cooling}.

\subsection{Conceptual design}

\begin{figure}
\centerline{\scalebox{0.6}{\includegraphics{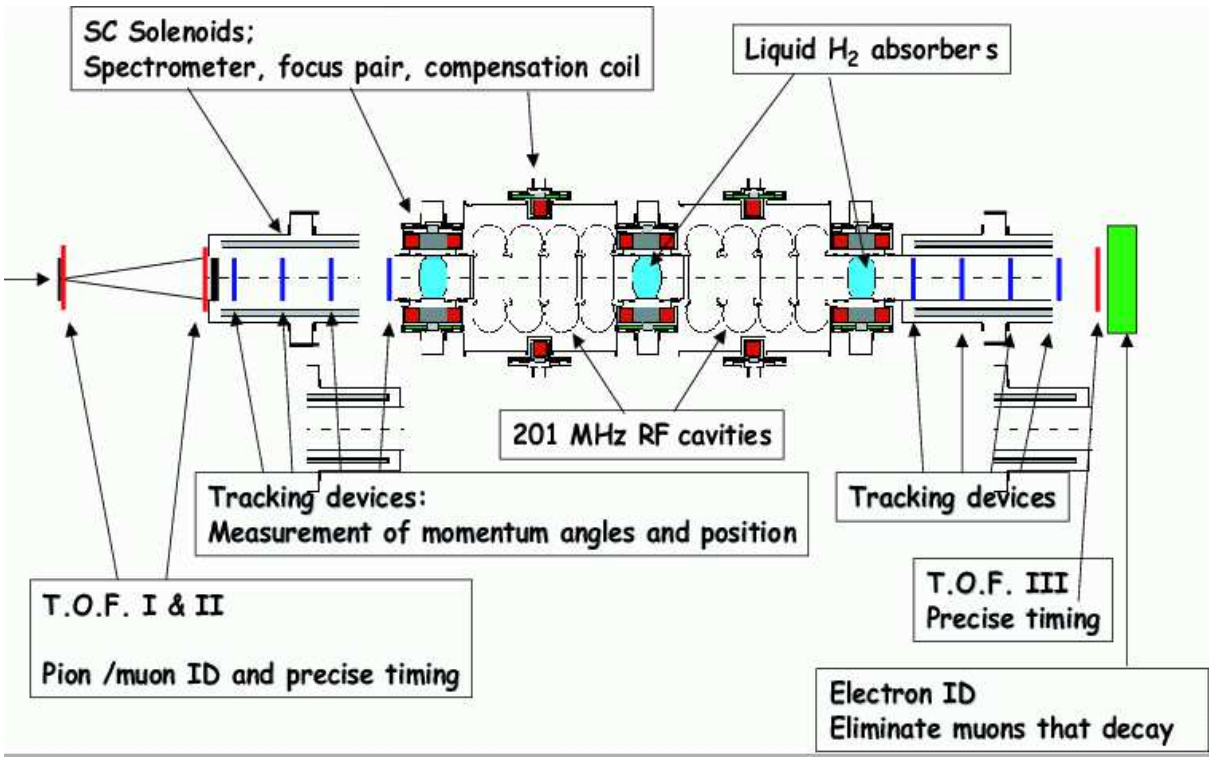}}}
\caption{(Color)Schematic layout of MICE apparatus.}
\label{fig:MICE}
\end{figure}

Figure~\ref{fig:MICE} shows the layout under consideration for MICE,
which is based on two cells of the Feasibility Study II ``Lattice 1"
cooling channel. The incoming muon beam encounters first a beam
preparation section, where the appropriate input emittance is
generated by a pair of high-$Z$ (lead) absorbers. In addition, a
precise time measurement is performed and the incident particles are
identified as muons. There follows a first measurement section, in
which the momenta, positions, and angles of the incoming particles are
measured by means of tracking devices located within a uniform-field
solenoid. Then comes the cooling section itself, with hydrogen
absorbers and 201 MHz rf cavities, the lattice optics being provided
by a series of superconducting coils; the pairs of coils surrounding
each absorber have opposite magnetic fields (``bucking" solenoids),
providing tight focusing. The momenta, positions, and angles of the
outgoing particles are measured within a second solenoid, equipped
with a tracking system identical to the first one. Finally, another
time-of-flight (TOF) measurement is performed together with particle
identification to eliminate those muons that have decayed within the
apparatus.

\subsection{Performance}

\begin{figure}
\centerline{\scalebox{0.7}{\includegraphics{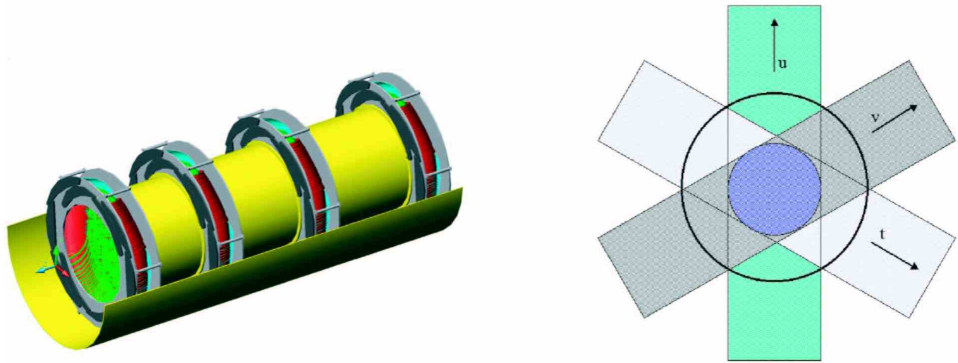}}}
\caption{(Color)A possible MICE tracking-detector configuration.}
\label{fig:MICE-tracking}
\end{figure}

\begin{figure}
\centerline{\scalebox{0.5}{\includegraphics{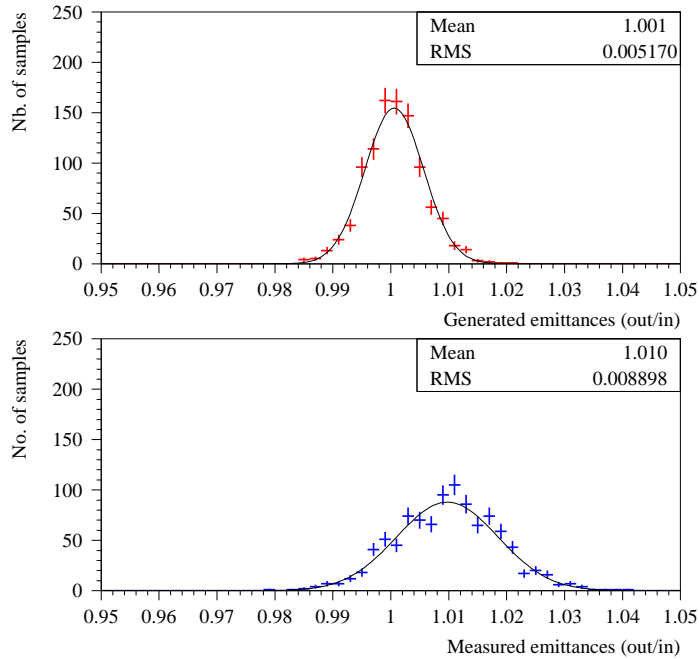}}}
\caption{(Color)Distribution of ratios of output to input six-dimensional 
emittance for 1000 simulated experiments, each with 1000 accepted
muons. The top figure shows the distribution of this ratio for the
emittances as generated by simulation; the bottom figure, as
``measured" in the simulated experiments. The curves are Gaussian fits
to the points.}
\label{fig:MICE-resolution}
\end{figure}

Simulations of MICE have been carried out for a configuration
including four tracking stations per spectrometer, each station
consisting of three crossed planes of 500-micron-thick
square-cross-section scintillating fibers
(Figure~\ref{fig:MICE-tracking}), immersed in a 5\,T solenoidal
field. Time of flight is assumed to be measured to 70\,ps rms. As
shown in Figure~\ref{fig:MICE-resolution}, measurement resolution and
multiple scattering of the muons in the detector material introduce a
correctable bias in the measured emittance ratio of only 1\%. (For
this study the effect of the cooling apparatus was ``turned off" so as
to isolate the effect of the spectrometers.)

\begin{figure}[htb!]
\begin{minipage}[b]{0.46\linewidth}
%emit-trans-vs-s.ps
\centering\includegraphics[width=\linewidth]{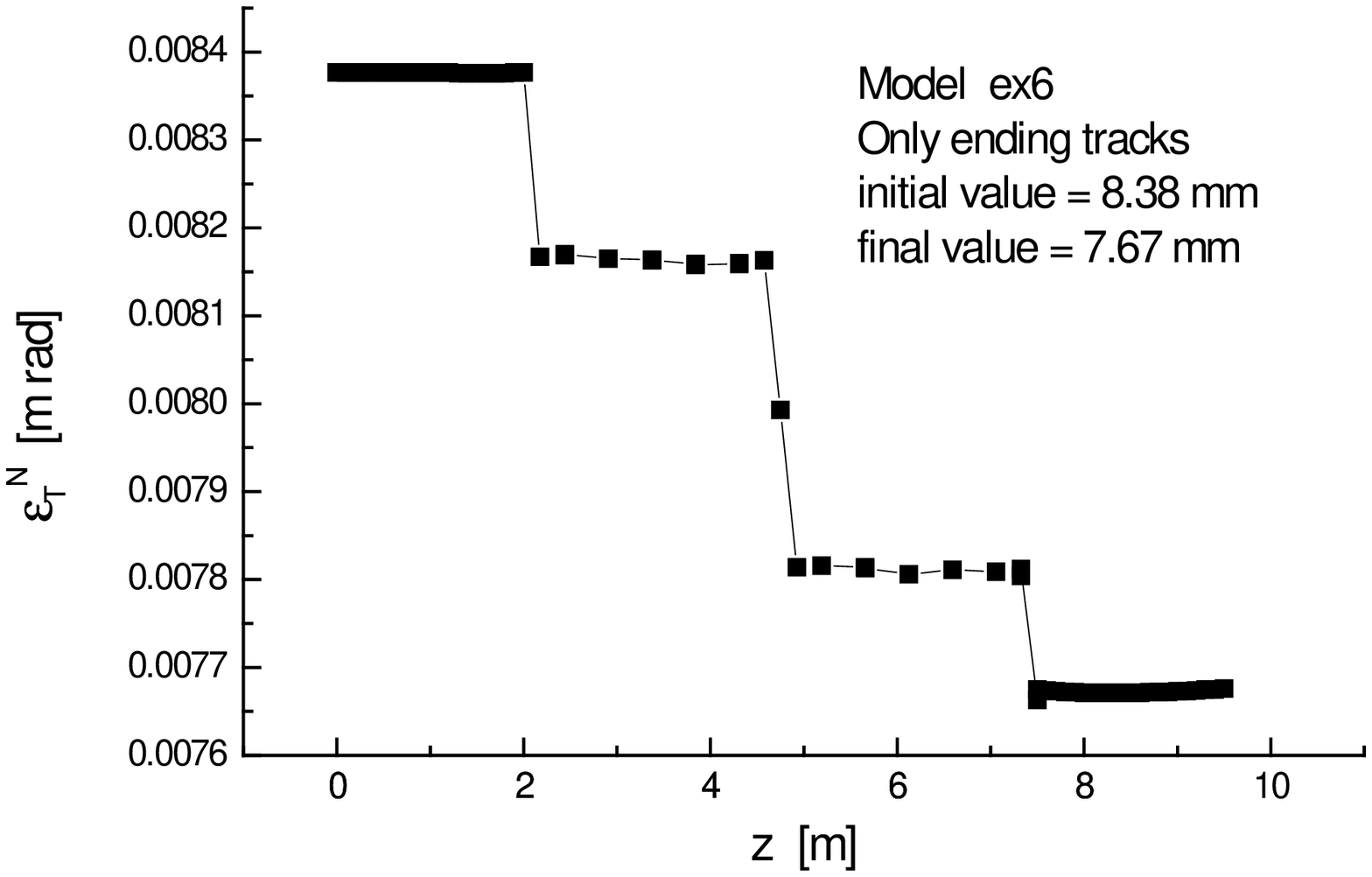}
\end{minipage}\hfill
\begin{minipage}[b]{0.46\linewidth}
%emit-long-vs-s.ps
\centering\includegraphics[width=\linewidth]{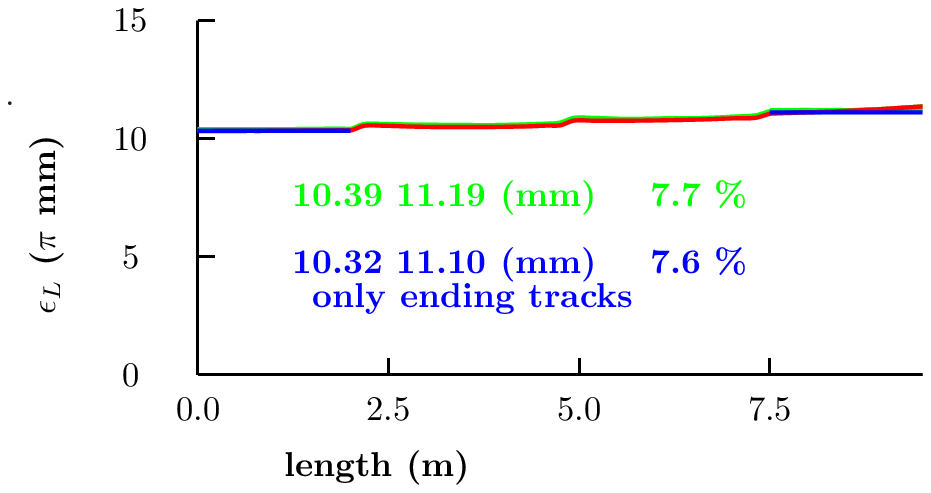}
\end{minipage}\hfill
\caption{(Color)Results from ICOOL simulation of MICE: normalized transverse
(left) and longitudinal (right) emittances {\em vs.}\ distance.}
\label{fig:MICE-sim1}
\end{figure}

Figure~\ref{fig:MICE-sim1} illustrates the muon-cooling performance of
the proposed MICE cooling apparatus. The normalized transverse
emittance of the incoming muon beam is reduced by about 8\%. The
longitudinal emittance increases by about the same amount, thus the
net cooling in six dimensions is also about 8\%. These are large
enough effects to be straightforwardly measured by the proposed
spectrometers.

\begin{figure}
\centerline{{\includegraphics[width=0.6\linewidth]{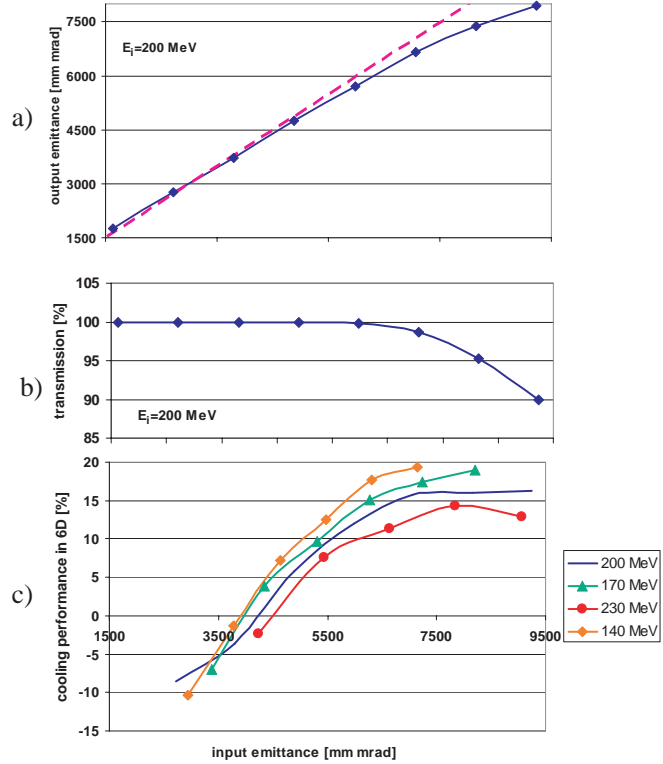}}}
\caption{(Color)Simulation results for 88-MHz variant of MICE apparatus: 
a) output emittance {\it vs.}\ input emittance, with 45$^\circ$ line (dashes) 
superimposed; b) beam transmission  {\it vs.}\ input emittance; c) 
cooling performance (see text) {\it vs.}\ input emittance for
 various beam kinetic energies (top to bottom: 140, 170, 200, 230 MeV).}
\label{fig:MICE-sim2}
\end{figure}

The CERN Neutrino Factory Working Group has studied a variant of the
proposed MICE cooling apparatus, in which 88-MHz rf cavities are
employed in place of the 201-MHz devices (the 88- and 201-MHz designs
have similar cooling
performance)~\cite{Hanke-cooling}. Figure~\ref{fig:MICE-sim2} (from the
CERN study) elucidates further experimental issues. As shown in
Figure~\ref{fig:MICE-sim2}a, for input emittance above the equilibrium
emittance of the channel (here about 3500\,mm$\cdot$mrad), the beam is
cooled, while for input emittance below equilibrium it is heated (and,
of course, for an input beam at the equilibrium emittance, the output
emittance equals the input emittance).  Figure~\ref{fig:MICE-sim2}b
illustrates the acceptance cutoff of the cooling-channel lattice; for
input emittance above 6000\,mm$\cdot$mrad, the transmission probability
falls below 100\% due to scraping of the
beam. Figure~\ref{fig:MICE-sim2}c shows the effect of varying the beam
momentum: cooling performance improves as the momentum is
lowered~\footnote{Despite the increased cooling efficiency at low muon
momentum, simulations of an entire muon production section and cooling
channel suggest that momenta near the ionization minimum represent the
global optimum for Neutrino Factory performance.}, as quantified here
in terms of the fractional increase in the number of muons within the
phase-space volume accepted by a hypothetical acceleration section
downstream of the cooling channel. The goal of MICE includes
verification of these effects in detail in order to show that the
performance of the cooling apparatus is well understood. Subsequent
running could include tests of additional transverse cooling cells,
alternative designs, or emittance exchange cells.

One critical aspect of this experiment is operation in the presence of
backgrounds due to dark currents from the rf cavities.  While it is possible to
operate the experiment using comparatively low rf gradients, it would be highly
desirable to produce cavities which would yield less dark current at higher
gradients.  This would permit more efficient use of the rf cavities and power
supplies.   We are trying to develop cavities with low dark currents.

%\input conclusions
%% new section from Mike, received Jan.9

\section{Conclusions}
\label{Summary}

In summary, the Muon Collaboration is developing the knowledge and ability
to create, manipulate, and accelerate muon beams. Our R\&D program will
position the HEP community such that, when it requires a Neutrino Factory or
a Muon Collider, we shall be in a position to provide it. A staged plan for
the deployment of a Neutrino Factory has been developed that provides an
active neutrino and muon physics program at each stage. The requisite R\&D
program is diversified over laboratories and universities and has
international participation.

The very fortuitous situation of having  intermediate steps along this path,
that offer a powerful and exciting physics program in their own right,
presents an ideal scientific opportunity, and it is hoped that the particle
physics community will be able to take advantage of it.


\begin{thebibliography}{100}
\bibitem{study2}Some of the text quoted here appears in the Study II document
\cite{EPP:studyii}. We would like to thank Robert Shrock for helpful comments.
\end{thebibliography}
\end{document}